\documentclass{elsart}
 \usepackage{graphicx}
\usepackage{amssymb}
\journal{Physica D}

\newcommand{\ii}{\mathrm{i}}
\newcommand{\ie}{i.e.}  \newcommand{\eg}{e.g.}
\begin{document}
\begin{frontmatter}
\title {Standing wave instabilities in a chain of nonlinear coupled 
oscillators}

\author[Saclay]{Anna Maria Morgante},
\corauth[cor]{Corresponding author. Present address: Department of Physics and 
Measurement Technology, Link\"oping University, S-581 83  Link\"oping, Sweden.}
\ead{morgante@llb.saclay.cea.fr}
\author[Saclay]{Magnus Johansson\corauthref{cor}},
\ead{mjn@ifm.liu.se}
\author[Saclay,Crete]{Georgios Kopidakis},
\ead{kopidaki@llb.saclay.cea.fr}
\author[Saclay]{Serge Aubry}
\ead{aubry@llb.saclay.cea.fr}

\address[Saclay]{Laboratoire L\'eon Brillouin (CEA-CNRS), CEA Saclay, 
F-91191 Gif-sur-Yvette Cedex, France}
\address[Crete]{Department of Physics, University of Crete, P.O. Box 2208, 
GR-71003 Heraklion, Crete, Greece}

\begin{abstract}
We consider existence and stability properties of nonlinear spatially 
periodic or quasiperiodic standing waves (SWs) in one-dimensional 
lattices 
of coupled anharmonic oscillators. Specifically, we consider Klein-Gordon (KG)
chains with either soft (\eg, Morse) or hard (\eg, quartic) on-site 
potentials, as well as discrete nonlinear Schr\"odinger (DNLS) chains 
approximating the small-amplitude dynamics of KG chains with weak inter-site 
coupling. The SWs are constructed as exact time-periodic multibreather 
solutions from the anticontinuous limit of uncoupled oscillators. In the 
validity regime of the DNLS approximation these solutions can be continued 
into the linear phonon band, where they merge into standard harmonic SWs. 
For SWs with incommensurate wave vectors, this continuation is associated 
with an inverse transition by breaking of analyticity. When the DNLS 
approximation is not valid, the continuation may be interrupted by 
bifurcations associated with resonances with higher harmonics of the SW. 
Concerning the stability, we identify one class of SWs which are  
always linearly stable close to the anticontinuous limit. However, 
approaching the linear limit all SWs with nontrivial wave vectors become 
unstable through oscillatory instabilities, persisting for arbitrarily small 
amplitudes in infinite lattices. Investigating the dynamics resulting from 
these instabilities, we find two qualitatively different regimes for wave 
vectors smaller than or larger than $\pi/2$, respectively. In one regime 
persisting breathers are found, while in the other regime the system rapidly 
thermalizes. 
\end{abstract}

\end{frontmatter}

\section{Introduction}

It is well-known that the dynamics of spatially periodic harmonic
systems is non-ergodic, since it can be completely described as a linear
combination (with time-independent coefficients) of travelling waves  
of the form $\cos (Qx-\omega(Q) t)$
with wave vector $Q$ and frequency $\omega(Q)$. When there are (even small) 
anharmonic
terms present, most harmonic solutions with amplitude zero are 
very
unlikely to be continuable to finite amplitude,  because it is generally 
believed (despite
the famous work of Fermi, Pasta and Ulam \cite{FPU}) that anharmonic
terms in general introduce some ergodicity in the system, which in the
long term should lead to its thermalization.

It is nevertheless an interesting question to understand
whether special harmonic solutions could persist as exact solutions
when the system becomes anharmonic.
Actually, it has been proven recently that small-amplitude periodic 
travelling waves (with the form $\cos (Qx - \omega t)$ in the harmonic limit) 
do exist for Hamiltonian chains of linearly coupled nonlinear 
oscillators (\emph{Klein-Gordon (KG) lattices}) \cite{IK00}  as well as for 
chains of mass particles with 
nonlinear nearest-neighbour interactions (\emph{Fermi-Pasta-Ulam (FPU) 
lattices}) 
\cite{I00}. 
However, for KG lattices the unique continuation of harmonic 
travelling
waves is possible only in a regime of weak inter-site coupling 
(i.e., close enough to the so-called \textit{anticontinuous} 
limit \cite{Aub97} of uncoupled oscillators);
beyond this regime, the situation is more complex and there are many 
bifurcations where new solutions appear 
(such as \textit{nanopterons}, which are travelling localized waves
with a noncancelling periodic tail at infinity) \cite{IK00,Lom99}. 
In the simpler regime close to the anticontinuous limit,
numerical techniques for finding
time-periodic propagating waves, 
based on the concept of {\em discrete breathers} 
(see e.g. \cite{Aub97,FW98} for reviews of recent
developments on discrete breathers), were implemented in \cite{CA97}.
The solutions were constructed as multi-site breathers 
('\emph{multibreathers}') 
from the trivial solution at the anticontinuous limit where all oscillators 
are oscillating in phase with the same frequency. An arbitrary spatially 
uniform phase 
torsion was then applied, and a Newton scheme used to continue  these 
solutions to non-zero coupling according to the theory described in 
\cite{Aub97}.

However, as was found already in \cite{KP92} from a discrete nonlinear 
Schr\"odinger (DNLS) approximation (generally valid for small-amplitude 
oscillations and small 
coupling, see also Sec. 2.2 below), propagating waves in KG
lattices may exhibit a 
\emph{modulational (Benjamin-Feir) instability} by which an initially 
uniform wave breaks up into an array of localized pulses. Depending on whether 
the KG 
on-site potential is soft or hard (i.e. whether the oscillation frequency 
decreases or increases with increasing oscillation amplitude for the 
individual oscillators), the modulational instability occurs for either 
small or large wave vectors of the travelling wave. 
(A similar result has been found also for 
FPU lattices \cite{DRT97}.)  
This instability also appears a consequence of the theory of effective action 
for multibreathers with phase torsion  as 
developed in \cite{Aub97} (see also further developments in \cite{MK99}).

In this paper, we are interested in the continuation of
{\em standing waves (SWs)}, which in the harmonic limit are superpositions
of two travelling waves with the same amplitude and opposite wave
vectors, and thus have the form  \linebreak
$\cos (Qx)  \cos (\omega t)$. We shall here  study such solutions for  
discrete KG systems, and for the related DNLS system describing 
small-amplitude solutions close to
the anticontinuous limit. 
Our approach for finding these solutions
is similar to that used in \cite{KA99} for disordered KG lattices: 
we construct appropriate multibreather solutions with frequency outside
the linear phonon band at the anticontinuous limit and continue them to 
nonzero coupling;  these solutions are then continued versus frequency
inside the phonon band. We shall propose an ansatz for
finding
the appropriate {\em coding sequence} \cite{Aub92,Aub97} for these
 multibreathers, allowing 
for a  smooth continuation of  the multibreather versus frequency 
down to zero amplitude as a linear SW. When the  spatial period of the SW 
is incommensurate with the 
lattice period, its envelope (\emph{hull}) function  undergoes a 
{\em 'transition by breaking of analyticity' (TBA)} \cite{Aub78,Aub80,AD83} 
at some critical value of 
its frequency.

Investigating the linear stability of these nonlinear SWs in infinite 
lattices, 
we find the rather striking result that although there exist linearly stable 
SWs close to the anticontinuous limit, all SWs with nontrivial 
wave vectors (i.e. different from 0 or $\pi$) 
become  {\em linearly unstable} through an 
{\em oscillatory instability bifurcation} when approaching the 
linear limit. Thus, small-amplitude nonlinear  SWs will be unstable also for 
wave numbers where the corresponding 
propagating waves are modulationally stable. These extended instabilities 
are a direct consequence of discreteness, and have not been found 
within earlier  continuum approximations \cite{KHS94}. However, similarly as 
found e.g. in \cite{MA98,JK99}, the 
instabilities are very sensitive to finite size effects, so that in a finite 
system linear stability will generally be recovered close to the linear limit.

Moreover, we show that when their hull function is non-analytic, 
these SWs can be viewed as arrays of 
discommensurations (see, \eg, \cite{AGARQ}) of a simple stationary phonon 
with wave vector $\pi$
or $0$, and that each discommensuration is unstable (as was found already in
\cite{JK99} for the DNLS model). The effect of
the instability on  a single discommensuration is analysed
with the technique developed in \cite{CAT96,AC98}, and it is found that, 
after 
an initial oscillatory regime, it moves spontaneously in the lattice with a 
rather well-defined velocity, analogously to the behaviour of the discrete 
dark mode in the DNLS model found in \cite{KKC94,JK99}. The 
long-time behaviour of the unstable SWs can then be interpreted  as the 
result of inelastic 
interactions between individual moving discommensurations. As for the DNLS 
model the SW with wave vector $\pi/2$ is found to exactly coincide with the 
line of phase transition recently discovered \cite{RCKG00} in this model, SWs 
with wave vectors larger or smaller than $\pi/2$ belong to different phases 
and therefore yield qualitatively different asymptotic dynamics: one phase 
corresponds to normal thermalization, while in the other phase persisting 
large-amplitude localized excitations occur. As we will find from 
numerical simulations, a similar scenario also occurs for KG chains with 
small inter-site coupling. 

The layout of this paper is as follows. In Section 2, we describe the KG model 
and derive a DNLS-type equation for its small-amplitude dynamics. This 
equation generally contains long-range nonlinear interactions, but reduces to
 the ordinary nearest-neighbour DNLS equation in the limit of small coupling. 
In Section 3, we describe the representation of nonlinear SWs as trajectories 
of a symplectic map, which in the DNLS approximation reduces to a 
cubic map in the 2-dimensional real plane \cite{Wan}. The breakup of 
Kol'mogorov-Arnol'd-Moser (KAM) tori for this map corresponds to a TBA for the 
hull function of incommensurate SWs, and the continuation of periodic and 
quasiperiodic map trajectories to the anticontinuous limit yields the SW 
multibreather coding sequences. In Section 4, we study numerically and 
analytically the linear stability 
properties of SWs within the DNLS approximation, where in particular the 
existence of oscillatory instabilities close to the linear limit can be proven 
using perturbation theory. In Section 5, SWs in the full KG model are analyzed 
numerically for two particular examples of soft (Morse) and hard (quartic) 
on-site potentials. Their linear stability properties are investigated using 
numerical Floquet analysis, and bifurcations determining the limit of 
existence for larger coupling are identified. In Section 6, we discuss the 
dynamics resulting from the SW instabilities over short, intermediate, and 
long time scales, and Section 7 concludes the paper. The more technical parts 
of the stability analysis for the DNLS case is deferred to the Appendix. 
Some of the results obtained in this paper 
have been presented in an abbreviated form elsewhere \cite{MJKA00}.

\section{The model}

\subsection{The Klein-Gordon system}

\begin{figure}[htbp]
\begin{center}
\quad
{\includegraphics[height=3cm]{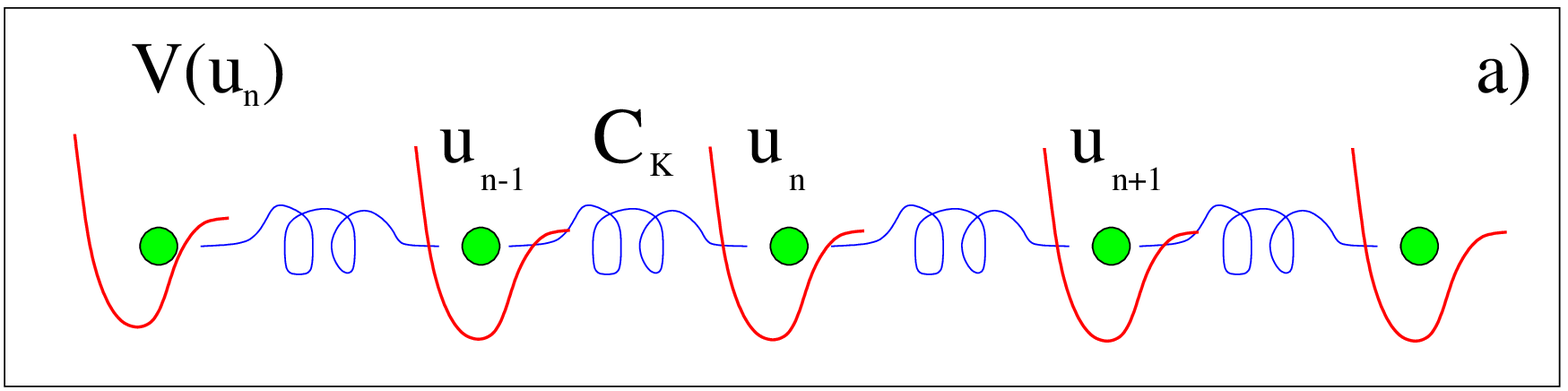}}\\ 
\vspace{.3cm}
    {\includegraphics[width=3cm]{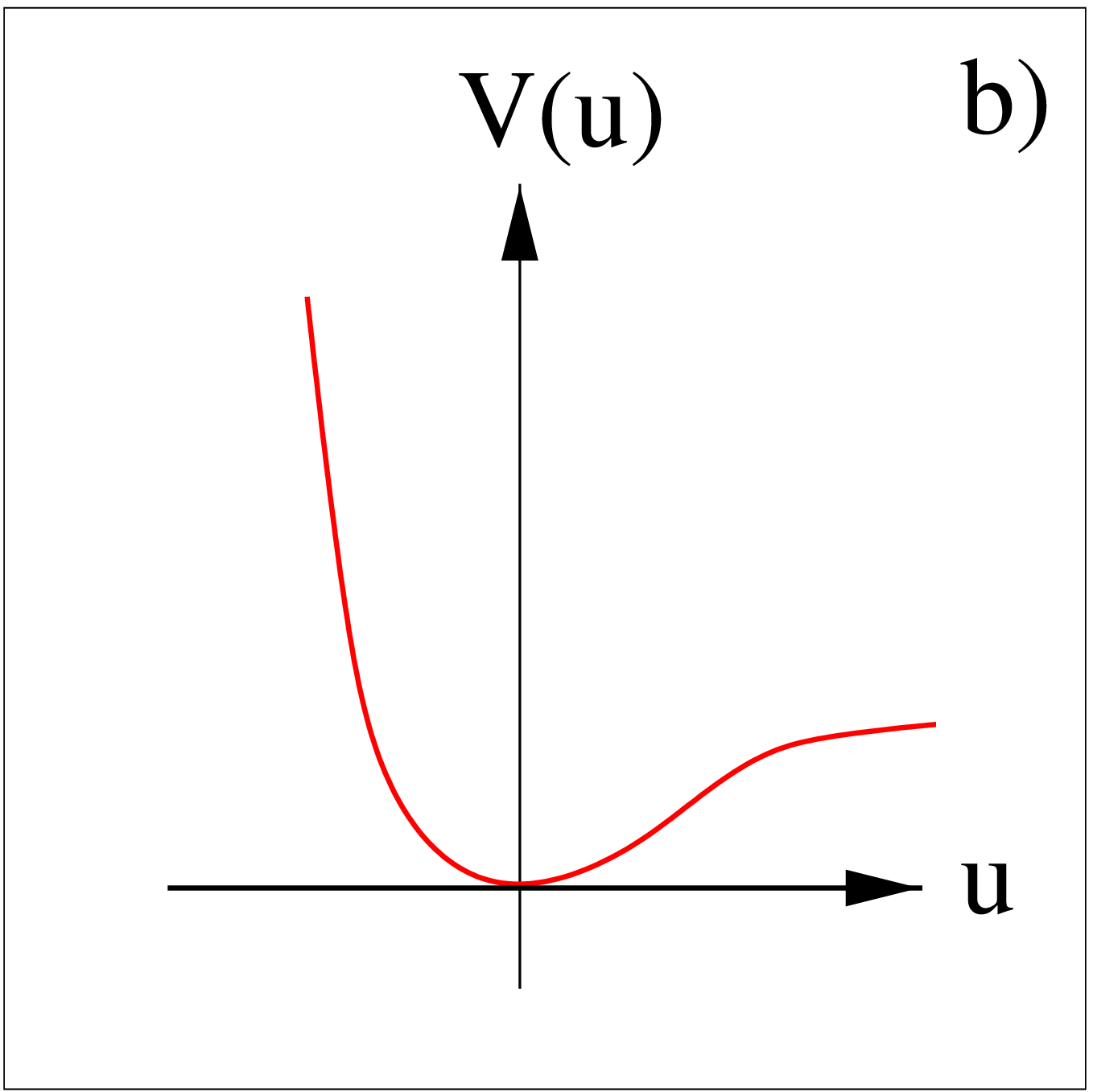}}\label{pMorse}\quad
    {\includegraphics[height=3cm]{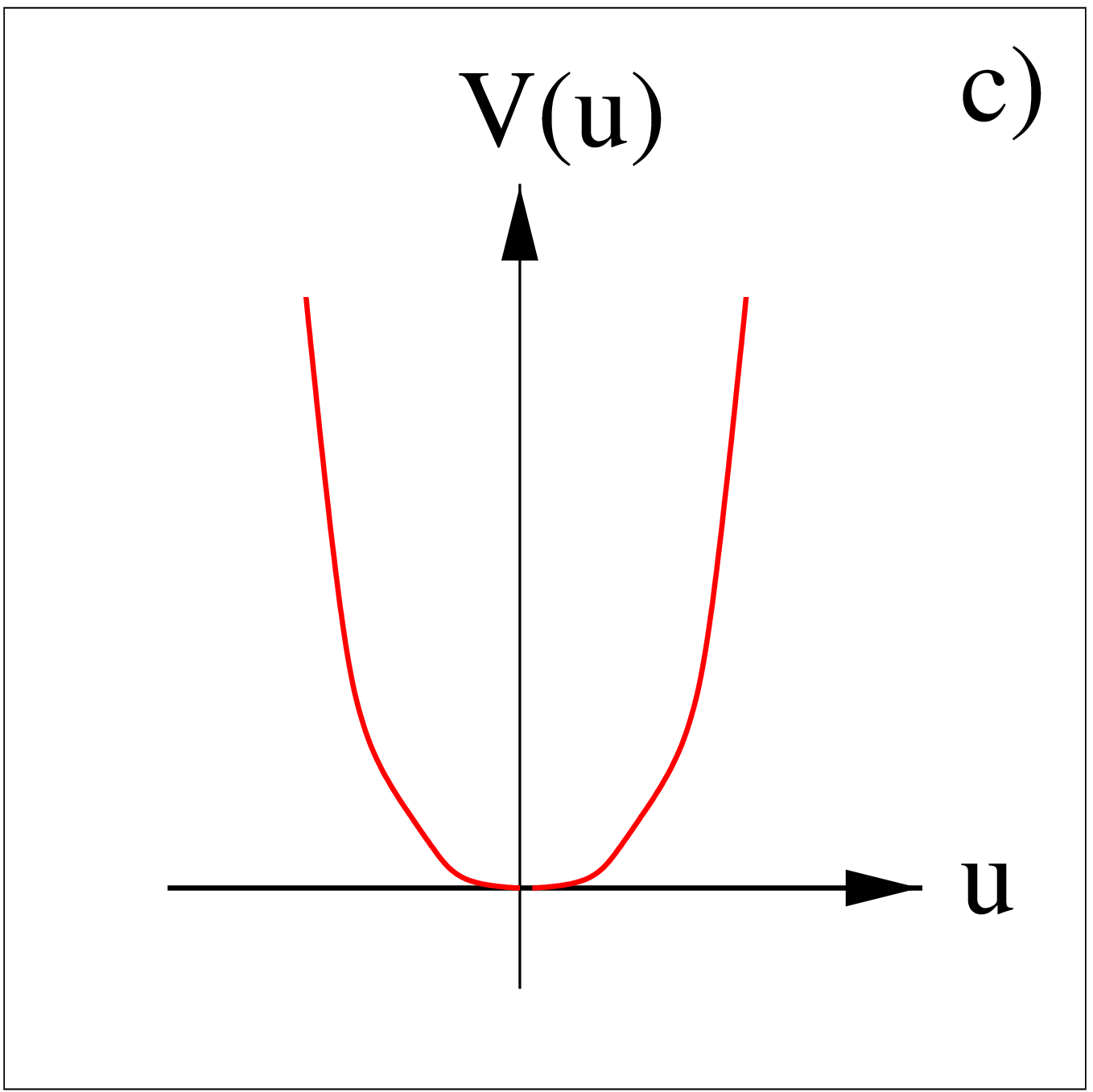}}\label{pquar}
\caption{  (a) Klein-Gordon chain of nonlinear oscillators with displacement 
$\{u\}$, coupled by a constant force $C_K$; (b) the soft Morse potential;  
(c) the hard quartic potential.} 
\label{fig_pot}
\end{center}
\end{figure}

We consider a one-dimensional chain of classical anharmonic oscillators 
(Klein-Gordon chain) (Fig. \ref{fig_pot}a) with Hamiltonian

\begin{equation}
        H= \sum_{n=1}^N \left[ \frac{1}{2} \dot{u}^2_n + V(u_{n})+
        \frac{1}{2} C_K (u_{n+1}-u_{n})^2 \right ] \equiv \sum_{n=1}^N H_n,  
        \label{hamilt1}
\end{equation}

where the total energy attributed to each site $n$ is defined as
\begin{equation}
H_n\equiv\frac{1}{2} \dot{u}^2_n + V(u_{n})+
        \frac{1}{4} C_K \left( (u_{n+1}-u_{n})^2 + (u_{n-1}-u_{n})^2\right) . 
\label{energy}
\end{equation}
The oscillators are submitted to a nonlinear on-site  potential $V(u)$ and 
are coupled by a harmonic inter-site coupling $C_{K}>0$. The nonlinear 
potential can be either a soft potential such as the Morse potential 
(Fig. \ref{fig_pot}b)
\begin{equation}
        V(u) = \frac{1}{2} \omega_0^2 (\e^{-u}-1)^2
        \label{morse}
\end{equation}
or a hard potential such as the quartic potential (Fig. \ref{fig_pot}c)
\begin{equation}
        V(u) = \omega_0^2 (\frac{u^2}{2} + \frac{u^{4}}{4}).
        \label{quartic}
\end{equation}
(Recall that for a soft [hard] potential the oscillation frequency decreases 
[increases] with the oscillation amplitude or, equivalently, with the 
action.) In the rest of the paper we will consider $V''(0)=\omega_0^2=1$.

The choice of the Morse potential is physically justified by the fact that it 
represents an appropriate description of hydrogen bonds, and it is then 
especially useful for the study of biological systems such as macromolecules 
and proteins.
   
The equations of motion for the particle displacements $u_n$ then take the 
form of  a discrete nonlinear KG equation, 
\begin{equation}
\label{DKG}
\ddot{u}_n + V'(u_n)  - C_K(u_{n+1} + u_{n-1} -2 u_n) = 0 .
\end{equation}
For a given solution $u_n(t)$, the corresponding linearized equations (Hill 
equations) are
\begin{equation}
\ddot{\epsilon}_n+\epsilon_{n} V''(u_n) -C_K(\epsilon_{n+1}+\epsilon_{n-1}
-2\epsilon_{n}) = 0,
\label{phonon}
\end{equation}
where $\epsilon_n$ denotes a small displacement with respect to the solution 
$u_n$. When $u_n=0$ (or, more generally, when $V''(u_n)$ is constant), 
the standard harmonic (plane-wave) small-amplitude solutions 
\begin{equation}
\label{linearW}
\epsilon_{n}(t) = \epsilon \cos {(Q n - \omega_l(Q) t)}
\end{equation}
 with wave vector $Q$ and frequency $ \omega_l(Q)$ are obtained.
By substitution of these solutions in  Eq.\ (\ref{phonon}), we find 
the (phonon) dispersion relation
\begin{equation}
\label{lin_disp}
\omega_l^2(Q) = 1 + 4 C_K \sin^{2} Q/2 
\end{equation}
for small-amplitude oscillations in a 
general potential.

\subsection{Small amplitude limit and DNLS equation}\label{sec_SAL}

We will now show how it is possible to describe the small-amplitude dynamics 
of general KG chains with small inter-site coupling $C_K$ by a DNLS 
approximation.
 Although similar results have been obtained earlier \cite{KP92,KHS94}, our 
approach is slightly different and offers a deeper understanding for the role 
of the different limits involved. Considering  small-amplitude solutions
to the nonlinear discrete KG system (\ref{DKG}), 
the anharmonic terms will be small, and the on-site anharmonic potential 
$V(u)$ can be expanded at the lowest significant order as
\begin{equation}
        V(u) = \frac{1}{2} u^{2} + \alpha \frac{u^{3}}{3} +  \beta
        \frac{u^{4}}{4} +\ldots
        \label{exppot}
\end{equation}
For example, in the case of the Morse potential, the expansion yields 
$\alpha=-3/2$ and $\beta= 7/6$, while in the case of the quartic potential 
$\alpha=0$ and $\beta=1$.

We then search for solutions of the nonlinear equation (\ref{DKG}) which are 
locally close to linearized phonons (\ref{linearW}) with wave vector $Q$ and 
small amplitude $\epsilon$.  If such a solution $u_n(t)$ is assumed to be 
time periodic with 
 period $T_b=\frac{2\pi}{\omega_b}$, where  $\omega_b$ is close to some 
phonon frequency $\omega_l(Q)$, it can be
expanded as a Fourier series, 
\begin{equation}
        u_{n}(t)=\sum_{p} a_{n}^{(p)} \e^{\ii p \omega_{b}t},
        \label{fseries}
\end{equation}
 where the complex coefficients
fulfill $a_{n}^{(p)}=a_{n}^{(-p) \ast }$. 
For a time-reversible solution, the coefficients $a_{n}^{(p)}$ will be  real.
Moreover, we require the Fourier series to be summable as a
smooth (say analytic) function of $t$  and thus
that the coefficients $a_n^{(p)}$ must decay asymptotically exponentially 
with $p$ for each $n$.

More generally, since for the linear solution the left-hand side of 
Eq.\ (\ref{DKG}) is not strictly vanishing but of order $\epsilon^2$, the 
perturbed solution will have a harmonic time dependence plus  corrections of 
order $\epsilon^2$. Thus, as in Refs.\ \cite{KP92,KHS94}, we search for 
general 
(not necessarily  time-periodic) 
small-amplitude solutions 
as series with the form (\ref{fseries}) but where the Fourier coefficients 
depend  slowly on time  as a function of $\epsilon^{2}t$.
Then, in the Fourier series (\ref{fseries}), $a_{n}^{(1) \ast}=a_{n}^{(-1)}$ 
are the leading order terms of order
$\epsilon$,  $a_{n}^{(0)}$ and
$a_{n}^{(2) \ast}=a_{n}^{(-2)}$ are smaller at order $\epsilon^{2}$, and, 
since the Fourier components must decay exponentially, it is natural to 
assume that $a_{n}^{(3) \ast}=a_{n}^{(-3)}$ are of order $\epsilon^{3}$,  
$a_{n}^{(4) \ast}=a_{n}^{(-4)}$ are of order $\epsilon^{4}$,  etc.
Then solving Eq.\ (\ref{DKG}) at order $\epsilon^{3}$ 
yields\footnote{Order $\epsilon^{3}$ is the lowest order taking into account
the anharmonicity of the local potential $V(u)$.}, with 
  $\omega_b^{2}-1 \equiv 2 \delta $, for each harmonic
\begin{eqnarray}
a_{n}^{(0)} +2 \alpha
|a_{n}^{(1)}|^{2}-C_K(a_{n+1}^{(0)}+a_{n-1}^{(0)}-2a_{n}^{(0)})
&=& 0\label{eqp=0} \\
2 i \omega_b \dot{a}_{n}^{(1)} - 2 \delta a_{n}^{(1)}+ 2 \alpha (a_{n}^{(1)}
a_{n}^{(0)}+ a_{n}^{(1)~*} a_{n}^{(2)})+
3 \beta |a_{n}^{(1)}|^{2} a_{n}^{(1)}&& \nonumber \\
- C_K(a_{n+1}^{(1)}+a_{n-1}^{(1)}-2 a_{n}^{(1)}) &=& 0
        \label{eqp=1} \\
(1-4 \omega_b^2) a_{n}^{(2)}+\alpha
a_{n}^{(1)~2}-C_K(a_{n+1}^{(2)}+a_{n-1}^{(2)}-2a_{n}^{(2)}) &=& 0
\label{eqp=2}\\
(1-9 \omega_b^2) a_{n}^{(3)} + 2\alpha a_{n}^{(1)}a_{n}^{(2)} + \beta
 a_{n}^{(1)~3} -C_K(a_{n+1}^{(3)}+a_{n-1}^{(3)}-2a_{n}^{(3)}) &=& 0. 
\label{eqp=3}
\end{eqnarray}

The term $a_{n}^{(0)}$ can be obtained explicitly from Eq.\ (\ref{eqp=0}) 
which yields
\begin{eqnarray}
        a_{n}^{(0)}& = & A \sum_{m} \eta^{|n-m|} |a_{m}^{(1)}|^{2} \quad
\mbox{with}
        \label{elim0} \\
        \eta &=& 1-\frac{1}{2C_K} (\sqrt{1+4C_K}-1) \quad \mbox{and} \quad
        A= - \frac{2\alpha}{\sqrt{1+4C_K}} . 
        \label{etaA}
\end{eqnarray}

Similarly, $a_{n}^{(2)}$ can be obtained explicitly from Eq.\ (\ref{eqp=2})
but only when $2\omega_b > \sqrt{1+4C_K} = \omega_l(\pi)$,
which means  that the second harmonic of the solution does not belong to the 
linear phonon band:
\begin{eqnarray}
                a_{n}^{(2)}& = & A^{\prime} \sum_{m} \eta^{\prime~|n-m|}
a_{m}^{(1)~2}
         \quad \mbox{with} \label{elim2} \\
        \eta^{\prime} &=&- 1+\frac{4 \omega_b^{2}-1-4C_K}{2C_K}
        (\sqrt{1+\frac{4C_K}{4 \omega_b^{2}-1-4C_K}}-1)
 \quad \mbox{and} \nonumber \\
        A^{\prime}&=&  \frac{\alpha}{\sqrt{(4 \omega_b^{2}-1-4C_K)(4
\omega_b^{2}-1)}} . 
        \label{etaAprime}
\end{eqnarray}
The substitution of Eqs.\ (\ref {elim0}) and (\ref {elim2}) in 
Eq.\ (\ref{eqp=1}) yields a time-dependent extended
DNLS equation with long range nonlinear interactions describing  the slow time 
evolution of the dominating Fourier 
coefficients $a_{n}^{(1)}$, where the nonlinearities are taken into account 
to leading 
order in the amplitude $\epsilon$. If the condition 
$2\omega_b > \sqrt{1+4C_K}$ is not fulfilled, $|\eta^{\prime}|=1$ 
($\eta^{\prime}$ complex, on the unit circle) and the sum (\ref {elim2}) 
defining the interaction 
range in general diverges.  
Thus, for a general potential with non-zero $\alpha$, this condition imposes 
an upper validity limit for the  coupling strength $C_K$ for this approach. 
(In the linear limit $\omega_b=\omega_l(Q)$, the condition is fulfilled for 
all $Q$ only when $C_K<3/4$.)

In  the particular case when  $\alpha = 0$ (e.g. the quartic potential), there 
are no long-range terms in Eq. (\ref{eqp=1}),\footnote{When $\alpha=0$, 
coupling to higher harmonics appears only to order
 $\epsilon^5$ and higher. In particular, for the quartic potential only odd 
harmonics couple to each other.} which immediately reduces to the 
ordinary DNLS equation with only nearest-neighbour interactions 
(Eq.\ (\ref{DNLS}) below). For a general potential, we 
consider the case of small $C_K$ (i.e. close to the anticontinuous limit 
 $C_K=0$). Then,  $\omega_b^{2}-1 = 2 \delta $ will be small of order of 
$C_K$, and $\eta = C_K + {\mathcal O} (C_K^2)$,
$\eta^{\prime} = - \frac{C_K}{4 \omega_b^2 -1} + {\mathcal O} (C_K^2)$. Considering the 
modified DNLS equation (\ref{eqp=1}) at order $C_K$ and neglecting terms of 
order $\epsilon^3  C_K$, the long-range terms disappear and it reduces to 
the standard DNLS equation which well describes the dynamics of 
small-amplitude oscillations in KG chains close to the anti-continuous limit: 
\begin{equation}
 i \dot{\psi}_{n}=\delta \psi_{n}-\sigma |\psi_{n}|^{2}
\psi_{n}+C(\psi_{n+1}+\psi_{n-1}-2 \psi_{n}) , 
 	\label{DNLS}
 \end{equation}

  where
$\psi_n=\sqrt{|\lambda|} a_n^{(1)}$, 
$\lambda$ $=$ $-5\alpha^2/3$ + $3\beta/2$, $\sigma=$ sign$(\lambda)$ and 
$C=C_K/2$. Substituting the corresponding values of the parameters $\alpha$ 
and $\beta$, we obtain that for soft potentials $\lambda < 0$ and then 
$\sigma=-1$, while for hard potentials $\lambda > 0$ and $\sigma=1$. Actually 
both cases $\sigma = \pm 1$ are formally equivalent at this
order of the expansion, since by changing
$\psi_{n}$ into $(-1)^{n} \psi^{*}_{n}$ we get a new equation for 
$\{\psi_{n}\}$
which has the same form as Eq.\ (\ref{DNLS}) but
with $\delta$  replaced by $(4C - \delta)$, $\sigma$ by $-\sigma$
and $C$ is unchanged. So within the DNLS approximation we can assume for 
convenience $\sigma=-1$, 
corresponding to a soft potential, without loss of generality. 

\section{Standing waves}
\subsection{Standing waves represented as map trajectories}\label{sect_SW}

SWs in the nonlinear KG model (\ref{DKG}) can be searched as time-reversible
and time-periodic solutions $\{u_{n}(t)\}$ with frequency $\omega_{b}$ as in 
Eq.\ (\ref{fseries}).
The Fourier coefficients $w^{(p)}$ of the time-periodic function
$V^{\prime}(u(t))
= \sum_{p} w^{(p)} \e^{\ii p \omega_{b}t}$ define nonlinear functions
$\{w^{(p)}(\{a^{(q)}\})\}$
of the Fourier coefficients $\{a^{(q)}\}$ of $u(t)$. Then, Eq.\ (\ref{DKG})
can be written as
\begin{equation}
        -p^{2}\omega_{b}^{2} a_{n}^{(p)} + w^{(p)}(\{a_{n}^{(q)}\})
        -C_K (a_{n+1}^{(p)}+a_{n-1}^{(p)}-2a_{n}^{(p)})=0.
        \label{fcoefrel}
\end{equation}
This equation  defines a symplectic map
\begin{equation}
        (\{a_{n+1}^{(p)}\},\{a_{n}^{(p)}\})={\tilde{\mathcal
S}}(\{a_{n}^{(p)}\},\{a_{n-1}^{(p)}\}),
        \label{sympmap}
\end{equation}
where the SWs should appear as bounded trajectories. Since this map operates 
in an infinite dimensional space, its study
is quite uneasy and moreover the numerous theorems about symplectic
maps available for finite systems do not hold for infinite systems. It
is convenient to reduce the dimensionality of this map, which can be
done using the DNLS approximation derived in Sec.\ \ref{sec_SAL} (generally 
valid in the limit of small-amplitude solutions and small inter-site 
coupling) to eliminate all $a^{(p)}_n $ with $p\neq 1$.

\begin{figure}[htbp]
\begin{center}
\includegraphics[height=8cm]{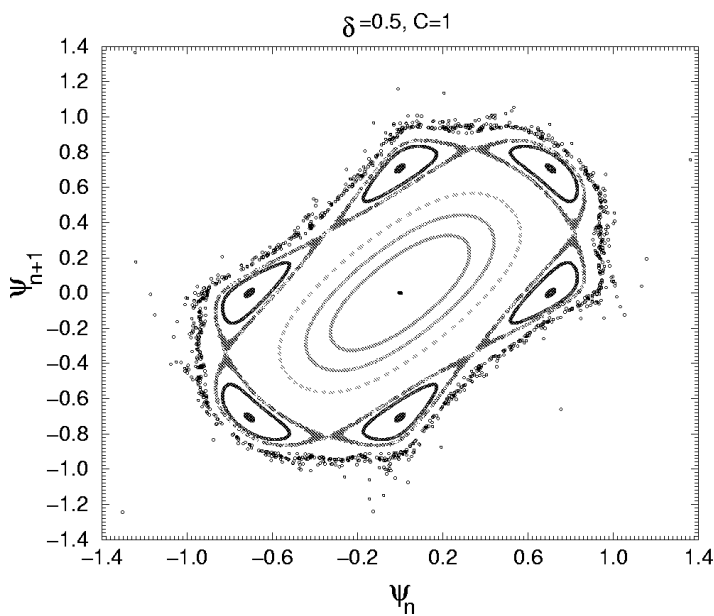}
\caption{ The symplectic map ${\mathcal S}$ for $\delta^{\prime} = 
\frac{\delta}{C} = 0.5$ ($\sigma =-1$). The visible chain of alternating 
elliptic and hyperbolic periodic points correspond to the two families of 
SWs with $Q=\pi/3$.} 
\label{figmappa}
\end{center}
\end{figure}

Since the SWs are time-periodic solutions, all the amplitudes $a^{(p)}_n $ of 
the Fourier expansion (\ref{fseries}), and therefore all the $\psi_n$, will 
be time independent, and they can be chosen real by restricting to 
time-reversible solutions. 
Then, vanishing the left-hand side of Eq.\ (\ref{DNLS}) yields
 \begin{equation}
 \delta \psi_{n} - \sigma  \psi_{n}^{3} + C(
 \psi_{n+1}+\psi_{n-1}-2 \psi_{n}) = 0.
        \label{DNLSde}
 \end{equation}
This equation defines a nonlinear symplectic cubic map ${\mathcal S}$ in the 
2-dimensional real plane, which  recursively determines $(\psi_{n+1},\psi_{n})$
from $(\psi_{n},\psi_{n-1})$. With the change of scale 
$\psi_{n} \rightarrow \sqrt{C}  \psi_{n}^\prime$, 
this map depends only on the parameter $\delta^{\prime} \equiv \delta/C$:
\begin{equation}
  \psi_{n+1}^\prime = (2-\delta^\prime) \psi_{n}^\prime +\sigma \psi_{n}^{\prime 3} -
\psi_{n-1}^\prime\;\; ,\;\;
  \psi_{n}^\prime=\psi_{n}^\prime .
\label{Map}
\end{equation}
As is well-known \cite{Wan},
this map exhibits a rich variety of orbits,
including elliptic and hyperbolic fixpoints and periodic cycles, KAM tori,
Aubry--Mather Cantor sets (cantori) \cite{AD83,Aub78,Aub86} and chaotic orbits 
(see Fig.\ \ref{figmappa}). Since we are searching for  nonlinear SWs in 
continuation of the linear SWs,  only bounded trajectories 
 which can be continued to zero amplitude by varying 
$\delta^\prime=\frac{\delta}{C}$ are of interest. 

When $\delta$ belongs to the phonon band, that is when $0 \leq \delta \leq 4C$,
the linearization of Eq.\  (\ref{DNLSde}) has the solution
$\psi_{n} = \epsilon_s \cos(Q n +\phi$),  where $Q$ fulfills
\begin{equation}
        \delta = 4 C \sin^{2} Q/2 \equiv \delta_0(Q),
        \label{ellipt}
\end{equation}
which is equivalent to the KG dispersion relation 
$\omega_{b}^{2}=\omega_l^{2}(Q)
= 1+4 C_K \sin^{2}\frac{Q}{2}$ (Eq.\ (\ref{lin_disp})). Then, the trivial fixpoint 
$F_0=(0,0)$ of the map
${\mathcal S}$ is linearly stable, that is elliptic with a rotation angle $Q$.
This solution  corresponds to a linear SW  and the rotation angle
is nothing but the wave vector $Q$ of the SW. When $\delta > \delta_0(\pi)=4C$ 
 the fixpoint $F_0=(0,0)$ becomes hyperbolic with
reflection, while it becomes hyperbolic without reflection for
$\delta<\delta_0(0)=0$.

When $\delta^{\prime} = \frac{\delta}{C} \geq \frac{\delta_0(\pi)}{C} = 4$ 
and $\sigma = -1$, it can be proven that the fixpoint $F_0=(0,0)$ corresponds 
to the only bounded trajectory of the map
and that all other trajectories $\{\psi_{n}\}$  go to infinity
for either positive or negative $n$ or for both. To demonstrate this 
assertion, let us assume that at some point a trajectory
fulfills the condition  $0 \neq |\psi_{n}| \geq |\psi_{n-1}|$. Then 
Eq.\ (\ref{Map}) yields that
\begin{equation}
|\psi^\prime _{n+1}| \geq (|2-\delta^{\prime}-\psi_{n}^{\prime 2}|-1)|\psi^\prime _{n}|,
        \label{ineqinfty}
\end{equation}
which for $\delta^{\prime} \geq 4$ yields $|\psi^\prime _{n+1}| >
(\delta^{\prime}-3)|\psi^\prime _{n}| \geq |\psi^\prime _{n}|$ and thus implies recursively 
that
$|\psi_{n}|$ goes to infinity for large positive $n$. If we have the
reverse inequality $|\psi_{n}| \leq |\psi_{n-1}|$, we obtain identically the 
same result but for $n$ going to $-\infty$.
Consequently, there are no spatially bounded time-periodic solutions with a
frequency above
the phonon band for a DNLS equation with a soft potential ($\sigma=-1$).
For a hard potential ($\sigma=1$), the same argument excludes non-trivial 
bounded time-periodic solutions below the phonon band $\delta^{\prime} \leq 0$.

When $\delta^{\prime} = \frac{\delta}{C}<4$ for soft potentials (or, 
equivalently,  $\delta^{\prime} > 0 $ for hard potentials), it can also be 
proven that the trajectories which escape too far from the origin go to 
infinity.
Indeed, for $\sigma=-1$, if $\psi_{n}^{\prime 2}>4-\delta^{\prime}$ and 
$|\psi_{n}| \geq |\psi_{n-1}|$, the inequality (\ref{ineqinfty})
implies that $\psi_{n+1}^{\prime 2} > \psi_{n}^{\prime 2} >4-\delta^{\prime}$
and recursively that $|\psi_{n}|$ goes to $+\infty$ for 
$n \rightarrow +\infty$. A similar result holds for the reverse inequality 
$|\psi_{n}| \leq |\psi_{n-1}|$ but then $|\psi_{n}|$ goes to $+\infty$ for 
$n \rightarrow -\infty$.
Nevertheless, when $\delta^{\prime}<4$ the nonlinear map ${\mathcal S}$ 
exhibits many bounded trajectories representing SWs. 
As mentioned above (see Fig.\ \ref{figmappa}), some of them are chaotic but 
there are also periodic cycles, KAM tori and cantori  for example around the 
elliptic fixpoint $F_0=(0,0)$.

As the nonlinear SWs we search for should be continuable to zero amplitude 
when varying $\delta^{\prime}=\frac{\delta}{C}$, the corresponding orbits 
 should merge into the 
elliptic 
fixpoint $F_0$ for the particular value $\delta=\delta_0(Q)$. As suggested in 
\cite{Aub92}, trajectories which have this property should be either periodic 
cycles or quasiperiodic trajectories. The quasiperiodic trajectories, 
corresponding to wave vectors $Q$ incommensurate with $2\pi$, should, 
for most irrational values of $Q/2\pi$, appear as KAM tori with rotation 
angle $Q$. Thus, given a rotation angle $Q$, there is at least one  
trajectory which 
appears with this rotation angle at the elliptic point $F_0$ for 
$\delta=\delta_0(Q)$, and which for $\delta$ close to  $\delta_0(Q)$ 
is located close to  $F_{0}$. This trajectory, which for rational $Q/2\pi=r/s$ 
is a periodic cycle with period $s$ and for irrational $Q/2\pi$ generally a 
KAM torus, represents a small amplitude SW with wave vector $Q$. 
For a general potential (soft or hard), the coefficient $\sigma$ of the cubic 
term in the DNLS equation (\ref{DNLSde})  determines the behaviour of the 
tori around the elliptic fixpoint $F_0$. 
 For a soft potential ($\sigma < 0$) the rotation angle for orbits rotating 
around $F_{0}$ increases with increasing  radius, and in particular 
for KAM tori with small radius $\epsilon_s$ the rotation angle 
can be calculated to lowest order in $\epsilon_s$ as\footnote{This follows
\eg \ from Eq.\ (\ref{eq:phsm}) in the Appendix using Eq.\ (\ref{epsilon}).}
\begin{equation}
\cos Q \simeq 1- \frac{\delta^{\prime}}{2} + \frac{3 \sigma \epsilon_s^2}{8C} .
\label{rotangle}
\end{equation}
Since the map ${\mathcal S}$ depends continuously on the map parameter 
$\delta^{\prime}$, the periodic cycles, KAM tori and cantori of 
${\mathcal S}$ can be continued as a function of $\delta^{\prime}$, and its 
orbits at fixed  rotation angle depend continuously on $\delta^{\prime}$ 
\cite{Aub92}. Then, as $\delta_0(Q)$ is a monotonously increasing function of 
$Q$, we can conclude that  SWs with wave number $Q$ exist only for 
$\delta \leq \delta_{0}(Q)$. Similarly, for a hard potential ($\sigma>0$) the 
rotation angle  decreases with increasing radius and SWs exist only for 
$\delta \geq \delta_0(Q)$.

\begin{figure}[htbp]
\begin{center}
\includegraphics[height=5cm]{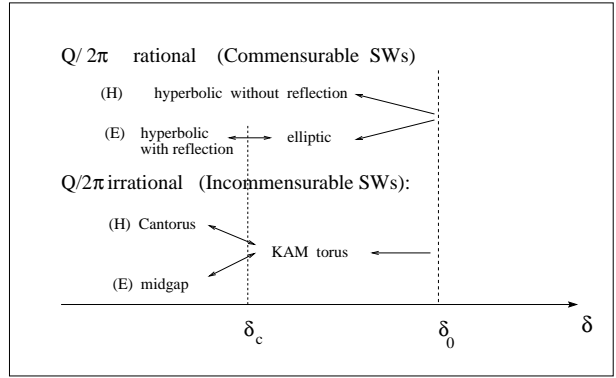}
\caption{Evolution of the trajectories for variable $\delta^{\prime}$ and 
fixed wave vector $Q$ ($\sigma=-1$).} \label{traj-evol}\quad
\end{center}
\end{figure}

To study the evolution of the trajectories representing  SWs (for  
$\sigma=-1$) when varying the 
map parameter $\delta^{\prime}$, we thus start from the linear limit at 
$\delta=\delta_0(Q)$ corresponding to the fixpoint $F_0$ and decrease the 
value of $\delta$. 
This evolution is schematically illustrated in Fig.\ \ref{traj-evol}.
For a rational value of $\frac{Q}{2\pi}=  r/s$ ($r$ and $s$ irreducible 
integers), which corresponds to  SWs  {\em commensurable} with the lattice 
periodicity, there are two families of SWs corresponding to the two 
$s$-periodic cycles which bifurcate in pair from $F_0$ at the linear limit 
$\delta=\delta_0(Q)$. 
We will call these  $s$-periodic cycles {\em 'h-cycle'\/} and
{\em 'e-cycle'}, respectively. The $h$-cycles are hyperbolic without reflection 
for all $\delta < \delta_0(Q)$. 
By contrast the $e$-cycles are elliptic trajectories for $\delta$ close to 
$\delta_{0}(Q)$, but decreasing the value of $\delta$ they become hyperbolic 
with reflection below a  critical value $\delta = \delta_{c}(Q)$ which can be 
calculated accurately numerically e.g.\ using Greene's residue \cite{Greene}. 

\begin{figure}[htbp]
 \centerline{\includegraphics[width=6cm,angle=0]{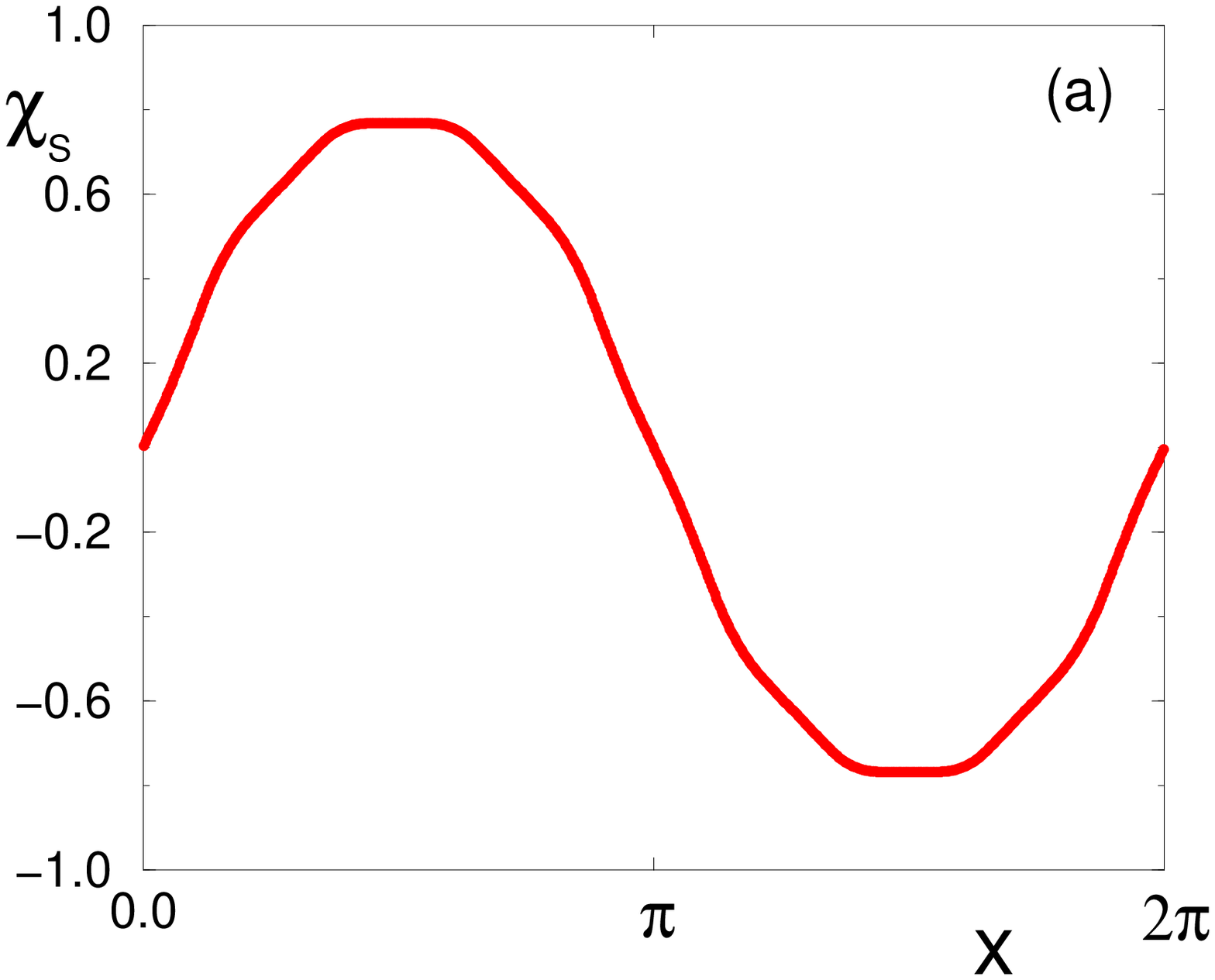}
  \includegraphics[width=6cm,angle=0]{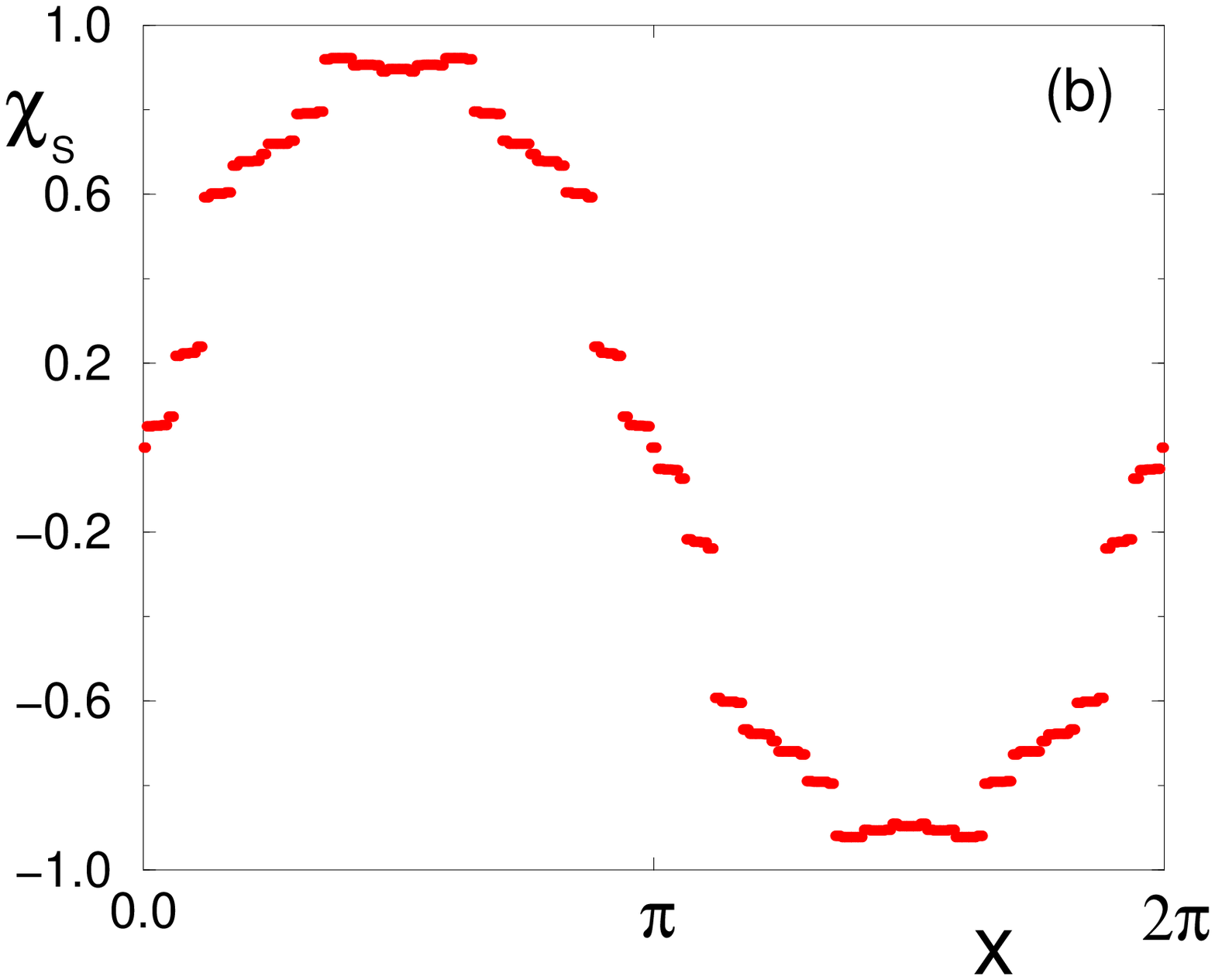}}
 \centerline{\includegraphics[width=6cm,angle=0]{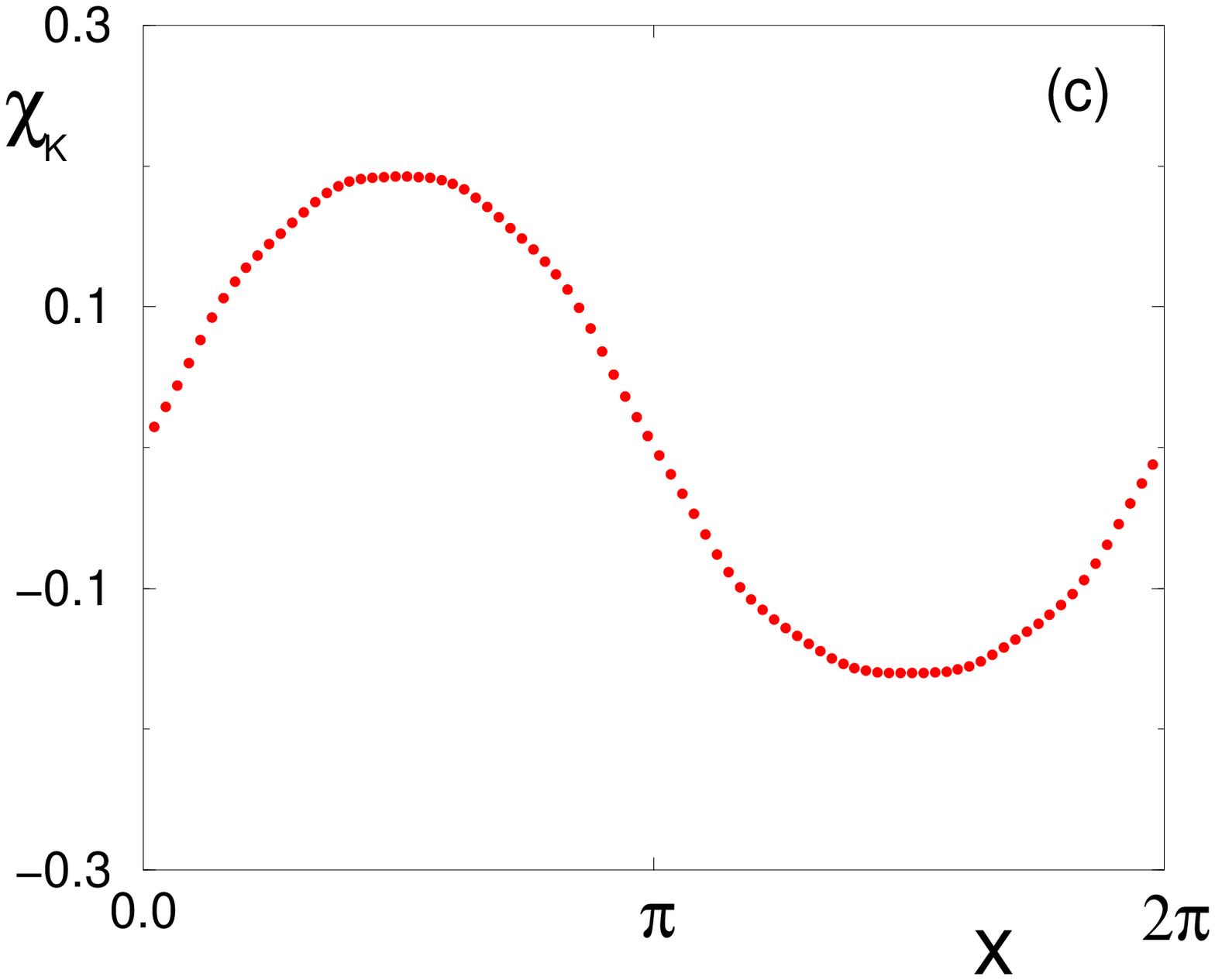}
 \includegraphics[width=6.0cm,angle=0]{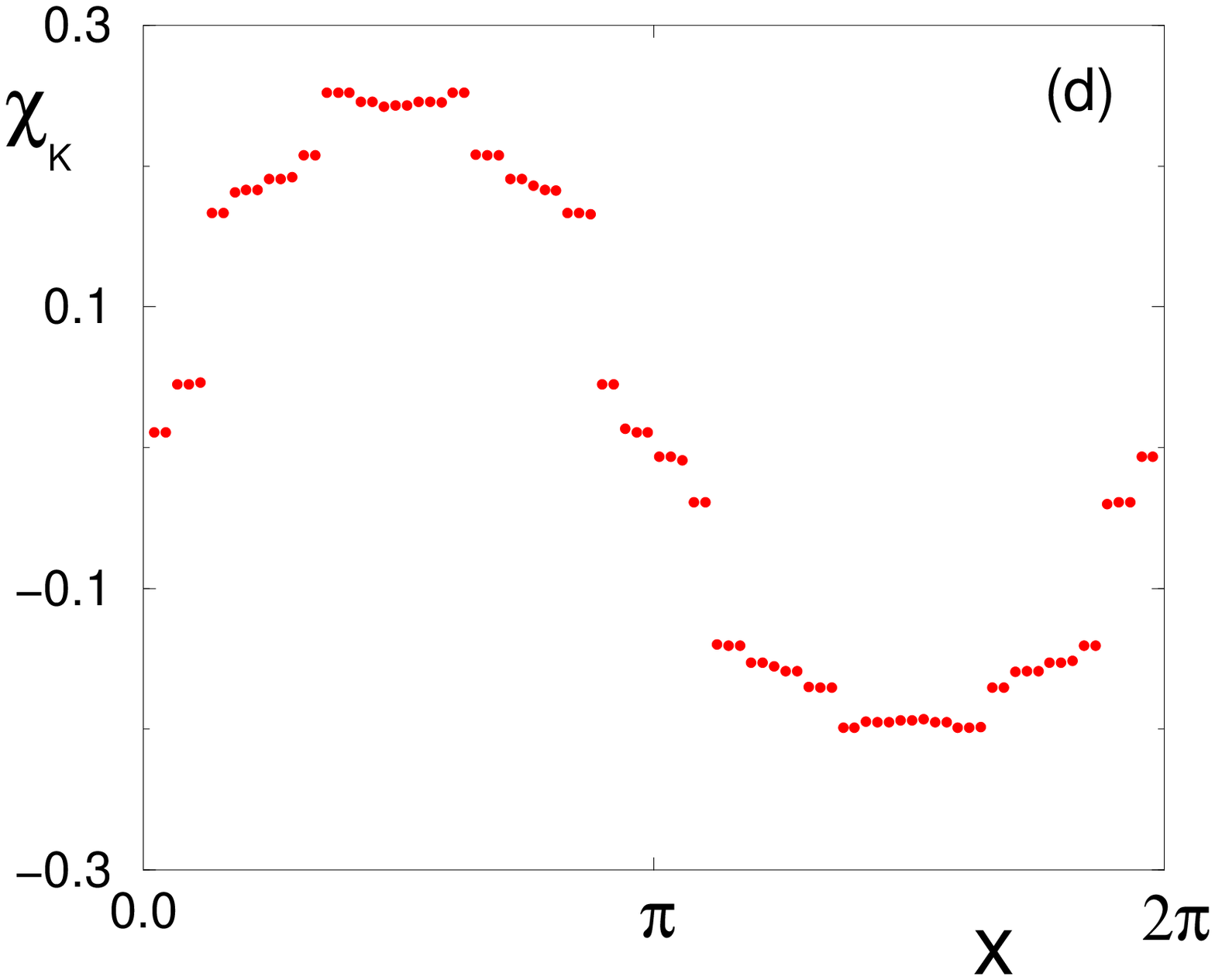}}
 \caption{(a) and (b): Hull functions $\chi_S (x)$ for SWs in a DNLS model  
(\ref{DNLSde}) with $Q/2\pi = 233/610 \simeq (3-\sqrt 5)/2 \equiv \sigma_G$ 
and (a) $\delta^{\prime}=3.0$ resp. (b) $\delta^{\prime} = 2.8 $ $(C=1)$. The 
analyticity is found to be broken at $\delta^{\prime} \simeq 2.92$ for 
$Q/2\pi = \sigma_G$.
 (c) and (d): Hull functions $\chi_K(x)$ for SWs in a KG chain (\ref{DKG}) 
with a Morse potential, with $C_K=0.05$, $Q/2\pi = 34/89 \simeq \sigma_G$ and 
(c) $\omega_{b}=1.072$ resp.\ (d) $\omega_{b}=1.065$.  
The breaking of analyticity occurs at  $\omega_{b} \simeq 1.070$ 
for $Q/2\pi = \sigma_G$.}
\label{transition}
\end{figure}

When $Q/2\pi$ is a generic irrational number (i.e., not a Liouville number), 
the trajectories representing the spatially quasiperiodic,  
{\em incommensurate}, SWs emerge from $F_{0}$ as 
the KAM torus with rotation angle $Q$, which is unique apart from a phase 
shift. 
Decreasing the value of $\delta$ below the critical value 
$\delta = \delta_{c}(Q)$, the KAM torus bifurcates into two types of 
trajectories, the cantorus 
and its associated  'midgap' trajectory \cite{midgap}. This midgap 
trajectory  consists of isolated, nonrecurrent  points in the gaps of the 
cantorus with 
rotation angle $Q$, to which it is homoclinic. As the KAM torus is an 
analytic trajectory while the cantorus and the midgap trajectory are 
non-analytic, the transition at the  critical point $\delta_{c}$ corresponds 
to a {\em transition by breaking of analyticity} (TBA)\cite {Aub78,Aub86}. 
The 
transition point  $\delta_{c}(Q)$ is determined as the limit of the critical 
values $\delta_{c}(Q^{(k)})$ for a sequence of periodic $e$-cycles whose 
rotation 
angles $Q^{(k)}$ are obtained from successive rational approximants of 
$Q/2\pi$  \cite{Greene}. 
With this transition  the {\em two families of non-analytic incommensurate 
SWs}, represented by the  cantori and the midgap trajectories, respectively, 
merge into the {\em unique family of analytic SW} defined by the KAM tori  
for $\delta \geq \delta_{c}(Q)$.
Defining the  \emph {hull} (or envelope) \emph {function} $\chi_S(x)$ for the 
SW as 
\begin{equation}
\psi_n = \chi_S(Qn+\phi) , 
\label{chi_S}
\end{equation}
which gives the shape of the wave rescaled 
inside one period $2 \pi$, the TBA for this hull function when decreasing  
$\delta$ is shown in Figs.\ \ref{transition}(a) and \ref{transition}(b). 
As the model symmetry implies two axis of symmetry for the
KAM tori (see Fig.\ref{figmappa}), the hull function $\chi_S(x)$ can be chosen
with the symmetries $\chi_S(x)=-\chi_S(-x)$ and $\chi_S(x)=-\chi_S(x+\pi)$. 
The 
nonanalyticity of the SW hull functions for large negative $\delta$ makes a 
description in terms of  {\em multibreathers} natural, as will be discussed 
in Sec. 3.2.

As the trajectories of the map {$\mathcal S$} represent approximate solutions 
for the KG model in the limit of small coupling, we should expect to find  
this TBA  also for SWs in the KG system (\ref{DKG}).  We can calculate 
numerically exact  KG SW solutions  using a Newton scheme as described,  
\eg,  in \cite{KA99,CA97,MA96} 
(see also Secs. 3.2 and \ref{sec_KG}). 
Defining the hull function $\chi_K(x)$ for these SWs as 
$u_{n}(0) = \chi_K(Qn+\phi)$ and continuing these solutions  versus 
$\omega_{b}$ to zero amplitude, a TBA is found for the function $\chi_K (x)$ 
as illustrated in Figs.\ \ref{transition}(c) and \ref{transition}(d). 
(Note however that for a non-symmetric potential such as the Morse potential, 
the hull function does not in general have the above-mentioned symmetries, 
\ie,  $\chi_K(x)\neq -\chi_K(-x)$ and $\chi_K(x) \neq -\chi_K(x+\pi)$.)

\subsection{Standing waves as multibreather solutions}
\label{swams}

When the SW frequency $\omega_{b}$ is below the phonon spectrum  for a soft 
potential  (that is $\omega_{b}<1$) or above for a hard potential 
($\omega_{b}> \sqrt{1+4C_K}$) and close enough to the \emph{anticontinuous 
limit}  $C_K = 0$, the SW solutions can be identified  as 
\emph{multibreathers} \cite{Aub97} with period 
$T_b=\frac{2\pi}{\omega_b}$. These are oscillatory states composed by 
nonlinear combinations of several\footnote{We will use the term 
multibreather to refer to combinations of either finitely or infinitely many 
individual discrete breathers.} time-periodic and spatially localised  
breathers in a single nonlinear system  \cite{Aub97,MA94}.
These multibreather solutions are characterized by a \emph{coding sequence} 
$\tilde{\sigma}_n$ \cite{Aub97,Aub92} which defines the particular 
oscillatory state of the oscillator $n$ in the case of zero coupling. 
As for general time-reversible time-periodic solutions with period $T_b$ each 
individual oscillator can oscillate with frequency $j \omega_b$ ($j$ integer) 
with two possible
 choices for the phase, we can, at $C_K=0$, characterize all such solutions 
for 
 the whole system by the following coding sequence:
\begin{itemize}
\item{$\tilde{\sigma}_n = 0$} when the oscillator $n$ is at rest;
\item{$\tilde{\sigma}_n =+1$} when the oscillator $n$ oscillates with a finite
amplitude at frequency $\omega_b$ and in phase with the chosen reference
solution of the single oscillator;
\item{$\tilde{\sigma}_n =-1$} when the oscillator $n$ oscillates at frequency
$\omega_b$ in antiphase  with the chosen reference
solution;
\item{$\tilde{\sigma}_n =+2$} when  the oscillator $n$ oscillates in phase at 
frequency
$2\omega_b$;
\item{...}\\
\end{itemize}
These solutions are proven to be continuable when $C_K$ varies from the 
anticontinuous limit  up  to a finite value of the coupling. At this value, 
which depends in general on the solution in consideration,  bifurcations are 
found and the solution can be lost. This will be numerically investigated in 
more details 
for SWs in Section \ref{sec_KG}. We  consider now only  the continuable 
regime of the multibreathers, and
in particular we search among these multibreather states the solutions which 
are in
continuation of the linearized SWs at low amplitude. 
For that purpose, we have to find an appropriate coding sequence 
$\{\tilde{\sigma}_{n}\}$
such that the multibreather which exists at $C_K \neq 0$ can be continued 
versus its
frequency inside the phonon band, as illustrated in Fig.\ \ref{fig_planeCw}.
\begin{figure}
\begin{center}
\includegraphics[width=6cm,angle=0]{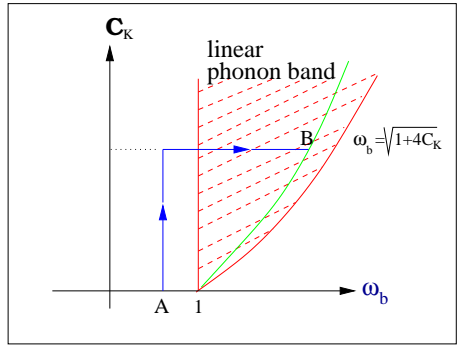}
\caption{Continuation of the multibreather solution with wave vector $Q$ into 
the linear phonon 
band for a soft potential, from the anti-continuous limit $C_K=0$ 
to the linear limit 
$\omega_b=\sqrt{1+4C_K\sin ^2\frac{Q}{2}}$ illustrated by the gray line. 
(For a hard potential, the continuation starts at $\omega_b>1$ and can be 
performed at constant frequency.)}
\label{fig_planeCw}
\end{center}
\end{figure}
 A similar problem has already been
studied in \cite{KA99} for a disordered model. 
Actually, it was found
that most of the multibreather solutions undergo bifurcations and disappear
before their frequency enters the phonon spectrum, although some of them can 
be continued while
their frequency belongs to the phonon spectrum. For this last type of 
solutions an ansatz was found for calculating specifically the multibreather 
states which are continuable as  linearized harmonic modes.
However, the method  used in \cite{KA99} was  explicitly based on the
property that the phonon spectrum is discrete with localized
eigenmodes. In our  translationally invariant model this property is not
fulfilled so that we cannot use the ansatz proposed in \cite{KA99}.
There are nevertheless two linearized solutions which can be trivially
obtained by continuation of multibreather states, namely the $Q=0$ phonon
from the multibreather with code
$... +1 +1 +1 +1 +1 +1 +1 +1...$, and the $Q=\pi$ phonon from the
multibreather with code $... +1 -1 +1 -1 +1 -1 +1 -1...$.
Our problem is to find the ansatz which gives the coding sequence for
a general SW with wave vector $Q$.
Once this code is obtained, it is possible to calculate the nonlinear SW for 
different values of the frequency and coupling by numerical continuation.

The method we propose to find the ansatz consists in passing from the KG 
model (\ref{DKG}) to the DNLS model (\ref{DNLS}) (where we for simplicity 
assume $\sigma=-1$ without loss of generality) by using the small-amplitude 
(and generally also small-coupling)
approximation (Sec. \ref{sec_SAL}), so to obtain the SW representations as 
trajectories of the 
symplectic  map {$\mathcal S$} (\ref{Map}). 
As the DNLS map {$\mathcal S$} only depends on the parameter 
$\delta^{\prime}=\frac{\delta}{C}$, the anticontinuous limit $C=0$ for fixed 
frequency $\delta<0$ is 
in the DNLS description equivalent to the large-amplitude limit 
$\delta \rightarrow - \infty$ for fixed coupling $C$. 
Thus, continuing the SW trajectories 
to this limit, they can be described as multibreathers and associated to 
a coding sequence. As for $C=0$ the stationary solutions to 
Eq.\ (\ref{DNLSde}) can only take 
the values $\psi_n=\pm \sqrt{-\delta}$ and $\psi_n = 0$, the code will be 
defined as $\tilde{\sigma}_n = +1$ when $\psi_n=\sqrt{-\delta}$, 
$\tilde{\sigma}_n = -1$ when $\psi_n=-\sqrt{-\delta}$ and 
$\tilde{\sigma}_n = 0$ when $\psi_n=0$. These codes 
can then be used as initial conditions in an iterative Newton scheme to 
calculate numerically the SWs in the original KG model (\ref{DKG}) to any 
desired accuracy.

By continuing the  SW solutions calculated for the DNLS map to large negative 
$\delta^{\prime}$
we guessed and tested, by inspecting the map trajectories, the following 
ansatz for their coding sequences\footnote{For a hard potential ($\sigma=+1$), 
Eqs.\ (\ref{code})-(\ref{hull}) yield the SWs with wave vector $\pi-Q$ via the 
transformation $\tilde{\sigma}_n \rightarrow (-1)^n \tilde{\sigma}_n$.}
\label{hard}
\begin{equation}
\tilde{\sigma}_{n}=\chi_0(Q n+\phi),
\label{code}
\end{equation}
 where $\tilde{\sigma}_{n}$ can be quasi-periodic or periodic, depending on 
whether $Q/2 \pi$ (with $0 \leq Q \leq \pi$) is irrational or rational, and 
the hull function
$\chi_{0}(x)$ is a $2\pi$-periodic odd function defined for 
$x \in [-\pi,\pi]$  as
\begin{equation}
\chi_0(x) = \left \{ \begin{array}{l} 1  \quad \mbox{for} \quad (\pi-Q)/2
\leq x \leq (\pi+Q)/2\\
-1 \quad \mbox{for} \quad -(\pi+Q)/2
\leq x \leq -(\pi-Q)/2\\
  0 \quad \mbox{elsewhere} \end{array} \right..
\label{hull}
\end{equation}
This function $\chi_0(x)$, rescaled by a factor $\sqrt{-\delta}$, is the 
limit for $C=0$ of the hull function $\chi_S(x)$ introduced  in Section 
\ref{sect_SW}.

\begin{figure}[htbp]
\begin{center}
\includegraphics[height=8cm]{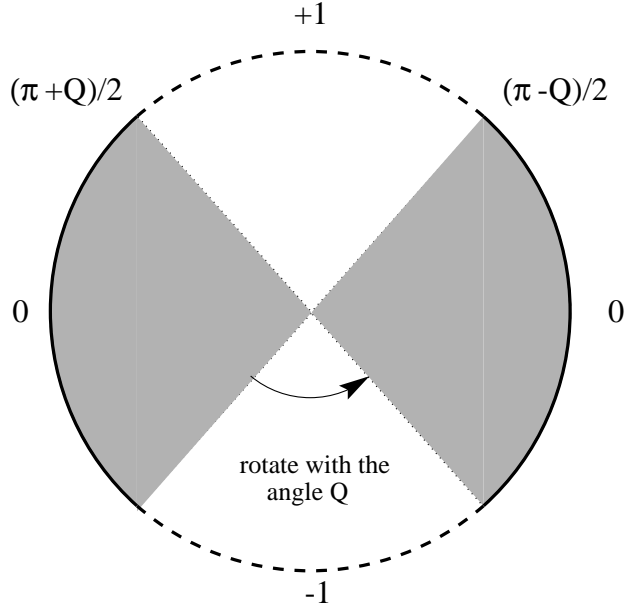}
\caption{Graphical representation of the generation of the code sequence for 
a general nonlinear SW with wave vector $Q$.} \label{fig_circle}
\end{center}
\end{figure}

This method for generating  the code can be  visualised by a graphical circle 
construction (see Fig. \ref{fig_circle}).
 To construct the code for a nonlinear SW with wave vector $Q$, the circle 
is divided in four parts to which the values $ 0, +1$ or $-1$ are associated. 
Rotating around the circle by the angle $Q$ yields the code sequence of the 
SW. The value of the phase $\phi$, corresponding in the circle construction 
to the initial point for the rotation, has particular importance.  
For all phases  $\phi \neq \phi_m \equiv  \pm(\pi-Q)/2 -mQ$ 
($m$ integer),  the coding sequence generated does not contain any 
consecutive codes $+1$ or $-1$. We call these SWs 
{\em 'type H'} and, with the terminology used in Sec. \ref{sect_SW},  they 
correspond either to the $h$-cycles (for rational $Q/2\pi$) or to the cantori 
(for irrational  $Q/2\pi$).
For the particular phases $\phi = \phi_m$,
$x=Qm+\phi$ is at a discontinuity point of $\chi_{0}(x)$, and, as a 
consequence, $\tilde{\sigma}_n$  has two consecutive $+1$ (or $-1$). These 
SWs, 
called {\em 'type E'}, correspond to the $e$-cycles (for rational $Q/2\pi$) or 
to midgap 
trajectories (for irrational $Q/2\pi$).
As  will be seen in Sec. \ref{sec_4}, this characteristic of the code will 
imply important consequences for the stability of nonlinear SWs. We note that 
the coding sequences for SWs of type $E$ will always be symmetric around a 
bond center, and can therefore be chosen to fulfill 
$\tilde{\sigma}_n=\tilde{\sigma}_{-n+1}$ by choosing $\phi=(\pi-Q)/2$. In the 
special cases when $Q=\frac{2k+1}{2k^\prime+1}\pi$ ($k,k^\prime$ integers), 
the codes for type $E$ SWs will in addition be antisymmetric around some 
lattice site, and can therefore alternatively be chosen to fulfill 
$\tilde{\sigma}_n=-\tilde{\sigma}_{-n}$ by choosing $\phi=0$. 
The coding sequences for type $H$ SWs are antisymmetric around a lattice site 
except when $Q=\frac{2k+1}{2k^\prime+1}\pi$, in which case they are 
antisymmetric around a bond center. In the former case we can choose $\phi=0$ 
so that $\tilde{\sigma}_n=-\tilde{\sigma}_{-n}$, while in the latter case 
$\tilde{\sigma}_n=-\tilde{\sigma}_{-n+1}$ by choosing  $\phi=-Q/2$.

Finally, we remark that the construction (\ref{hull}) to generate the SW 
coding sequence can, for type $H$ SWs with $Q$ close to $\pi$,  be regarded as 
a periodic or quasiperiodic repetition of defects or {\em discommensurations} 
added to the $Q=\pi$ phonon, where each discommensuration consists of 
inserting one additional site with code $\tilde{\sigma}_n=0$ in the coding 
sequence for the $Q=\pi$ phonon. In the limit $Q\rightarrow\pi$, the 
distance between the discommensuration sites goes to infinity, so that we 
recover the case of one single defect site which was studied for the DNLS 
model in \cite{JK99} and shown to correspond to the discrete counterpart of 
the dark NLS soliton. As will be discussed later, this point of view is useful 
in order to understand the origin of the instabilities of the SWs for 
$Q>\pi/2$.

\section{Stability of nonlinear standing waves in the DNLS approximation} 
\label{sec_4}

We now turn to the main topic of this paper, namely the characterization of 
the dynamical stability properties of nonlinear SWs. We will here first 
consider small-amplitude solutions to the KG 
equation (\ref{DKG}) with small $C_K$, so that the DNLS approximation 
(\ref{DNLS}) is well justified. In the next section we will study 
numerically the original KG model for larger coupling to  investigate the 
limit of validity for the DNLS approximation. Here, we assume for convenience 
$\sigma=-1$ without loss of generality (Sec. 2.2). 
With the substitution  $\psi_n \rightarrow \psi_{n}+\epsilon_{n}(t)$,  where 
$\epsilon_{n}$ is small, 
the linearization of Eq.\ (\ref{DNLS}) yields a standard Hill equation 
for $\epsilon_{n}$ \cite{Eilbeck}, 
\begin{equation}
i \dot{\epsilon}_{n} = \delta \epsilon_{n} + 2|\psi_{n}|^{2}
\epsilon_{n}
+ \psi_{n}^{2} \epsilon_{n}^{*}
+C(\epsilon_{n+1}+\epsilon_{n-1}-2\epsilon_{n}).
        \label{DNLSHill}
\end{equation}
A solution $\{\psi_{n}\}$ of Eq.\ (\ref{DNLS}) is said to be linearly stable 
when there is no solution $\{\epsilon_{n}(t)\}$ of Eq.\ (\ref{DNLSHill})
which diverges exponentially at large time.

In the case when $\psi_{n}$ is a real and time-independent solution of
Eq. (\ref{DNLS}), the general solution of (\ref{DNLSHill}) 
can be searched as linear combinations of solutions with the form
  \begin{equation}
\epsilon_{n}(t) = a_{n}\e^{\ii \omega t} + b_{n}^{\ast} \e^{-\ii \omega t}
\label{harmdec}
\end{equation}
where the eigenfrequencies  $\omega=\omega_s$ and 
eigenmodes $\{a_{n}, b_{n}\}$ are determined
by the eigenequation
\begin{eqnarray}
(2C-\delta- 2 \psi_{n}^{2}) a_{n} - C(a_{n+1}+a_{n-1}) -
\psi_{n}^{2} b_{n} &=& \omega_s a_{n} \nonumber \\
\psi_{n}^{2} a_{n} + C(b_{n+1} + b_{n-1}) - (2C -\delta -2
\psi_{n}^{2}) b_{n} &=& \omega_s b_{n} .\label{eigeneqns}
\end{eqnarray}
Then, we note that if $\omega_s$ is an eigenvalue for the eigenvector
$\{a_{n}, b_{n}\}$,  then $-\omega_s$ and $\omega_s^{\ast}$ are eigenvalues
for the eigenvectors  $\{b_{n}, a_{n}\}$ and $\{a_{n}^{\ast},
b_{n}^{\ast}\}$, respectively\footnote{The exponential of this operator
is symplectic.}.
As a result, linear stability is achieved if and only if all the eigenvalues
of Eq. (\ref{eigeneqns}) are real.
Thus, for a large but finite system with $N$ sites, we should have
$2N$ real eigenvalues $\omega_s$ for a linearly stable solution. 

It is convenient to redefine new variables $U_{n}=a_{n}+b_{n}$
and $W_{n}=a_{n}-b_{n}$, so that Eqs.\ (\ref{eigeneqns}) become \cite{Eilbeck}
\begin{eqnarray}
{\mathcal{L}}_0 W_n \equiv  (2C-\delta - \psi_{n}^{2}) W_{n}-C(W_{n+1}+W_{n-1})
&=&  \omega_s U_{n} \nonumber \\
{\mathcal{L}}_1 U_n \equiv (2C-\delta -3 \psi_{n}^{2}) U_{n}
-C(U_{n+1}+U_{n-1})&=&
 \omega_s W_{n} .
\label{eigen_prob}
\end{eqnarray}
The operators ${\mathcal{L}}_0$ and ${\mathcal{L}}_1$ are 
self-adjoint, but not the whole operator  
$\mathcal{L}\equiv \left ( \begin{array}{c} 0\;\;\;{\mathcal{L}}_0 \\ 
{\mathcal{L}}_1\;\;\;0  \end{array} \right)$. Note that $\omega_s=0$ is 
always an eigenvalue of $\mathcal{L}$ for  $U_n \equiv 0$, since the first 
equation in 
(\ref{eigen_prob}) then has the explicit solution $W_n=\psi_n$, corresponding 
to the time-independent {\em phase mode} $\epsilon_n=i\psi_n$ resulting from 
the invariance of the model under global phase rotations. Moreover, for an 
incommensurate SW with analytic hull function (\ref{chi_S}) there is also 
an additional  solution to (\ref{eigen_prob}) with $\omega_s=0$ and 
$W_n \equiv 0$,  
namely the {\em sliding mode} solution $U_{n}=\partial \chi_S/\partial \phi$ 
to the second equation,  
resulting from the invariance of analytic SWs under spatial translations. 
See also Appendix \ref{subsec2} for a further discussion of the spectrum of 
the operators ${\mathcal{L}}_0$ and ${\mathcal{L}}_1$. 

We also define the {\em Krein signature} \cite{Arnold,Aub97} associated with 
a pair of  eigenvalues $\pm \omega_s$ as \cite{PRE61}
\begin{equation}
{\mathcal K}(\omega_s) =  {\rm sign} \sum_n \left[|a_n|^2-|b_n|^2\right] = 
 {\rm sign}  \sum_n  \left[ U_n W_n^* + W_n U_n^*\right],
\label{Krein_sig}
\end{equation}
where $\{a_{n}, b_{n}\}$ and $\{U_{n}, W_{n}\}$ are  the 
eigenvectors of (\ref{eigeneqns}) resp. (\ref{eigen_prob})
with eigenvalue $+\omega_s$. In general, when varying the parameter $\delta^\prime$, 
instabilities may 
occur only through  collisions between eigenvalues with different Krein 
signatures. 
 
When $\delta <0$ and $C \rightarrow  0$ (anticontinuous limit), 
$\psi_n^2=-\delta$ or $0$ (codes $\tilde{\sigma}_n=\pm 1$ or 0, 
respectively), and all eigenfrequencies  of (\ref{eigen_prob}) are easily 
seen to be real. For a finite system with $P$ sites where 
$|\tilde{\sigma}_n|=1$ and $Z=N-P$ remaining sites where $\tilde{\sigma}_n=0$, 
there are $2P$ eigenvalues degenerated at $\omega_s=0$, while the remaining 
eigenvalues are located as $Z$ pairs at $\omega_s=\pm \delta$.
Increasing the value of the coupling, the eigenvalues will move losing their 
degeneracy and, if a pair of eigenvalues goes out in the imaginary plane, the 
nonlinear SW may  become unstable. In particular, when $C$ is increased from 
zero,  the eigenvalues at $\omega_s=\pm \delta$ (which all have Krein 
signature $+1$) will move only along the real axis causing no instabilities 
for the SWs. 
On the other hand, the behaviour of the eigenvalues originating from 
$\omega_s=0$ 
will be different for the $E$-type and $H$-type SWs (as defined in 
Sec. {\ref{swams}). 
As the coding sequences for the $E$-type SWs contain pairs of consecutive 
$+1$ (or $-1$) and the anharmonicity is soft, it follows from a general 
result \cite {Aub97} that some pairs of eigenvalues originally at 
$\omega_s=0$ leave 
the real axis and move out on the imaginary axis as soon as 
$C\neq 0$. As a consequence, the corresponding SWs become {\em dynamically 
unstable}  for $\delta<0$ and small $C$. 
Moreover, we find numerically that {\em this instability persists for
all $\delta < \delta_{0}(Q)$}, so that SWs of type $E$ are unstable for any 
nonvanishing amplitude.\footnote{This non-oscillatory instability 
is also proven in the Appendix for  commensurate SWs of type $E$ in the 
small-amplitude limit.}
By contrast,  for $H$-type SWs, for every site with code 
$\tilde{\sigma}_n=1$ the nearest surrounding sites with nonzero code have
$\tilde{\sigma}_n=-1$, and reciprocally (i.e., there is a phase shift of 
$\pi$ between any two consecutive sites with nonzero oscillation at $C=0$, 
neighboring or not). Then, 
the theory of effective action developed in Ref.\ \cite{Aub97} yields, 
with similar calculations as in \cite{JA97,JK99}, that all eigenvalues 
originating from $\omega_s=0$
(having Krein signature $-1$ opposite to those originating from 
$\omega_s = \pm \delta$) move away from the origin along the real axis, 
except one pair remaining at $\omega_s=0$ (with Krein signature 0) 
corresponding to the global 
phase invariance. Thus, {\em the stability of these SWs 
is preserved for $C$ not too large} (c.f.\ Fig.\ \ref{collision} (a)).

\subsection{Commensurate SWs}
\label{CSW}

For commensurate SWs with $Q= \pi r^\prime/s^\prime$ 
($r^\prime$ and $s^\prime$ irreducible integers) in an infinite system,
the lattice periodicity of $\psi_{n}^{2}$ in (\ref{eigen_prob}) implies
that the degenerate eigenvalues on the real axis form bands as $C$ is 
increased from zero
(see Fig.\ \ref{collision}).
As there are $s'-r'$ sites with code $\tilde{\sigma}_n=0$ in a lattice period 
of $s'$ sites, there  will be $s'-r'$ symmetric pairs 
of bands with Krein signature $+1$ originating from $\omega_s=\pm \delta$ and 
separated by gaps. 
The remaining eigenvalues form  $r'$ symmetric  pairs of bands with Krein 
signature $-1$  originating from $\omega_{s}= 0$. As the eigenvalue 
$\omega_s=0$  
always remains in the spectrum,  there is no gap at zero which thus always 
remains at a band edge.
For fixed $C$ and increasing $\delta$, the bands originating from 
$\omega_s=0$ and those from $\omega_s= \pm \delta$ broaden and  the gap 
separating these two classes of bands  will shrink. At a  critical value 
$\delta$ $=$ $\delta_{K}(Q)$  this middle gap shrinks to zero and, as shown 
in Fig.\ \ref{collision}(b), symmetric pairs of eigenfrequencies with 
opposite Krein signature start to collide and move out from the real axis 
into the complex plane.
Thus, \emph{oscillatory (Krein) instabilities\/} (\ie, caused by complex 
eigenvalues) of the nonlinear SW are induced.

\begin{figure}[htbp]
\centerline{\includegraphics[height=7cm,angle=270]{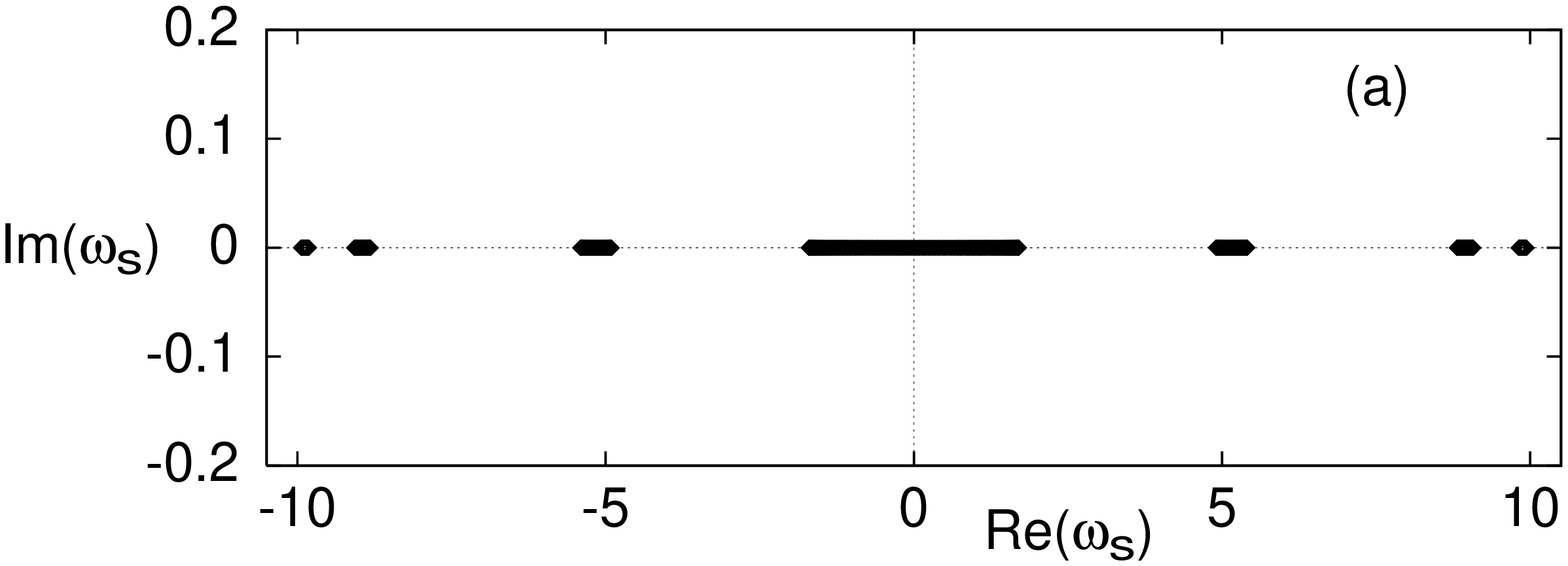}
\includegraphics[height=7cm,angle=270]{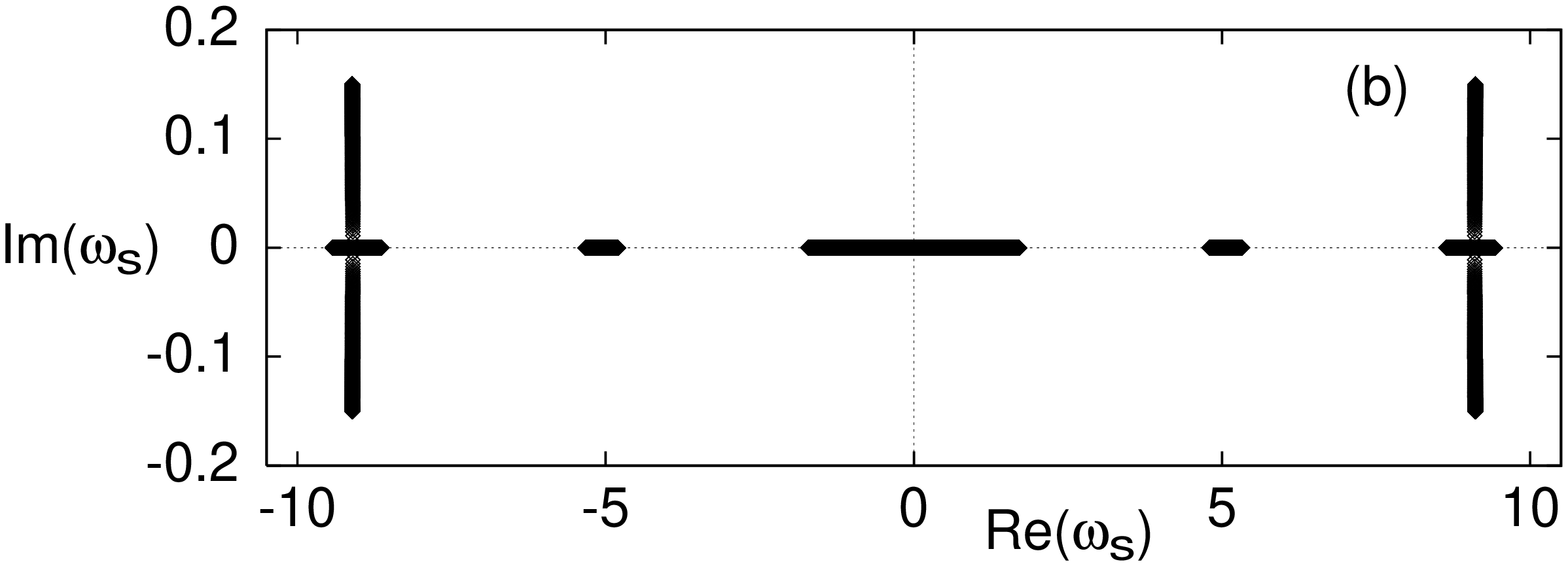}}
\centerline{\includegraphics[height=7cm,angle=270]{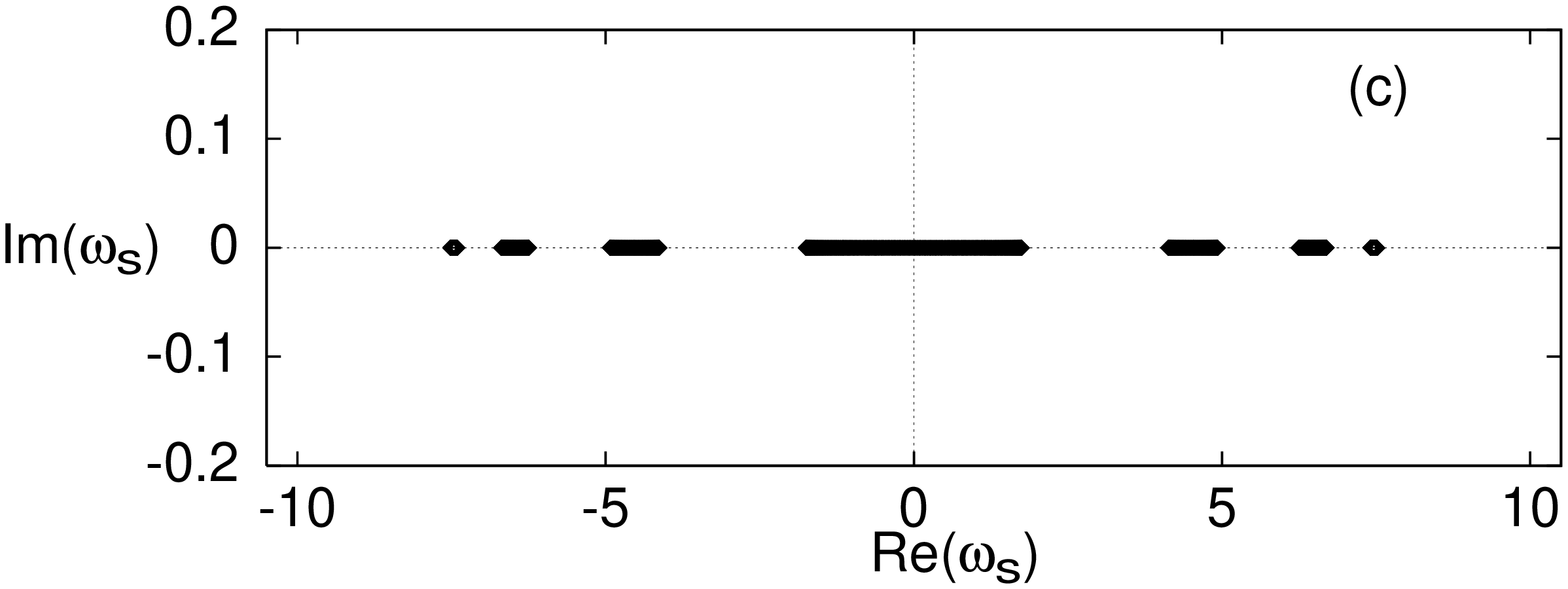}
\includegraphics[height=7cm,angle=270]{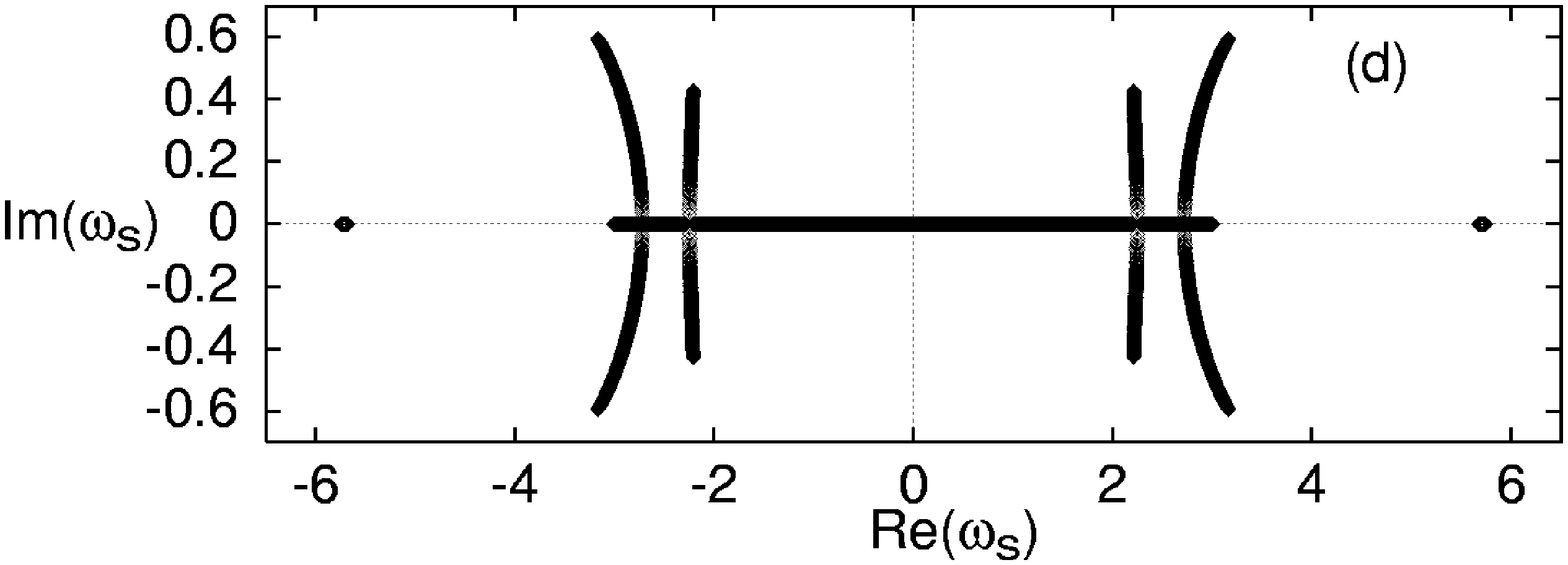}}
\caption{Eigenvalues of Eq.\ (\ref{eigen_prob}) for a type $H$ SW with 
$Q=3 \pi / 4 $ and (a) $\delta^\prime=-8.0$, (b) 
$\delta^\prime=-7.5$, (c) $\delta^\prime=-4.5$ resp. (d) $\delta^\prime=-0.5$ 
($C=1$). }
\label{collision}
\end{figure}

This instability mechanism is similar to that previously observed 
\cite{JK99} for the 'discrete dark solitons', where the anticontinuous coding 
sequence consists of one single zero-amplitude 'defect' site added to a 
background wave with $Q=\pi$ (the latter being always stable \cite{KP92}). In 
fact, as was mentioned in Sec.\ \ref{swams}, for $Q$ close to $\pi$ the SWs
generated from (\ref{hull}) can be viewed 
as a periodic (or quasi-periodic for incommensurate SWs) repetition of 
defects or discommensurations added to the stable wave with $Q=\pi$. Thus, 
the SW can be regarded as  a lattice of 
weakly interacting  discommensurations (``dark solitons''), each of which 
shows an oscillatory instability due to a 
resonance between a translational mode localized at the discommensuration and 
extended modes \cite{JK99}. 
This oscillatory instability detected  numerically for the DNLS model can 
also be confirmed  by a direct numerical Floquet analysis of the 
KG Eq.\ (\ref{DKG}) (see Section \ref{sec_KG}).

Increasing further $\delta$ above the critical value  $\delta_{K}(Q)$, 
bands continue to overlap and generate new instabilities. 
For many commensurate values of $Q$, it is found that when $\delta$ increases 
beyond $\delta_K(Q)$, the SW always  remains unstable until it vanishes with 
zero amplitude at the linear limit $\delta=\delta_0(Q)$.
However, for some rational values $Q/2\pi=\frac{r}{s}$ 
with small $s$, in some intervals of $\delta$  the band overlap disappears, 
and consequently  the stability is temporarily regained 
(e.g. for $Q = 3\pi / 4$, the $H$-type SW is stable for
 $-4.9 \lesssim \delta^{\prime} \lesssim -3.9$, see 
Fig.\ \ref{collision}(c)). In any case,  for {\em all} commensurate SWs 
with $Q\neq\pi$, 
there will always be a final interval  $\delta_1(Q)\le
\delta<\delta_0(Q)$ where
bands with opposite Krein signature overlap (Fig.\ \ref{collision}). As a 
consequence 
{\em all commensurate  SWs with $Q\neq\pi$ are unstable for small but nonzero 
amplitude}. 
This result  can be simply understood from the remark that at the linear 
limit  
$\delta=\delta_{0}(Q)$ where $\psi_{n}^{2}=0$, the frequencies of the 
(gapless) spectrum of (\ref{eigen_prob}) are given by  
\begin{equation}
\omega_{s}=\pm 2 C |\cos Q -\cos q|, 
\label{linspect}
\end{equation} 
$0 \leq q <  2\pi$,
and the 
corresponding Krein signatures are given by $\mbox{sign}(\cos Q -\cos q)$. 
Thus, 
 in the linear limit there is an interval around $\omega_s = 0$ where 
eigenvalues with opposite Krein signature overlap, although there is of course no 
instability since there is no coupling between modes with opposite Krein 
signature in this limit. 
By a continuity argument,  since in the  commensurate case  when  
$\psi_{n}^{2} \neq 0$ a finite number of bands with nonzero widths appear for
$\delta<\delta_{0}(Q)$,  bands with opposite Krein signature would, 
assuming the absence of mode coupling, continue to overlap on the real axis 
close to $\omega_s = 0$ until $\delta$ reaches some 
value $\delta_1(Q)$ which must be strictly smaller than $\delta_0(Q)$. 
However, as in general there is coupling (\ie, resonances) between 
modes of different Krein signature, the overlapping bands move out in the 
complex plane and instabilities result. This is proven in the 
Appendix using perturbation theory and the method of 'band analysis' 
outlined in Sec.\ 4.2 below. 

Let us also remark that in the case of finite-size systems (\eg,  with 
periodic 
boundary conditions), the eigenvalues 
$\omega_s$ will evidently be separated in a discrete spectrum without 
forming  continuous bands.  The initial scenario when increasing 
$\delta^\prime$ from 
$-\infty$ will however be qualitatively the same as described above, and 
oscillatory instabilities will result similarly as illustrated in 
Fig.\ \ref{collision}(b) (although the bands are replaced by discrete 
eigenvalues). However, the above argument showing instability close to the 
linear limit for infinite systems is obviously not valid for finite systems, 
as 
there are no continuous bands around $\omega_s=0$ but only discrete 
eigenvalues. Thus, overlap (resonances) can be avoided and, in fact, we find 
that for finite systems stability will generally be regained in some 
interval close to the linear 
limit. The scenario by which the stability is recovered is 
through 'reentrant instabilities' similarly as in \cite{MA98,JK99}, and 
appears in an analogous way also for the KG model as will be described 
in Sec.\ \ref{sec_KG}. 

\subsection{Incommensurate SWs}
\label{ISW}

For incommensurate SWs ($Q/2\pi$ irrational), oscillatory instabilities 
will occur at some critical value $\delta_K(Q)$ just as for the commensurate 
case described above, although the spectrum of (\ref{eigen_prob}) now 
has a Cantor set structure with infinitely many gaps. However, for 
small-amplitude (analytic) 
incommensurate SWs, the Cantor set nature of the spectrum (with 
'bands' of zero width) implies that the argument used in the previous section 
cannot 
be used to prove the instability of the nonlinear SWs close to the linear 
limit, since it would only yield
$\delta_1(Q)=\delta_0(Q)$. Indeed, numerical analysis of finite-size 
approximants to incommensurate analytic SWs with  $|Q| > \pi/2$ typically 
yields a scenario with reentrant 
instabilities which become very weak and practically invisible for $\delta$ 
close to $\delta_0(Q)$, and extrapolating this scenario to infinite systems 
could lead to the (erroneous) guess that analytic SWs  with 
$|Q| > \pi/2$ could be stable for some $\delta$ (possibly constituting a 
Cantor set as observed in disordered systems \cite{KA99}). 
This scenario would mean that in the absence of mode coupling,  the 
(Cantor-like) subspectrum of (\ref{eigen_prob}) with positive Krein signature 
would, for some $\delta$, exactly fit into the gaps of the other (Cantor-like) 
subspectrum with negative Krein signature, and thus instabilities could be 
avoided. However, as $\delta \rightarrow \delta_0(Q)$ both subspectra acquire 
full measure, and if their measure is assumed to be continuous with respect 
to $\delta$ overlap cannot be avoided for $\delta$ close enough to 
$\delta_0(Q)$, and instabilities should result also in this case. 
In fact, this intuitive argument is confirmed by the more sophisticated 
analysis described below (some technical details are carried out in the 
Appendix), 
which shows that also the incommensurate SWs are
unstable for $\delta$ close to 
$\delta_0(Q)$, but when $|Q| > \pi/2$ only through  higher order 
instabilities, which are thus more difficult to detect numerically or
 experimentally.

To prove the instability of analytic SWs we use here the method of 
'band analysis' \cite{Aub97}.\footnote{The same approach can be used for 
small-amplitude commensurate SWs, see Appendix.}
Its basic idea is to embed the non-Hermitian eigenvalue 
problem  (\ref{eigen_prob}) into  a wider eigenvalue problem but for a 
Hermitian operator, which has more properties. Then, according to
\cite{Aub97}, we associate this
problem with an extended eigenvalue  problem
\begin{equation}
i \dot{\epsilon}_{n} - \delta \epsilon_{n} - 2|\psi_{n}|^{2}
\epsilon_{n}
- \psi_{n}^{2} \epsilon_{n}^{*}
-C(\epsilon_{n+1}+\epsilon_{n-1}-2\epsilon_{n}) = E \epsilon_{n} .
\label{DNLSBand}
\end{equation}
The eigenvalues and the corresponding eigensolutions of
this eigenequation can be searched again with the form
(\ref{harmdec}), where $\omega$ is a real parameter. Then,
the eigenvalues $E$ with eigenvectors $\{a_{n}, b_{n}\}$ are
those of a self-adjoint operator:
\begin{eqnarray}
( - \omega+2C-\delta- 2 \psi_{n}^{2}) a_{n} - C(a_{n+1}+a_{n-1}) -
\psi_{n}^{2} b_{n} &=& E a _{n} \nonumber \\
- \psi_{n}^{2} a_{n} - C(b_{n+1}+b_{n-1}) + (\omega +2C -\delta -2
\psi_{n}^{2}) b_{n} &=& E b_{n}. \label{eigeneqharm}
\end{eqnarray}
Writing as before $U_{n}=a_{n}+b_{n}$
and $W_{n}=a_{n}-b_{n}$, (\ref{eigeneqharm}) becomes
\begin{eqnarray}
(2C-\delta -3 \psi_{n}^{2}) U_{n}-C(U_{n+1}+U_{n-1}) - \omega W_{n} &=& E
U_{n}
\nonumber \\
  - \omega U_{n} + (2C-\delta - \psi_{n}^{2}) W_{n}-C(W_{n+1}+W_{n-1}) &=&
E W_{n} , 
\label{eigeneqharm2}
\end{eqnarray}
or
\begin{equation}
\label{eigen_prob_her}
\left ( \begin{array}{c} {\mathcal{L}}_1 \;\; \; -\omega \\ -\omega
\;\; \;
{\mathcal{L}}_0  \end{array} \right) \left ( \begin{array}{c} \{U_n\}\\\{W_n\}
\end{array}
\right)  = E \left ( \begin{array}{c} \{U_n\}\\\{W_n\} \end{array}
\right) ,
\end{equation}
with  ${\mathcal{L}}_0$ and ${\mathcal{L}}_1$ defined by 
Eqs.\ (\ref{eigen_prob}). 
Then, each  eigenvalue $E_{\nu}(\omega)$ of the problem 
(\ref{eigen_prob_her}) becomes a 
smooth, real 
function of $\omega$ called band 
(symmetric with respect to $\omega=0$), 
which also depends smoothly on the other model 
parameters (as before, we only need to consider variations of the 
parameter $\delta^{\prime}=\frac{\delta}{C}$). 
If we assume that the parameter $\omega$ 
represents the eigenfrequencies for the initial non-Hermitian problem 
(\ref{eigen_prob}), the problem 
(\ref{eigen_prob_her}) reduces to the problem (\ref{eigen_prob}) 
for the particular value $E=0$. Thus the 
real eigenfrequencies
$\omega_{s}$ of (\ref{eigen_prob}) are determined by the intersections 
$E_\nu(\omega_s)=0$ of the bands with the axis $E=0$, and their Krein 
signature by sign$(-\frac{dE_{\nu}}{d\omega}$) at $ \omega=\omega_s >0$ 
\cite{Aub97}. 
Varying the model parameter $\delta^\prime$ the bands will generally move in 
the $(\omega,E)$-plane, and if a band loses a pair of intersections with 
the zero-energy axis (i.e., if a 'gap' opens around this axis) an 
instability occurs for the initial problem 
(\ref{eigen_prob}).\footnote{For an 
oscillatory (Krein) instability, two symmetric pairs of intersection are 
simultaneously lost.}

At the linear limit ($\psi_n \rightarrow 0$ 
and $\delta \rightarrow \delta_0(Q)$), the eigenvalue problem 
(\ref{eigen_prob_her}) is easily solved by introducing a wave vector $q$ and 
 writing 
\begin{equation}
U(q)= \sum_{n} U_{n} \e^{\ii qn} , \qquad W(q) = \sum_{n} W_{n} \e^ {\ii qn} . 
\label{uq}
\end{equation}
At this limit, there are two sets of plane wave solutions 
$U_n = \pm W_n=\e^{\ii qn}$ with corresponding eigenvalues 
\begin{equation}
E_{0\pm}(q;\omega)=2C-\delta_0(Q) -2C \cos q \mp \omega = 
2C (\cos Q - \cos q) \mp \omega . 
\label{eq:bdh} 
\end{equation}
For a finite system with $N$ sites, 
$q=2 \pi k/N$ with $0 \leq k <N$ integer, and thus there are $2N$ bands 
described by straight lines with slope $\mp 1$ and indexed by the wave 
vector $q$ (Fig.\ \ref{scheme} (a)).  
Each band $E_{0\pm}(q;\omega)$ intersects $E=0$ at $\omega_{\pm}(q)
= \pm (2C- \delta_0(Q) - 2C \cos q) \equiv \omega_s(q)$, so that we recover 
the result (\ref{linspect}) that Eq.\ (\ref{eigen_prob}) has $2N$ eigenvalues 
on the real axis so that the solution 
$\psi_{n} \equiv 0$ is linearly stable as expected, and remains stable also
in the limit of large $N$.

When $\psi_n$ is small but  non-zero, it can be treated as a perturbation
for the operator in the left-hand side of Eq.\ (\ref{eigen_prob_her}). 
Then, gaps will generally open in the
continuous band spectrum of this operator at points or lines of degeneracy.
Such a gap might suppress a band intersection 
$E_{0\pm}(\omega)=0$ 
with the real axis,  and thus generate a linear instability. There are two 
kinds of degeneracies. The first kind occurs for both branches of
eigenenergies $E_{0+}(q;\omega)$ and $E_{0-}(q;\omega)$ for all $\omega$,  
since $E_{0\pm} (q; \omega) = E_{0\pm} (2 \pi -q; \omega)$). The second kind 
occurs at transverse intersection points, where bands with opposite slope 
intersect if
$E_{0-} (q; \omega)= E_{0+} (q^\prime; \omega)$ for some $q$, $q^\prime$. 

For an analytic SW, the hull function (c.f. Eq.\ (\ref{chi_S})) 
$\chi_S^{2}(x)$ of
$\psi_{n}^{2}$ can be chosen even and $\pi$-periodic, and thus $\psi_n^2$ 
can be 
expanded in a Fourier series as 
\begin{equation}
\psi_{n}^{2} = \sum_{p}f_{p} \e^{\ii 2p(n Q +\phi)}, 
\label{psi_esp} 
\end{equation}
where $f_{p}=f_{-p}$. We can choose the arbitrary phase as $\phi=0$. 
Since, as for conventional perturbation theory, the total sum
of the gap openings will be  proportional to $\sum_{p}|f_{p}|$, 
perturbation theory will be consistent in principle only when this sum
is convergent and small. Thus, perturbation theory is applicable for 
incommensurate SWs only when 
 $\chi_S(x)$ is analytic, in which case $f_{p}$ decays exponentially for 
$p\rightarrow \infty$, and 
not when $\psi_{n}$ is represented by cantori 
because then this sum is divergent.
Then, considering only the regime close to the linear limit where the SW 
is analytic 
with small amplitude $\epsilon_s$, 
$\chi_S(x)$ becomes close to a harmonic function $\epsilon_s \cos x$, and 
we have 
\begin{equation}
f_{0} \approx  2 f_{1}=2 f_{-1} \approx \epsilon_s^{2}/2 >0 , 
\label{epsilon}
\end{equation}
while the other Fourier coefficients $f_p$ will be small of 
order $\epsilon_s^{2|p|}$. Using the space Fourier transform (\ref{uq}), 
Eq.\ (\ref{eigeneqharm2}) becomes without approximations
\begin{eqnarray}
(2C-\delta-2C \cos{q}-3 f_{0}) U(q)-3 \sum_{p\neq 0}
f_{p}U(q+2pQ)
- \omega W(q) &=& E U(q)
  \nonumber  \\
- \omega U(q) + (2C-\delta-2C \cos{q}- f_{0}) W(q)- \sum_{p\neq 0}
f_{p}W(q+2pQ) &=& E W(q) .\nonumber \\
\label{eigeneqfour}
\end{eqnarray}
Thus, this equation couples the space of solutions $(U(q),W(q))$ 
with wave vector $q$ of the unperturbed problem 
(\ref{eigen_prob_her}) with those with wave vector $q +2pQ$, 
and the coupling strength is proportional to $f_p$. The matrix of the 
operator corresponding to the left-hand side of 
Eq.\ (\ref{eigeneqfour}) can be decomposed into blocks of 
$2 \times 2$-matrices, which we denote as 
$\mathbf{D}(q)$ for the diagonal part acting on the subspace with
wave vector $q$, and $- f_{p}\mathbf{A}$ for the parts which couple
subspace $q$ with subspace $q+2pQ$:
$$
%\begin{equation}
\mathbf{D}(q) = \left(\begin{array}{cc}
2C-\delta-2C \cos{q}-3 f_{0} & -\omega  \\
-\omega & 2C-\delta-2C \cos{q}-f_{0}
\end{array}\right) ; 
\mathbf{A} =
\left(\begin{array}{cc}
3~ & 0  \\
0~ & 1  \end{array}\right) .
$$
%\label{blocks}
%\end{equation}
The eigenvalues of the diagonal part $\mathbf{D}(q)$  are given by 
\begin{equation}
E_{\pm}(q; \omega) = 2C-\delta -2C \cos{q}-2 f_{0} \mp 
\sqrt {\omega^{2}+f_{0}^{2}} ,  
\end{equation}
which determines the downward shift of the unperturbed eigenvalues 
$E_{0\pm}(q; \omega)$ (cf. Eq.\ (\ref{eq:bdh})) caused by 
the static term  $f_0$ in the expansion (\ref{psi_esp}) (which has the same 
order of magnitude as $f_1$). 
The perturbation due to $|f_{p}|$, $|p| \ge 1$, will then 
act mostly by raising the degeneracies of these eigenvalues. 
The first kind of degeneracy,  
\begin{equation}
E_{+}(q;\omega) = E_{+}(2\pi-q;\omega) \;\;
\mbox{and} \;\;  E_{-}(q;\omega) = E_{-}(2\pi-q;\omega) , 
\label{degen1}
\end{equation}
is raised when $q+2pQ=-q \; (\mathrm{mod} \; 2 \pi) $, or 
\begin{equation}
	q=-p Q \;  (\mathrm{mod} \; \pi) , 
	\label{eq:qres}
\end{equation}
  with $p$ integer. Then, a gap of width proportional to
$|f_{p}|$ appears
between the two curves $E_{+}(-p Q;\omega)$ and $E_{+}(pQ;\omega)$, 
and an identical gap opens between the
curves  $E_{-}(-pQ;\omega)$ and $E_{-}(pQ; \omega)$ (Fig.\ \ref{scheme} (b)). 
However, the existence of 
these gaps does not change the number of
intersections $E_{\pm}(q;\omega)=0$ which exist
in the linear limit, and thus
does not induce any instabilities for the SWs.

\begin{figure}[tbp]
\centering
\includegraphics[width=0.7 \textwidth]{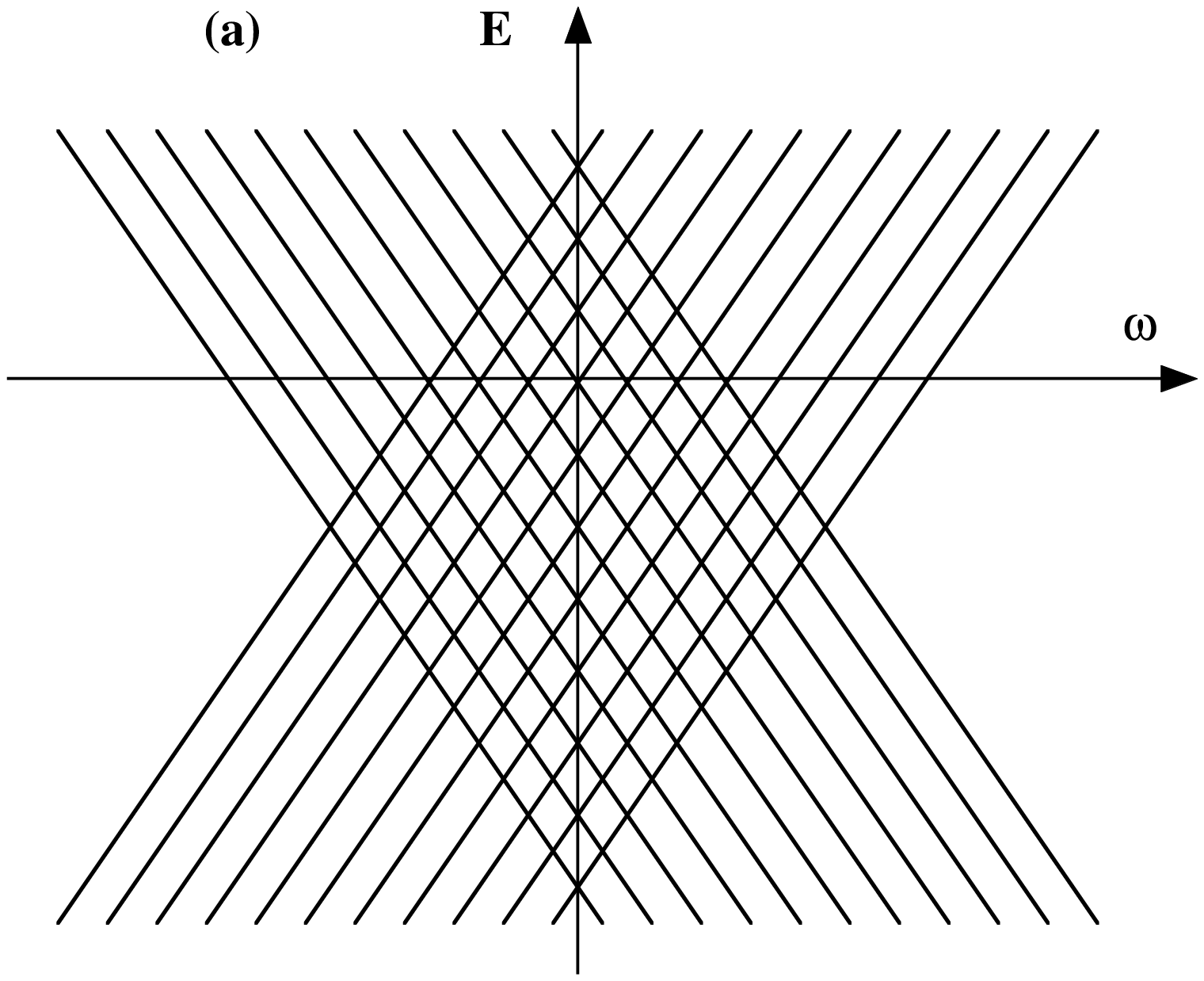}\\
\includegraphics[width=0.45 \textwidth]{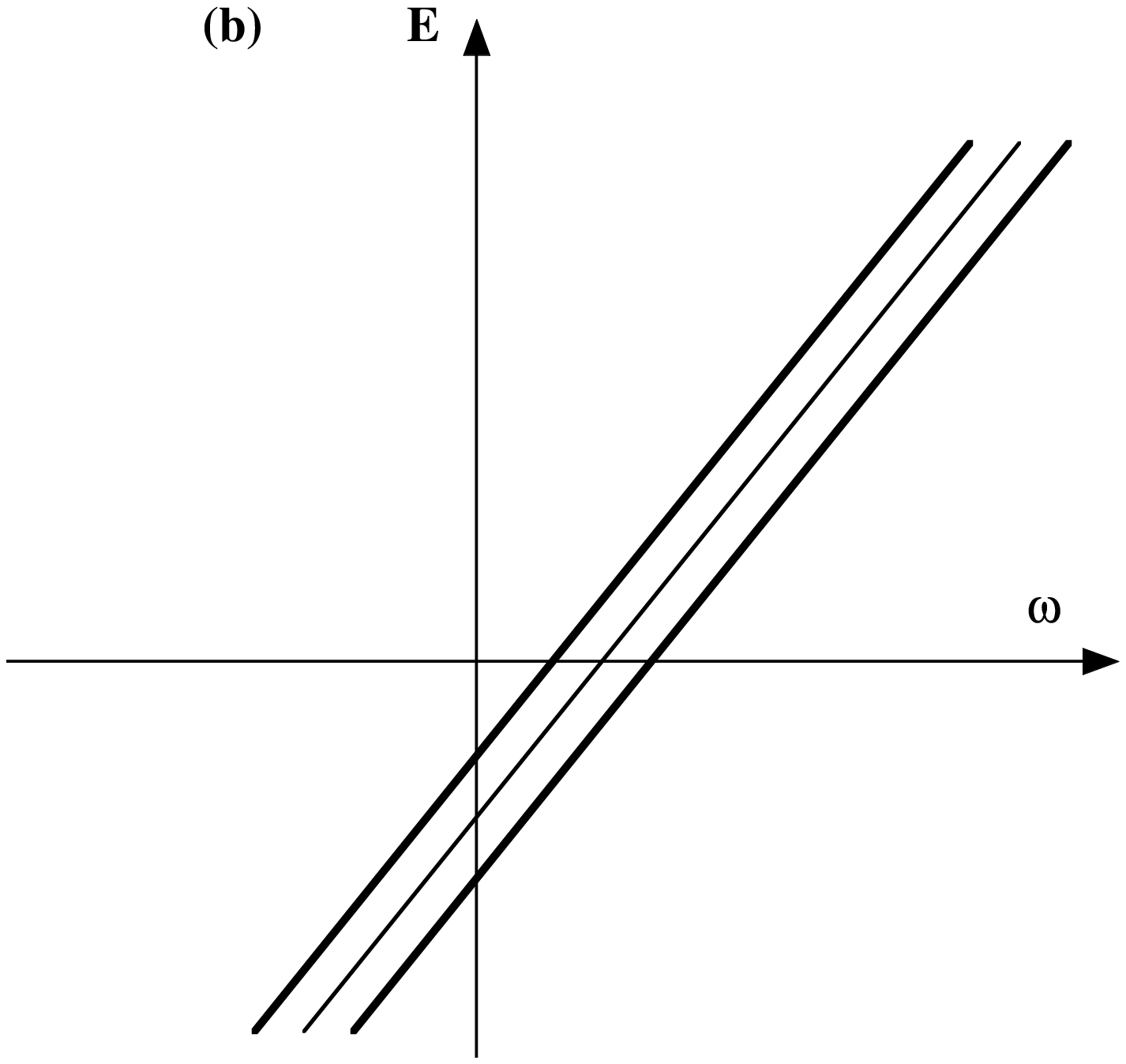}
\includegraphics[width=0.45 \textwidth]{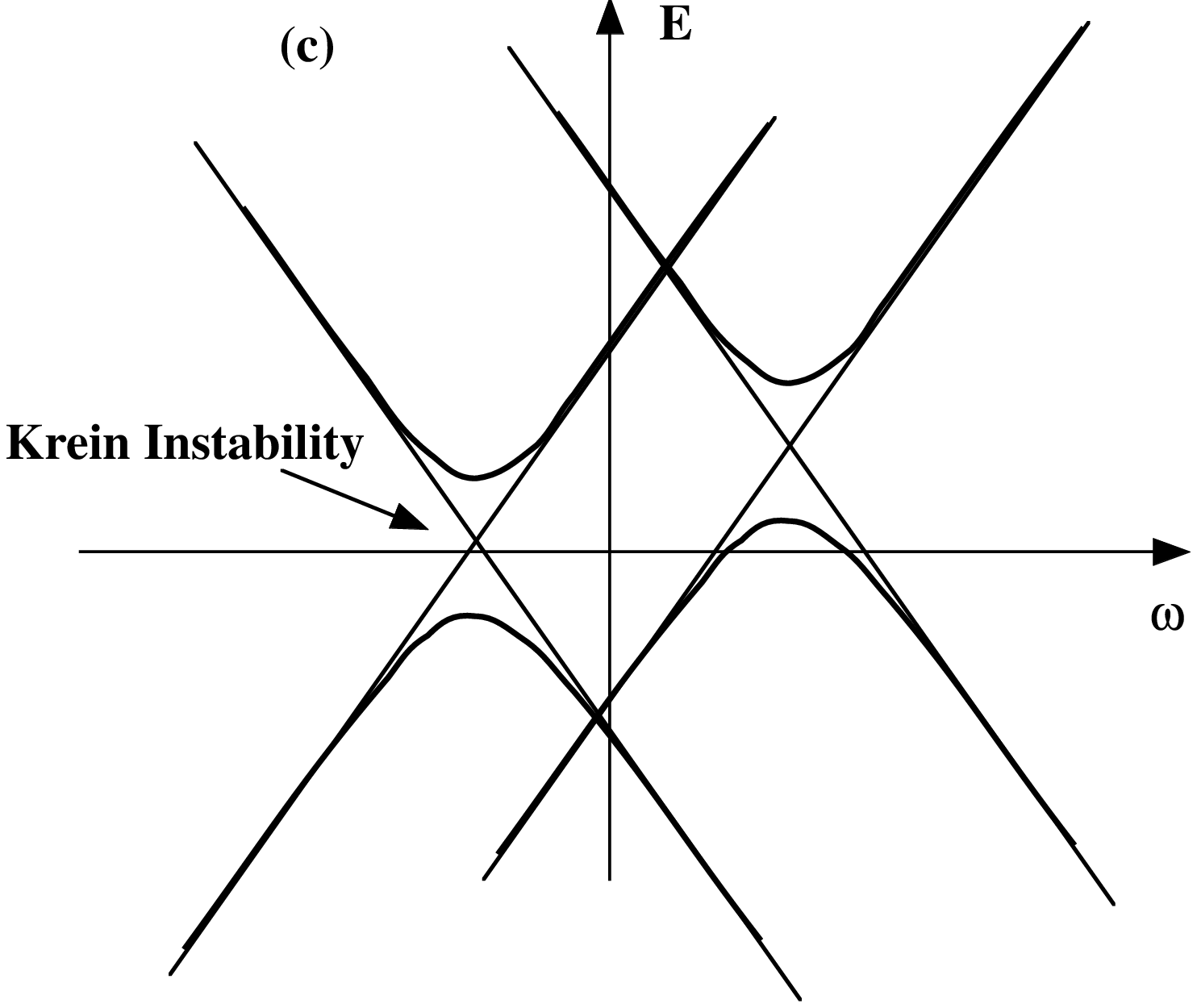}
\caption{(a) Schematical picture of the unperturbed band structure 
$E(\omega)=E_{0\pm}(q;\omega)$ 
(Eq.\ \ref{eq:bdh});
(b) Gap
opening for the first kind of degeneracy (\ref{degen1}) (no instability);
(c) Gap opening for the second kind of degeneracy (\ref{degen2}).
An instability is generated if the gap overlaps the axis $E=0$.}
\label{scheme}
\end{figure}

For the second kind of degeneracy, at transverse intersection points, a gap 
with width proportional to $|f_{p}|$ opens if 
\begin{equation}
E_{+}(q;\omega_p) = E_{-}(q+2pQ;\omega_p) \equiv E_p (\omega_p), 
\label{degen2}
\end{equation}
and a straightforward calculation yields the perturbed eigenvalues at 
$\omega_p$ to order 
$f_p$:\footnote{Here, the possible interference with resonances of the first 
kind (\ref{eq:qres}) has been neglected. The case of simultaneous 
resonances, which occurs at special points, is discussed in the Appendix.}
\begin{equation}
E^\prime_{p \pm} (\omega_p)=E_p (\omega_p) \pm 
\frac{|f_p \omega_p|} {\sqrt{\omega_p^2+f_0^2}} .  
\label{Eprime}
\end{equation}
If $E=0$ belongs to one of these gaps, the SW becomes
unstable (see Fig.\ \ref{scheme} (c)).
These intersections occur when 
\begin{eqnarray}
\frac{1}{C}\sqrt{\omega_{p}^{2}+f_{0}^{2}} & = & \cos{(q+2pQ)}
- \cos{q}
\label{omegcond}  \\
  E_{p} &=& 2C- \delta -2f_{0}-C \cos{(q+2pQ)}
  -C\cos{q} . 
\label{Econd}
\end{eqnarray}
It comes out readily from Eqs.\ (\ref{omegcond}) and (\ref{Econd}) that 
for a given  $p$, these intersection points are 
located on an ellipse  in the $(\omega,E)$-plane given by
\begin{equation}
\frac{\omega_p^2}{4 C^2 \sin^2 pQ-f_0^2}
+\frac{\left(2C-\delta-2f_0-E_p\right)^2} {4C^2 \cos^2 pQ-f_0^2 \cot ^2 pQ}=1 ,
 \label{ellips}
 \end{equation}
which exists for $| f_{0}| < 2C |\sin pQ| $, 
with its center at $\omega=0$ and $E=2C-\delta -2 f_{0}$.
Thus, this ellipse represents the locus of the middle of the gaps
which are opened by the perturbation of order $f_{p}$, and 
an instability of the SW is generated when this ellipse
intersects the axis $E=0$, that is when
\begin{equation}
(2C- \delta -2f_{0})^2 < 4 C^2 \cos^{2} p Q -f_{0}^{2} \cot^2 pQ .
\label{interEeq0}
\end{equation}
The existence of these ellipses of gaps is confirmed by numerical 
calculations of the set of bands $E(\omega)$ (see Fig.\ \ref{crossings}).
In the incommensurate case, $|p|$ may run from $1$ to $+\infty$, and thus there
are infinitely many ellipses which cover densely the rhombic area where the
set of curves $E_{-}(q;\omega)$ intersects the set of curves
$E_{+}(q^{\prime};\omega)$
for arbitrary $q$ and $q^{\prime}$. When $E=0$ crosses this area,
there are surely infinitely many ellipses intersecting this axis, but
the instabilities which are generated at these intersections become very
weak and numerically invisible when $p$ becomes large. 

\begin{figure}[htbp]
\begin{center}
\centerline{\includegraphics[height=8cm,angle=270]{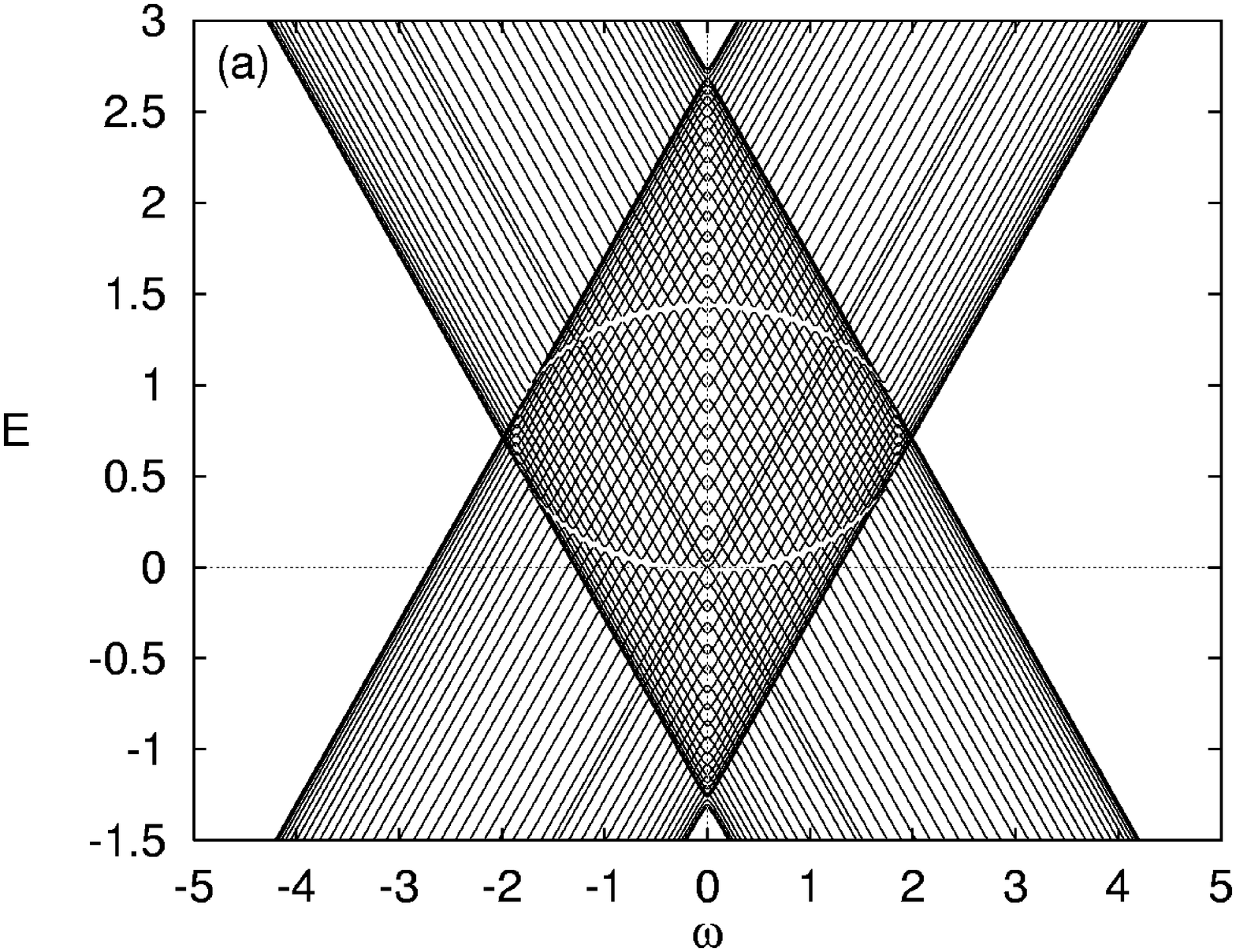}
\includegraphics[height=8cm,angle=270]{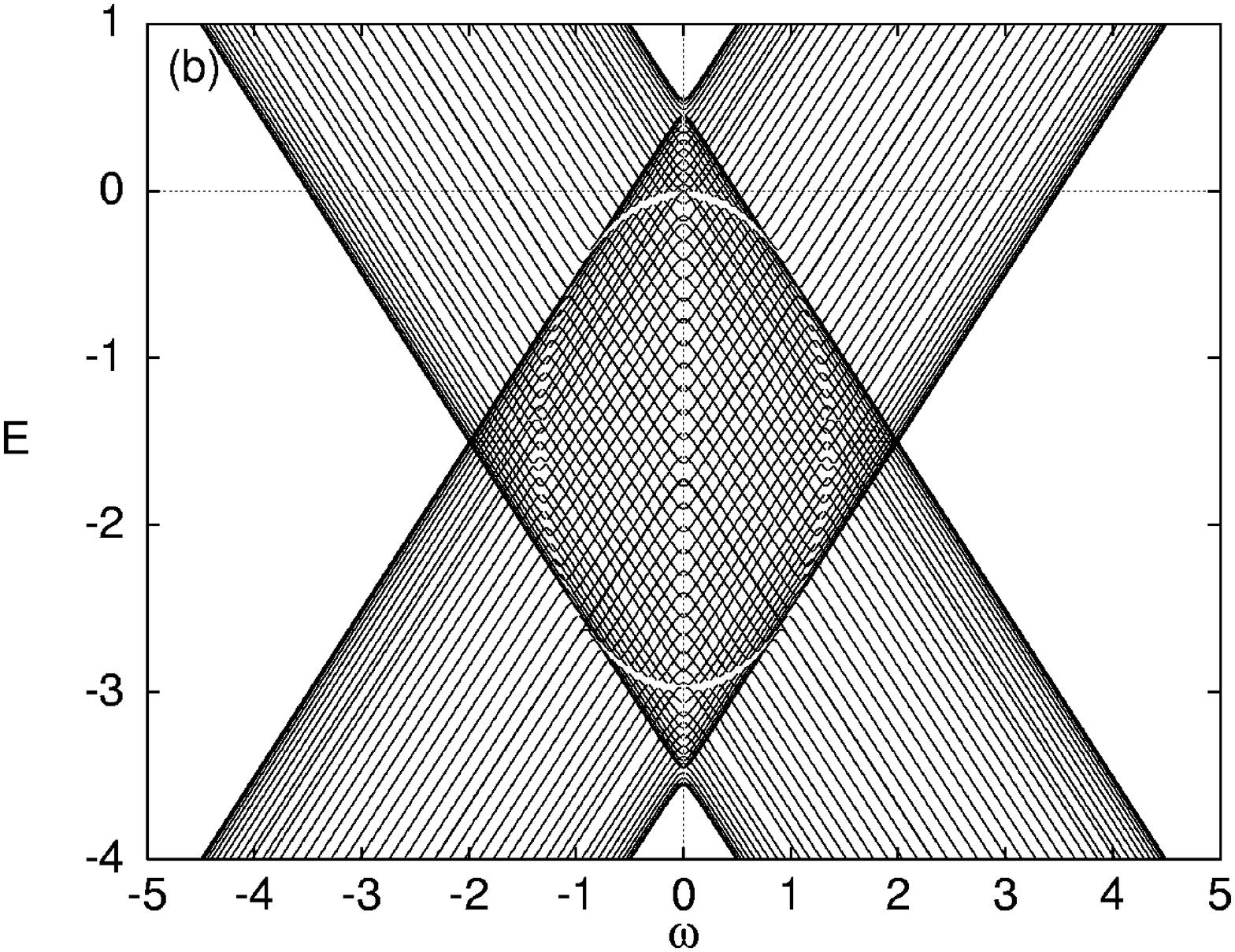}}
\caption{Band spectrum of (\ref{eigen_prob_her}) for a SW 
with (a) $Q/2\pi = 17/89  \simeq \sigma_G/2 $ ($Q < \pi/2$),
$\delta^\prime=1.23$, and (b) $Q/2\pi = 34/89 \simeq \sigma_G $ ($Q > \pi/2$),
$\delta^\prime=3.4$ ($C=1$). The visible gap
openings occur along the ellipse (\ref{ellips}) with  $p=1$.}
\label{crossings}
\end{center}
\end{figure}

As $p=1$ 
corresponds to the ellipse with the largest gaps, it yields the strongest 
instabilities if these gaps are opening around $E=0$. 
For $p=1$ and in the limit of zero amplitude SWs 
($f_{0}=0, \delta=\delta_0(Q)$), the particular 
ellipse (\ref{ellips}) for the intersections  ($\omega_{1},E_1$)
has the equation:
\begin{equation}
\frac{\omega_{1}^{2}}{4 C^2\sin^{2} Q}
+ \frac{(2C \cos Q-E_{1})^{2}}{4 C^2 \cos^{2}Q} = 1 . 
\label{ellipeq}
\end{equation}
Thus, it is tangent to the axis 
$E=0$ at $\omega=0$, and it lies above $E=0$ for wave vectors $0<Q<\pi/2$ 
and below for  $\pi/2<Q<\pi$. When the SW amplitude $\epsilon_s$ is small but 
nonzero, we obtain using Eqs.\ (\ref{rotangle}) and (\ref{epsilon}) that, 
to lowest order in $\epsilon_s$, the 
'midgap' ellipse (\ref{ellips}) with $p=1$ cuts the axis $\omega=0$ at 
$E=-\epsilon_s^2/4$. Then, using Eq.\ (\ref{Eprime}) to calculate the 
perturbed eigenvalues for small but nonzero 
$|\omega_1|$ and $|f_0|\ll|\omega_1|$ we obtain, again by using 
(\ref{epsilon}), that 
$E_{1-}^\prime \approx -\epsilon_s^2/2$ and  
$E_{1+}^\prime \approx 0 $ to order $\epsilon_s^2$ 
when $|\omega_1|$ is small but nonzero. Thus, we should expect that 
when $0<Q<\pi/2$ (so that the unperturbed ellipse is above the axis $E=0$), 
gaps opening in the lower part of the
ellipse will surround the axis $E=0$ for some interval of $|\omega|$ close to  
$\omega=0$.\footnote{A similar situation with first order instability 
occurs also for all commensurate SWs with $Q \ne \pi$, see Appendix.} 
As a consequence, 
{\em SWs with $0<Q<\pi/2$
become unstable through first-order oscillatory instabilities} for 
arbitrarily small amplitude 
$\epsilon_s$ (Fig.\ \ref{crossings} (a)).  
By contrast, as  for $\pi/2<Q<\pi$  the unperturbed ellipse 
lies below the axis $E=0$, we expect 
all first-order gaps outside the immediate neighbourhood of $\omega=0$ 
to open {\em strictly below} $E=0$. Moreover, as for analytic SWs there are 
two explicit eigensolutions with $E=0$ at $\omega=0$ (the phase mode and the 
sliding mode, see Appendix \ref{subsec2}), there is no gap around $E=0$
(and consequently no instability) for $\omega=0$. We also  
confirmed by numerical calculations (Figs.\ \ref{crossings} (b) and 
\ref{ellipsecd} (a)) that no first-order gaps open around $E=0$, 
so that for SWs with $\pi/2<Q<\pi$
{\em no first-order instabilities} develop.
However, it is important to remark that the above estimate based on 
Eq.\ (\ref{Eprime}) is not valid in 
the immediate neighbourhood of $\omega=0$. Actually, in the zero-amplitude 
limit $\epsilon_s=0$, the point 
$E_{1}=0,\omega_{1}=0$
where the unperturbed first-order ellipse (\ref{ellipeq}) is tangent to the 
axis $E=0$ is such that
$E_{0+}(0)=E_{0-}(0)=0$, which implies by Eqs.\ (\ref{eq:bdh})
that $q=\pm Q$ so that also the first resonance condition (\ref{eq:qres}) 
is fulfilled 
for $|p|=1$. Thus, close to the origin $E=0,\omega=0$ the two different kinds 
of gap openings strongly {\em interfere} with each other at the same order 
($p=1$), and thus must be considered simultaneously as described in the 
Appendix.

\begin{figure}[htbp]
\begin{center}
\centerline{\includegraphics[height=8cm,angle=270]{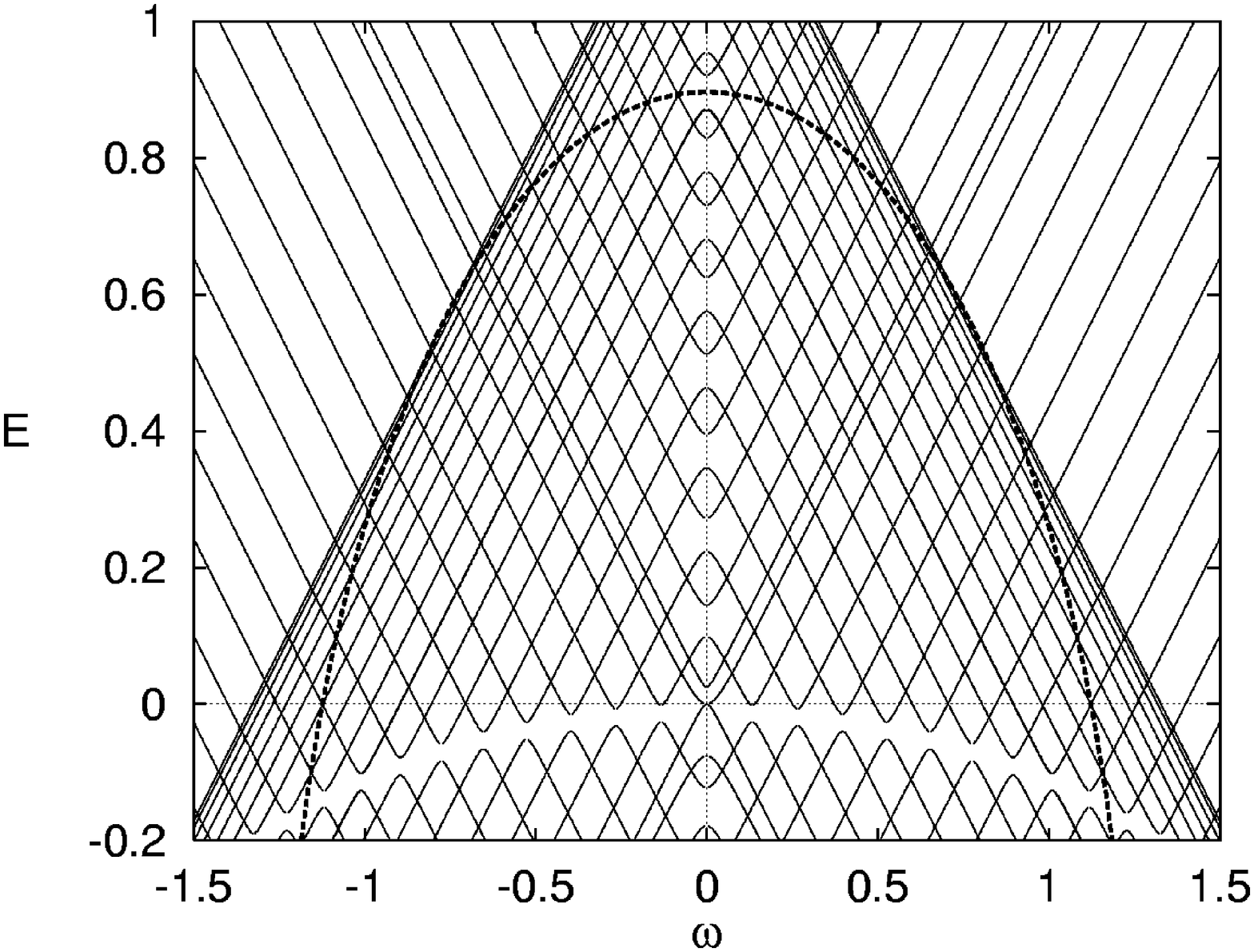}
\includegraphics[height=8cm,angle=270]{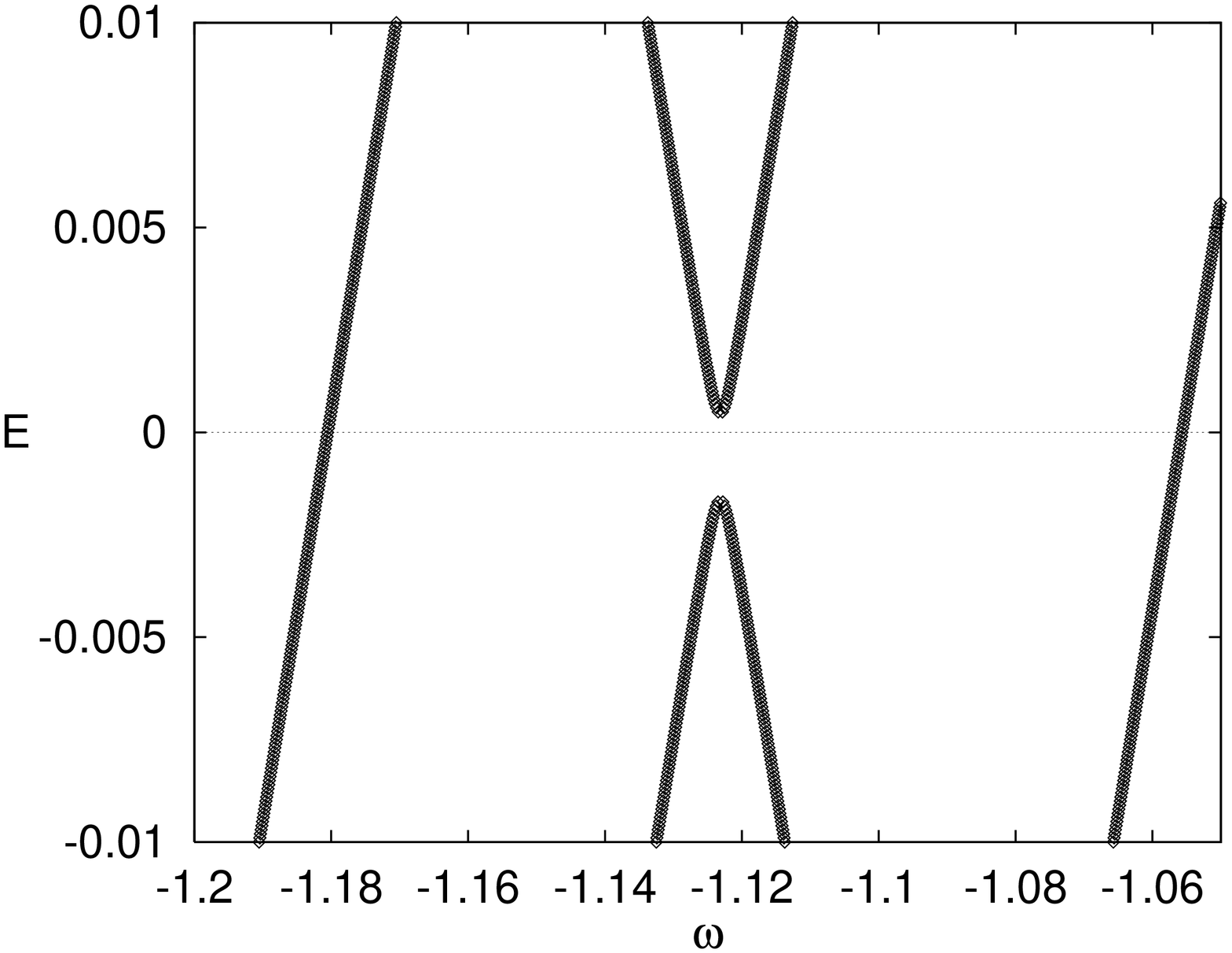}}
\caption{Part of the band spectrum of (\ref{eigen_prob_her}) for a SW 
with  $Q/2\pi = 27/89$ ($Q > \pi/2$) and 
$\delta^\prime=2.62$. The 
thick dashed line in the left figure is the second-order ellipse 
($p=2$ in Eq.\ (\ref{ellips})). Right figure shows a magnification of the part 
around $E=0$ where a second-order gap opens.}
\label{ellipsecd}
\end{center}
\end{figure}

Thus, when $\pi/2<Q<\pi$, the strongest instability appears for the smallest 
$p>1$ such that the condition (\ref{interEeq0}), which 
in the zero-amplitude limit reduces to 
$|\cos pQ|$ $>$ $|\cos Q|$, is fulfilled. An example with $p=2$ is illustrated 
in Fig.\ \ref{ellipsecd}.
Then, {\em there always exists a $p$ such that, for arbitrarily small 
amplitude, the SW is unstable to order $p$}. Anyway, since the width of the 
gap openings, 
and therefore also the maximum instability growth rates, are proportional to 
$|f_p|$,
the instabilities  become very weak for $Q$ close to $\pi$ as the smallest 
$p$ yielding an instability 
becomes large,  and they could be practically invisible in numerical 
simulations for finite-size systems. We
might therefore describe these analytic SWs as 'quasi-stable'.

\section{Numerical analysis of SWs in KG chains}\label{sec_KG}

Let us now return to the original KG model (\ref{DKG}), for which we will 
investigate numerically the validity of our previous results when the 
conditions of small amplitude and small coupling are not necessarily 
fulfilled. In general, when the SW amplitude is non-negligible also 
harmonics $a_n^{(p)}$ with $|p|\neq1$ in the Fourier expansion (\ref{fseries}) 
will affect 
the dynamics. Moreover, as already mentioned in Sec.\ \ref{sec_SAL} 
in connection with 
Eqs.\ (\ref{elim0})-(\ref{etaAprime}), when the coupling $C_K$ is increased 
also the small-amplitude dynamics will in general be affected as the 
long-range nonlinear interaction terms in the extended DNLS equation become 
non-negligible, and in particular resonances may 
occur also for small-amplitude SWs if higher harmonics of the solution 
enter the linear phonon band. 
We will consider the particular cases of the soft Morse 
potential (\ref{morse}) and the hard quartic potential (\ref{quartic}). 
As described in Sec.\ \ref{swams}, the SWs are obtained by numerical 
continuation of the multibreathers with coding sequences defined by 
Eqs.\ (\ref{code})-(\ref{hull}) (modified according to footnote \ref{hard} in 
the case of a hard potential), using a modified Newton scheme 
(described in detail in Ref.\ \cite{KA99}) to find numerically exact 
time-reversible solutions $\{u_n(t)\}$ with a given time period 
$T_b=2\pi/\omega_b$. The 
linear stability 
of these solutions is then determined by numerical integration of the Hill 
equations (\ref{phonon}) over one period $T_b$, defining a linear symplectic 
$2N$-dimensional map given by the Floquet matrix ${\bf T_0}(\{u_n\})$: 
\begin{equation}
\left(\begin{array}{c}\{\epsilon_n(T_b)\} \\ \{\dot{\epsilon}_n(T_b)\} 
\end{array} \right) = {\bf T_0} 
\left(\begin{array}{c}\{\epsilon_n(0)\} \\ \{\dot{\epsilon}_n(0)\} \end{array} 
\right) .
\label{Floquet}
\end{equation}
As ${\bf T_0}$ is symplectic, for each eigenvalue $\Lambda$ there are also 
eigenvalues at $\Lambda^\ast$ and $1/\Lambda$. Then, the solution 
$\{u_n(t)\}$ is linearly stable if and only if all the
eigenvalues of ${\bf T_0}(\{u_n\})$ lie on the unit circle. The Krein 
signature associated to a complex conjugated pair of eigenvalues 
$\Lambda=r\e^{\pm \ii \theta}$ 
is defined as \cite{Arnold,Aub97}
\begin{equation}
{\mathcal K}(\theta) = {\rm sign} \left( \ii \sum_n  
\left[ \epsilon_n \dot{\epsilon}_n^\ast - 
\epsilon_n^\ast \dot{\epsilon_n}\right] \right) ,
\label{KreinKG}
\end{equation}
where $\{\epsilon_n, \dot{\epsilon}_n\}$ is the eigenvector of ${\bf T_0}$ 
with eigenvalue $r\e^{+ \ii \theta}$. As before, instabilities can only occur 
through collisions of eigenvalues with opposite Krein signature. 

At the 
anticontinuous limit $C_K=0$, each site with code  $|\tilde{\sigma}_n|=1$ 
yields a degenerate pair of eigenvalues at +1, while each site with code  
$\tilde{\sigma}_n=0$ yields a complex conjugated pair of eigenvalues at 
$\e^{\pm \ii 2\pi/\omega_b}$. When $C_K$ is increased from zero, the 
degeneracies are raised and the spreading of the eigenvalues is just 
analogous to the DNLS case discussed in the previous section. For type $H$ 
SWs all eigenvalues will initially remain on the unit circle 
(constituting bands for commensurate SWs and Cantor-spectra for 
incommensurate SWs as for the DNLS case), preserving the linear stability 
until the first oscillatory instability appears through a collision on the 
unit circle between eigenvalues of different Krein signature originating 
from $+1$  and $\e^{\pm \ii 2\pi/\omega_b}$, respectively. For type $E$ SWs, 
in addition pairs of eigenvalues originating from +1 will 
immediately go out on the real axis when $C_K$ becomes nonzero. Thus, in 
general, the instability scenario we observe is as expected identical to that 
of the DNLS model, as long as the amplitude and coupling
remain small, and no higher-order resonances occur. 
The scenario outside this regime is however quite different 
for the hard quartic and the soft Morse potentials, and we will therefore 
discuss them separately below. 

\subsection{The quartic potential}
\label{sec_quartic}

Let us first discuss SWs with wave vector $Q$ in the hard quartic potential 
(\ref{quartic}), 
having frequencies $\omega_b>\omega_l(Q)>1$ (c.f.\ Eq.\ (\ref{lin_disp}) 
and Fig.\ \ref{fig_planeCw}). This case is particularly simple 
since the on-site potential is 
spatially symmetric, $V(u)=V(-u)$, and consequently the time-periodic 
solutions with 
fundamental period $T_b=2\pi/\omega_b$ fulfill $u_n(t+T_b/2)=-u_n(t)$ 
(\ie, the time-reversible solutions are also time antisymmetric with respect 
to $t=T_b/4$), so that we can restrict our numerical Newton algorithm to 
search for solutions fulfilling this condition. Note that this condition is 
not fulfilled for any even higher harmonic of the solution (\ie, with period 
$T_b/p$, $p$ even), so that there can be no resonances with even harmonics 
disturbing the unique continuation of the SW. Actually, we 
already remarked in Sec.\ \ref{sec_SAL} that there is no coupling between 
even and odd harmonics of the solutions for the quartic potential. Moreover, 
as $V''(u_n)$ in the Hill equations (\ref{phonon}) in this case has the 
fundamental period $T_b/2$ rather than $T_b$, it is convenient to 
redefine the Floquet matrix from (\ref{Floquet}) as 
\begin{equation}
\left(\begin{array}{c}\{-\epsilon_n(T_b/2)\} \\ \{-\dot{\epsilon}_n(T_b/2)\} 
\end{array} \right) = {\bf T_0^\prime}
\left(\begin{array}{c}\{\epsilon_n(0)\} \\ \{\dot{\epsilon}_n(0)\} \end{array} 
\right) .
\label{Floquetq}
\end{equation}
Then, the eigenvalues of ${\bf T_0^\prime}$ corresponding to sites with code 
$\tilde{\sigma}_n=0$ will at the anticontinuous limit be located at 
$\e^{\pm \ii \pi/\omega_b}$, \ie, at half the angle of the corresponding 
eigenvalues of $ {\bf T_0}$. 
 
\begin{figure}
\centerline{\includegraphics[height=7cm,angle=270]{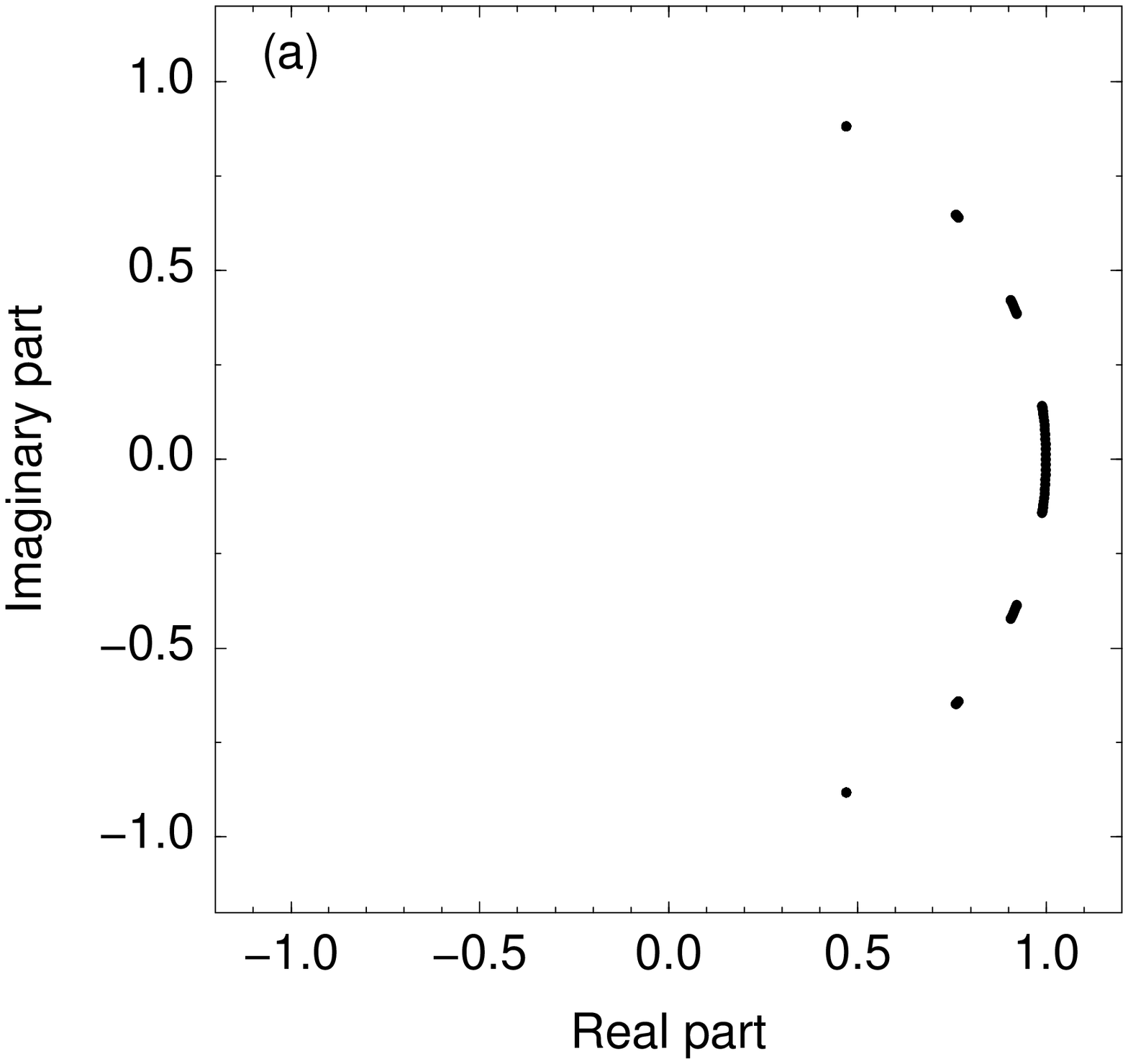}
\includegraphics[height=7cm,angle=270]{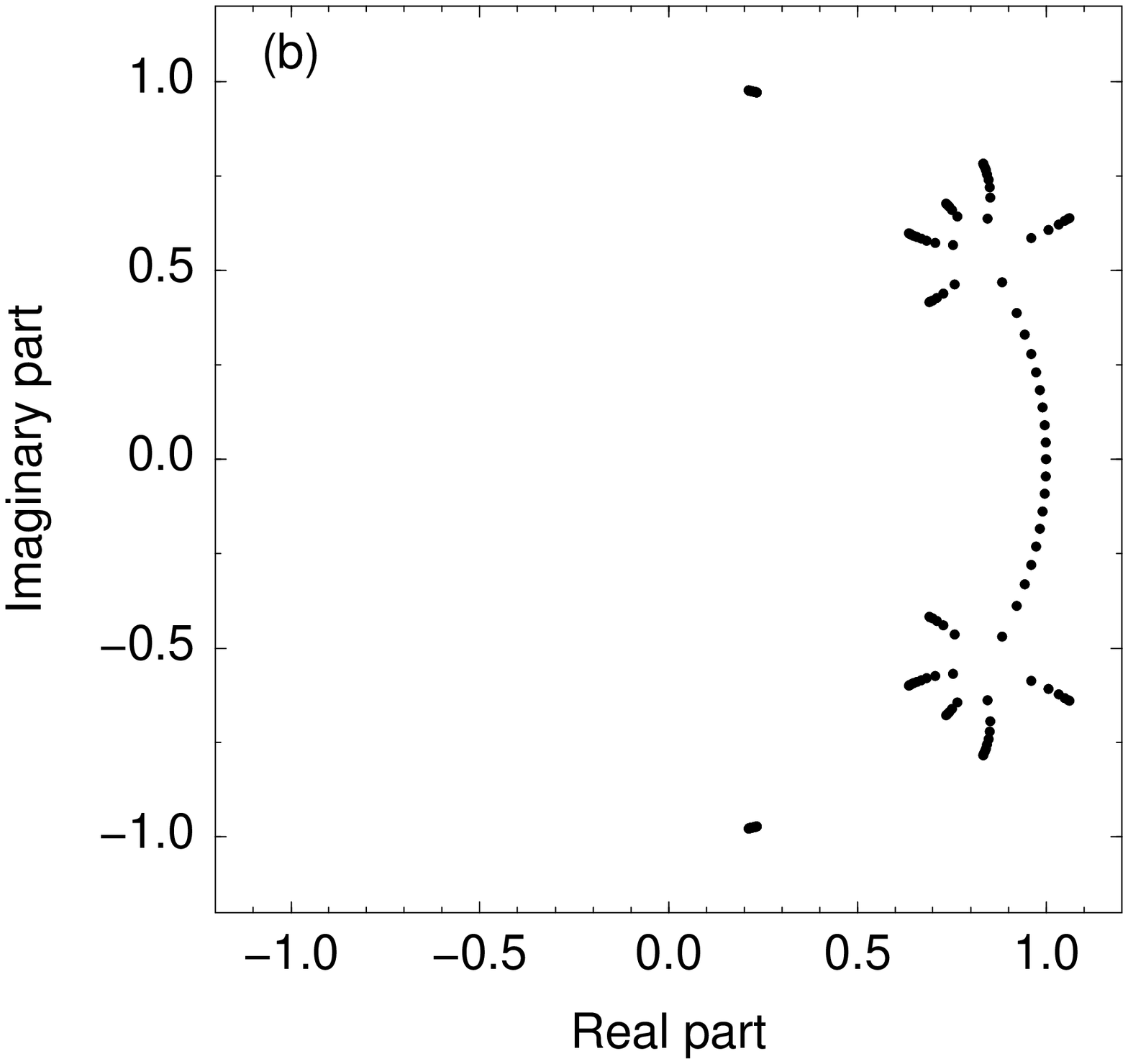}}
\centerline{\includegraphics[height=7cm,angle=270]{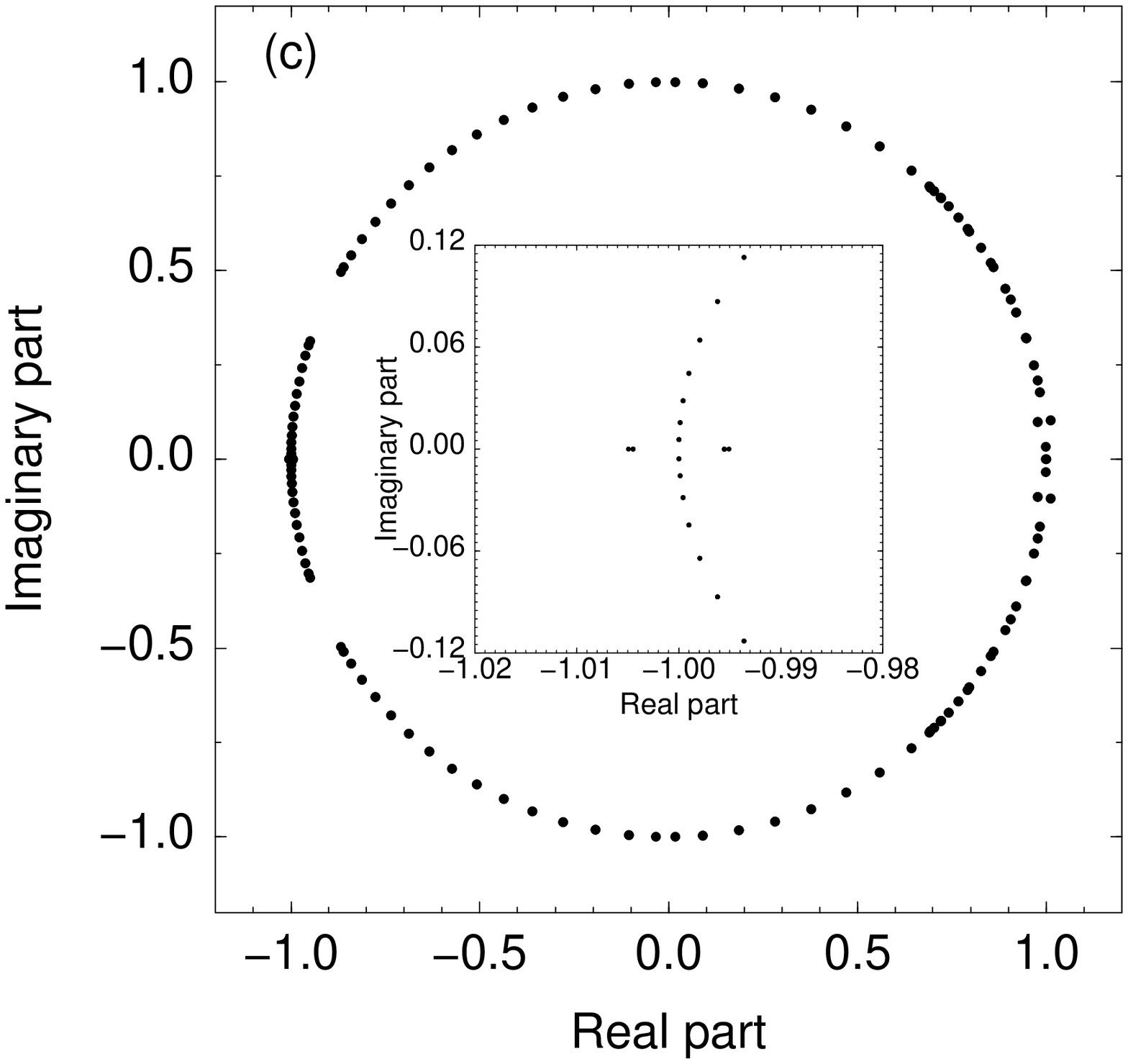}
\includegraphics[height=7cm,angle=270]{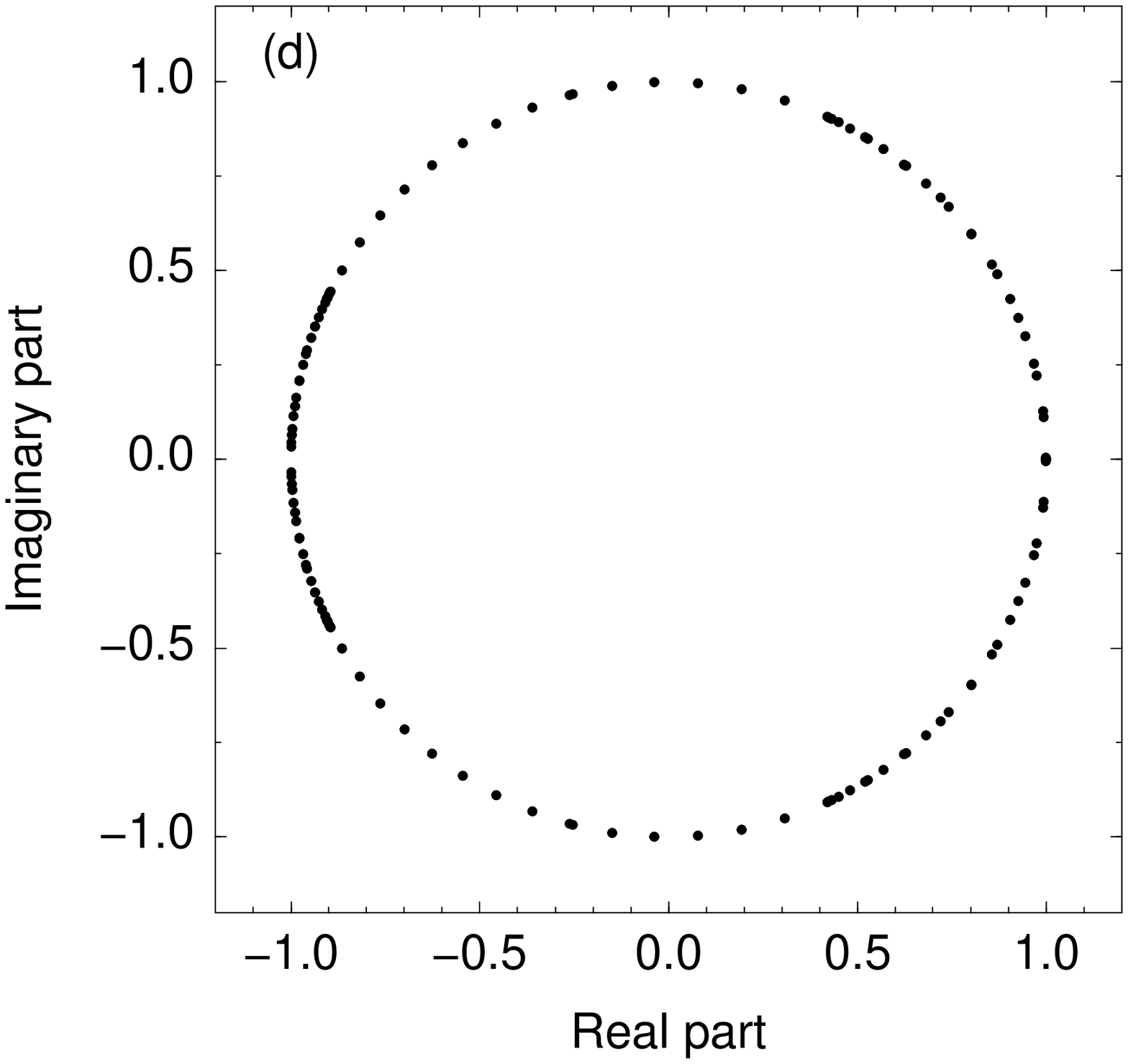}}
\caption{Eigenvalues of the Floquet matrix ${\bf T_0^\prime}$ for a type $H$ 
SW 
with wave vector $Q=\pi/4$ in 
a finite-size KG lattice ($N=120$) with the  
quartic potential (\ref{quartic}), $\omega_b=1.8$, and increasing $C_K$: 
(a) $C_K=0.199$ 
(stable wave); (b) $C_K=0.699$ (oscillatory instabilities); (c) $C_K=2.773$ 
(additional non-oscillatory instabilities due to collisions at $-1$ 
resulting from second-harmonic resonances, see inset); and 
(d) $C_K=3.399$ (return to 
stability close to the linear limit for finite-size system).}
\label{fig_Floq1}
\end{figure}

Considering now small-amplitude solutions with increasing 
coupling $C_K$, we find for different wave vectors $Q$ three qualitatively 
different scenarios, depending on whether the second and/or third harmonic 
of the SW frequency $\omega_b$ enters the linear phonon band. In general, 
the condition that the $p$th harmonic of a SW with wave vector $Q$ 
has entered the phonon band can be written as 
\begin{equation}
\sqrt{1+4C_K\sin^2{\frac{Q}{2}}} < \omega_b < \frac{1}{p}\sqrt{1+4C_K} .
\label{3omega}
\end{equation}
This condition is fulfilled only when
\begin{equation}
\sin{\frac{Q}{2}} < \frac{1}{p} \quad \mbox{and} \quad
C_K > \frac{p^2-1}{4\left(1-p^2\sin^2{\frac{Q}{2}}\right)} .
\label{Qlimit}
\end{equation}
Thus, for SWs with wave vector $Q>\pi/3$ there are no higher order 
resonances. 
In this case, the SWs can for fixed frequency $\omega_b$ be continued 
versus $C_K$ until they reach their linear limit at 
$C_K=\frac{\omega_b^2-1}{4 \sin^2{\frac{Q}{2}}}$ 
(where $\omega_b=\omega_l(Q)$ according to (\ref{lin_disp})), and the 
instability scenario during this continuation is completely analogous to 
that observed in the DNLS approximation discussed in the previous section, 
also when $\omega_b$ is not close to 1. On the other hand, when $Q<\pi/3$ 
the second harmonic enters the phonon band before the SW reaches its linear 
limit if $\omega_b$ is large enough. As discussed above, this resonance does 
not affect the unique 
continuation of the 
SW which still can be continued to its linear limit, but causes  
additional  non-oscillatory instabilities which are signalled by collisions 
of eigenvalues of ${\bf T_0^\prime}$ at $-1$. An example 
of the instability scenario observed in this regime is illustrated  in 
Fig.\ \ref{fig_Floq1}.  

\begin{figure}
\centerline{\includegraphics[height=8.5cm,angle=270]{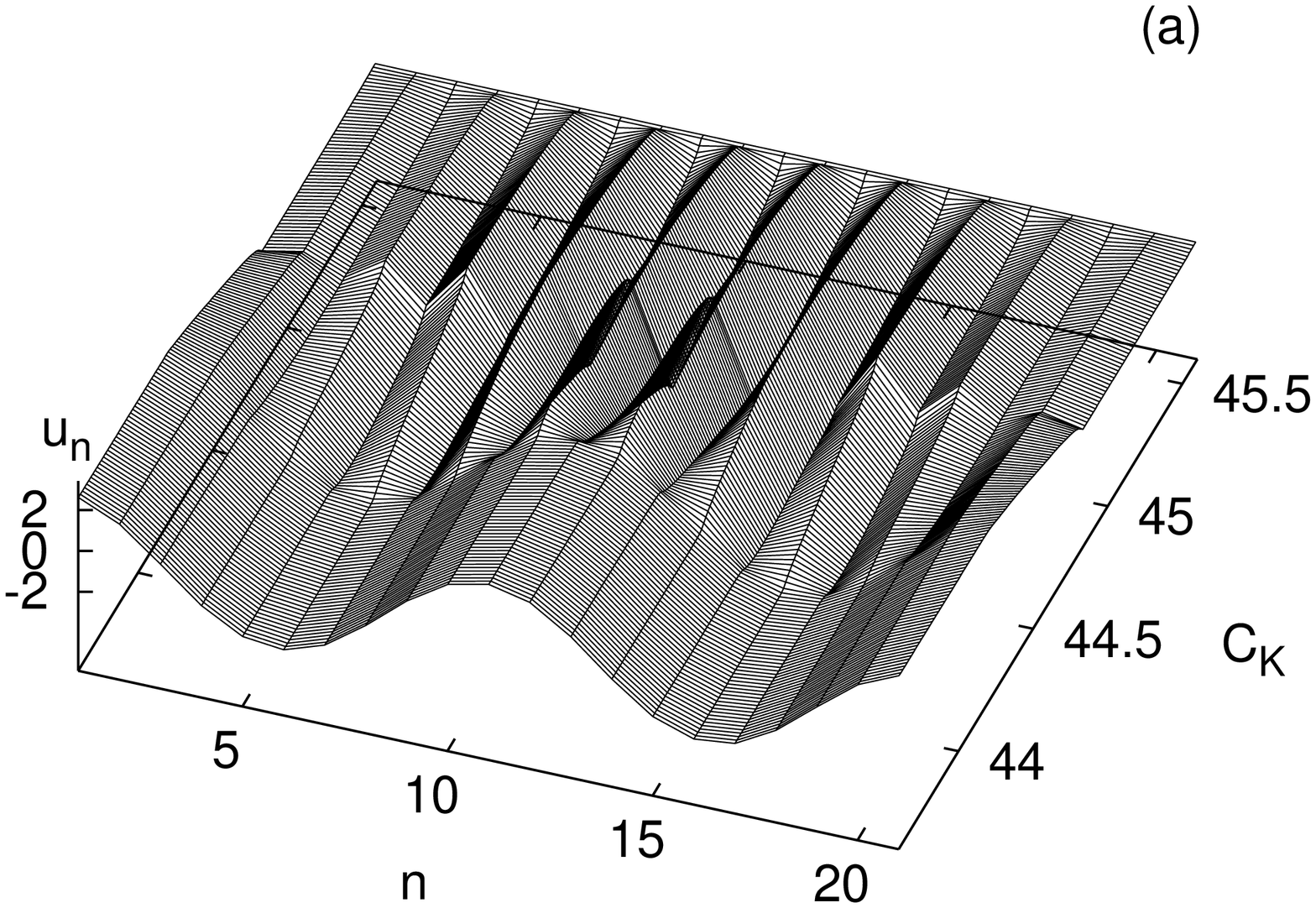}
\includegraphics [height=8.5cm,angle=270] {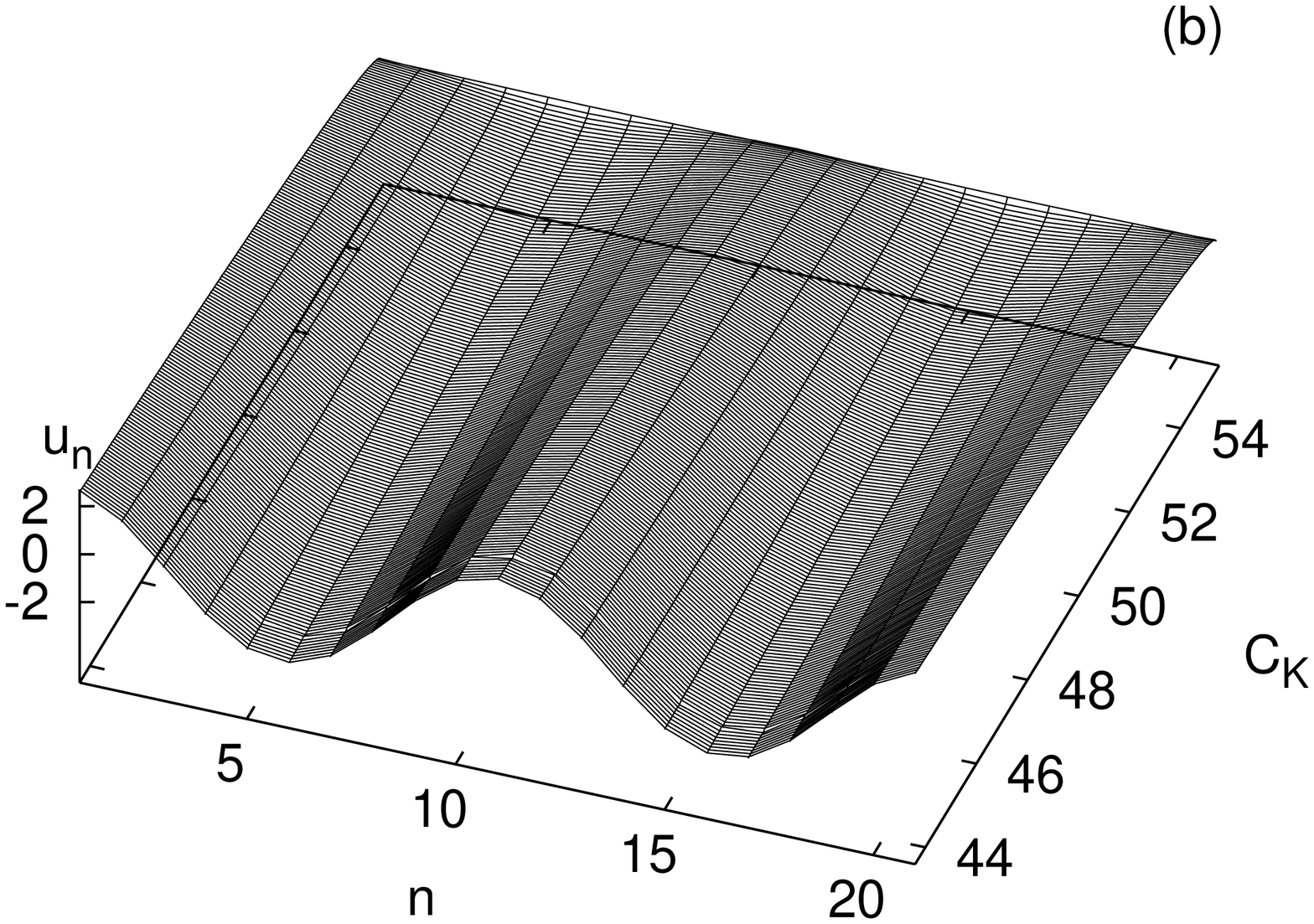}}
\centerline
{\includegraphics[height=7cm,angle=270]{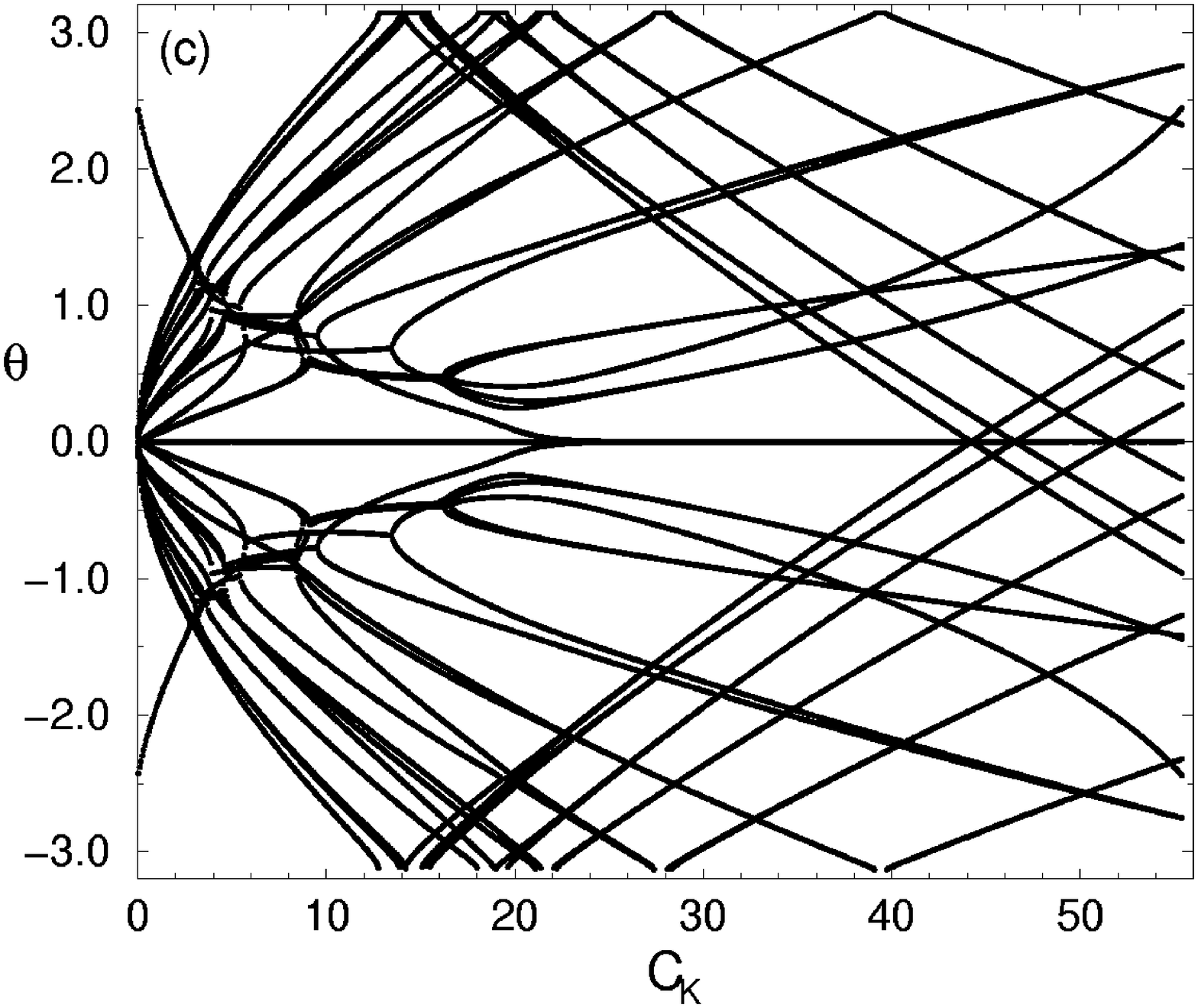}
\includegraphics[height=7cm,angle=270]{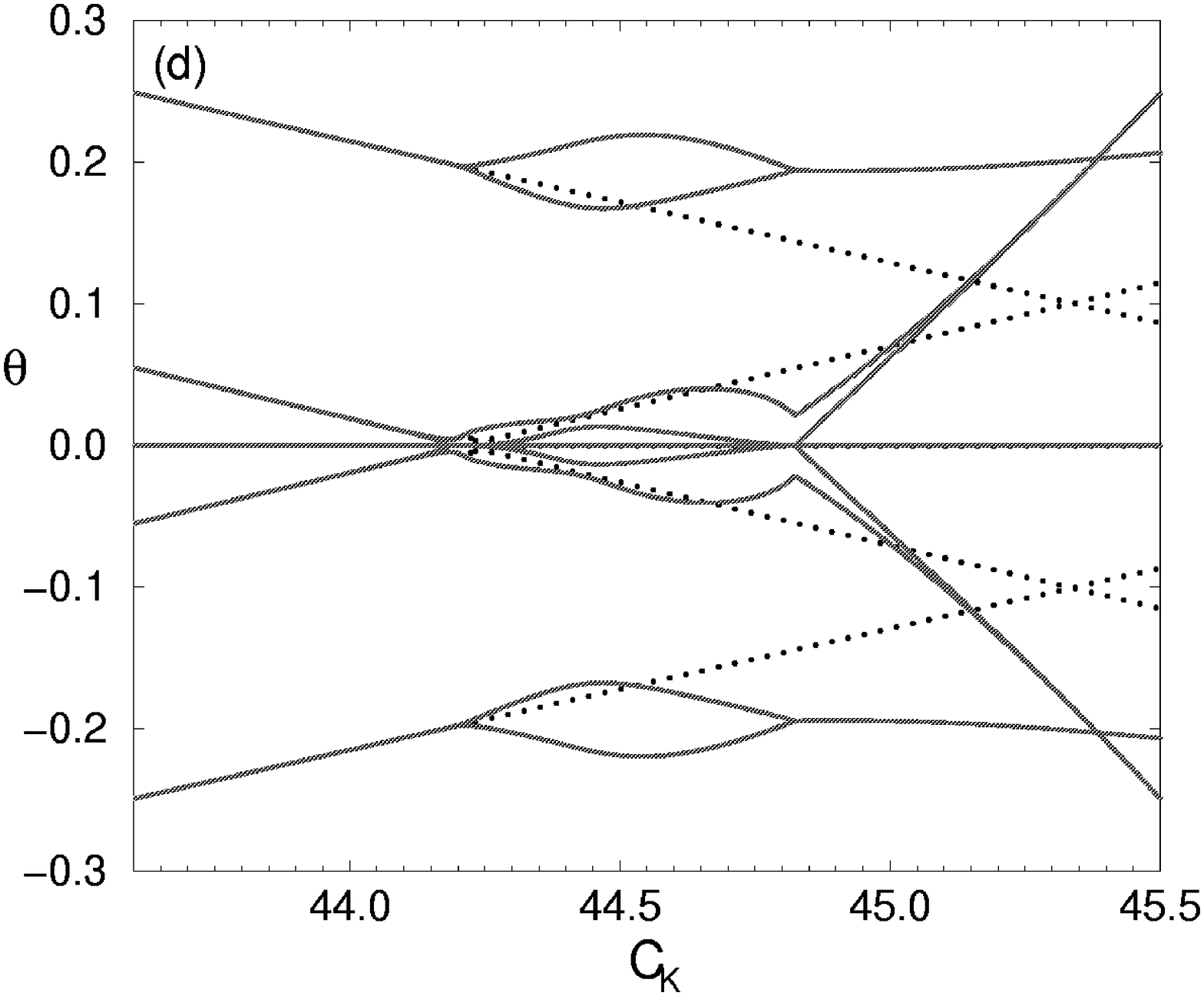}}
\centerline{\includegraphics[width=5cm,angle=270]{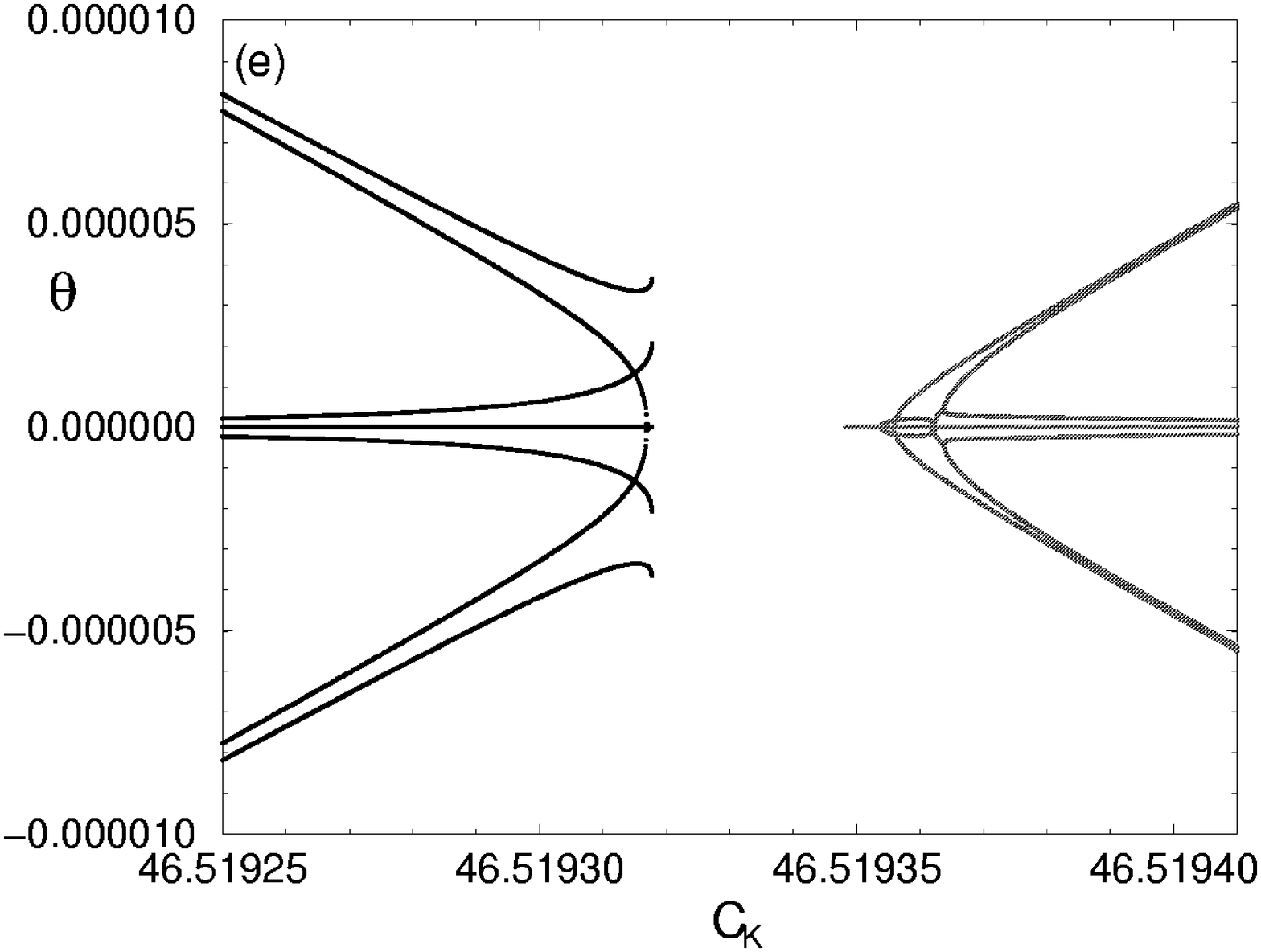}
\includegraphics [width=5cm,angle=270] {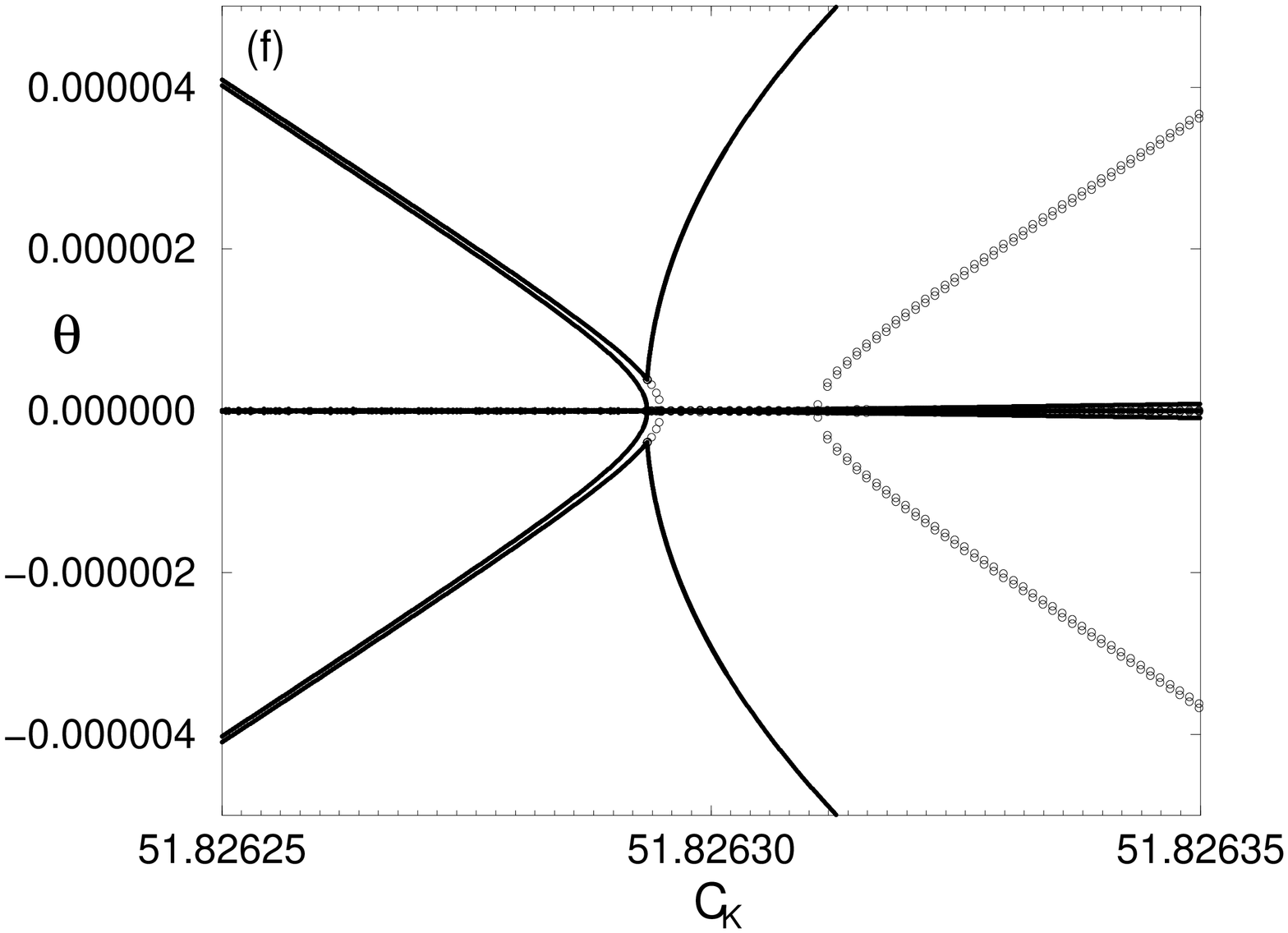}}
\caption{Bifurcation scenario resulting from third-harmonic resonances in 
the 
quartic model. The initial type $H$ SW has wave vector $Q=4 \pi/21$ and 
frequency  
$\omega_b=4.5$. (a) Smooth continuation for increasing $C_K$ of a SW with 
wave vector $Q$ generated from the anticontinuous limit into a type $H$ SW 
with 
frequency $3\omega_b$ and wave vector $Q^\prime=20\pi/21$. (b) Intentionally 
'careless' continuation from the linear limit for decreasing $C_K$ of a SW 
with wave vector $Q$, where the step size in $C_K$ has been chosen large 
enough to jump over all resonance regimes (note the small visible gap at 
$C_K\approx 44.2$ corresponding to the resonance in (a)). (c) Arguments of 
all eigenvalues $r\e^{\ii \theta}$ of the Floquet matrix ${\bf T_0^\prime}$ 
versus coupling for one unit cell ($N=21$) of the SW with wave vector $Q$ in 
(b). (d) Magnification of 
the 
first resonance in (c). The solid line represents the smooth continuation 
from below as in (a), the dotted line the continuation from above as in (b) 
which stops at $C_K\approx 44.22$. (e), (f) Magnification of the second and 
third resonances, respectively (see text for explanation). }
\label{fig_3res}
\end{figure}

The third regime occurs when $\sin{\frac{Q}{2}}<\frac{1}{3}$, or 
$Q\lesssim 0.216 \pi$. Then, if $\omega_b$ is large enough so that the 
conditions (\ref{3omega})-(\ref{Qlimit}) are fulfilled for $p=3$, 
the SW with frequency $\omega_b$ will generally bifurcate with solutions 
with frequency $3\omega_b$ before reaching the linear limit. These 
bifurcations are signalled by collisions at $+1$ in the Floquet eigenvalue 
spectrum of ${\bf T_0^\prime}$. We will here not attempt 
to give any rigorous description of the possible nature of these (sometimes 
rather complicated) bifurcations, 
but rather make some qualitative general observations (more details are 
given in \cite{AnnaPhD}). 
A typical example of a scenario resulting from such 
bifurcations is illustrated in  Fig.\ \ref{fig_3res}. In this case, the 
continuation from the anticontinuous limit of the SW with wave vector 
$Q=4\pi/21$ and $\omega_b=4.5$ yields the first collision at +1 resulting from 
a third-harmonic resonance at $C_K\approx 44.18$ (see Fig.\ \ref{fig_3res} (c) 
and the magnification in (d)). This is slightly before 
the third harmonic enters the linear phonon band (Eq.\ (\ref{3omega}) yields 
$C_K=45.3125$), since the wave amplitude is non-negligible and the resonating 
wave is a nonlinear rather than a linear SW. As illustrated in 
Fig.\ \ref{fig_3res} (a) and (d), it is possible to continue smoothly the 
solution through this bifurcation, but the nature of the solution changes 
drastically. In fact, this scenario typically occurs in two steps: at the 
first bifurcation point the solution changes into an intermediate 
(unstable) solution, and later, at a second bifurcation point 
($C_K\approx 44.8$ in Fig.\ \ref{fig_3res}) into the final solution. The 
latter 
is the SW whose wave vector $Q^\prime$ (close to $\pi$) coincides with the 
wave vector of the resonating eigenmode\footnote{A necessary
 resonance 
condition for the quartic model is $Q^\prime=mQ$ mod $2\pi$, 
$m$ odd. The strength of the resonance generally decreases with 
$|m|$.}\label{note12}, and it corresponds 
to an anticontinuous coding sequence generated by 
Eqs.\ (\ref{code})-(\ref{hull}) but with the coding sequence 
$\tilde{\sigma}_n $ taking the values $\pm 3$ instead of $\pm 1$. Then, 
the smooth continuation versus $C_K$ for fixed $\omega_b$ follows this new 
solution without further bifurcations, until it reaches its linear limit 
at $C_K=\frac{9\omega_b^2-1}{4 \sin^2{\frac{Q^\prime}{2}}}$. 

However, it is 
important to note that there generally exists also a solution with the 
original wave vector 
$Q$ close to the linear limit (in smooth continuation of the linear SW with 
wave vector $Q$), 
but due to the third-harmonic resonances
there should in this case be no continuous path in parameter space 
connecting this  
linear SW with the SW with wave vector $Q$ generated from the anticontinuous 
limit. However, if the numerical continuation is performed with less care, 
the resonances might be missed and an apparently (but falsely!) smooth 
continuation of the SW with wave vector $Q$ from the linear to the 
anticontinuous limit can be observed as illustrated in 
Fig.\ \ref{fig_3res} (b)-(c). The more careful numerics yields the behaviour 
close to the three resonances illustrated in Fig.\ \ref{fig_3res} (d)-(f). 
Thus, we find that there exists a SW with the original wave vector $Q$ 
also immediately after the first resonance, but attempts to continue it 
towards smaller $C_K$ (dotted line in Fig. \ref{fig_3res} (d)) fail for 
$C_K\approx 44.22$. Similarly, the attempts to continue it towards larger 
$C_K$ fail when approaching the second resonance at $C_K\approx 46.51932$ 
(left part of Fig. \ref{fig_3res} (e))\footnote{However, a smooth (but 
non-monotonous) 
continuation into a SW with wave vector $Q''=18\pi/21$, analogous to that in 
Fig.\ \ref{fig_3res} (a), can be obtained in the following way 
\cite{AnnaPhD}: at $C_K\approx 46.51932$, there is a bifurcation with a 
second solution, which can be followed smoothly by {\em decreasing} $C_K$ 
until (at $C_K\approx 46.51914$) it bifurcates with a third solution, which, 
again by {\em increasing} $C_K$, can be followed as it smoothly changes
into the SW with wave vector $Q''$. This 'Z-type' behaviour, with the 
simultaneous existence of three different solutions for the same $\omega_b$
and $C_K$ in some regime, typically occurs in systems with effective 
ultra-long-range interactions yielding length-scale competition 
\cite{Mingaleev}.}\label{note13}, so that its existence region 
apparently is 
limited by these two resonances. A similar behaviour is observed 
also between the second and the third resonances. Approaching the second 
resonance from above the continuation of the SW with wave vector $Q$ stops 
at $C_K\approx 46.51935$ (right part of Fig. \ref{fig_3res} (e)), while 
approaching  
the third resonance from below it transforms, via an intermediate solution 
appearing through a collision at +1 at $C_K\approx 51.826293472$ 
(solid line in Fig. \ref{fig_3res} (f)),  into a SW with wave vector 
$Q'''=16\pi/21$ similarly as in Fig. \ref{fig_3res} (a), (d). Finally, the 
smooth continuation of the linear SW with wave vector $Q$ towards smaller 
$C_K$ stops when approaching the third resonance from above (open circles in 
Fig. \ref{fig_3res} (f)) at $C_K\approx 51.826293474688$.
(As the third resonance is extremely weak, its correct numerical resolution 
requires working with quadruple precision and using a step size in $C_K$ 
smaller than $10^{-11}$.)

\begin{figure}
\centerline{\includegraphics[height=11cm,angle=270]{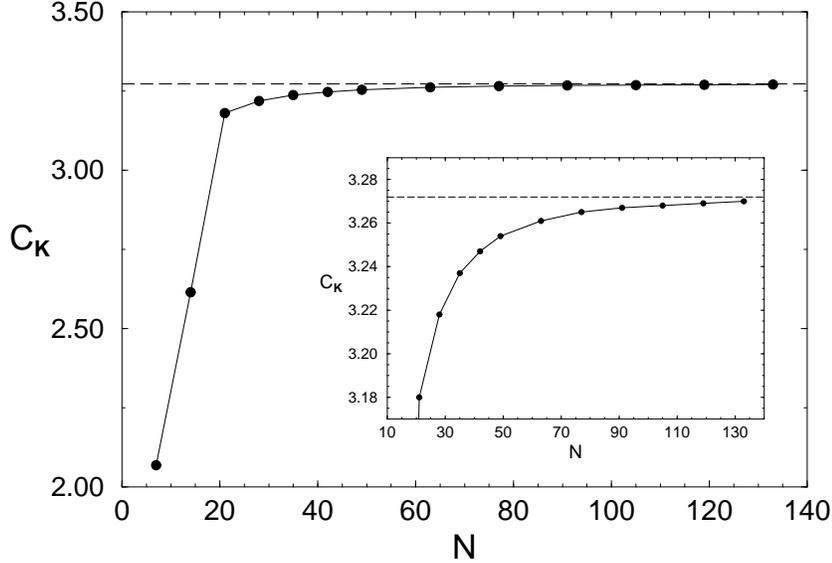}}
\caption{Lower limit for the interval of stability close to the linear limit 
as a function of system size, for a type $H$ SW with $Q=4\pi/7$ and 
$\omega_b=3.0$ in 
a quartic KG lattice. The dashed line is the linear limit according to 
Eq.\ (\ref{lin_disp}).}
\label{fig_stab}
\end{figure}

Let us finally note that in all cases, we find that for type $H$ SWs in 
finite systems with 
periodic boundary conditions, stability will always be recovered in an 
interval close to the linear limit (c.f. Fig.\ \ref{fig_Floq1} (d)), which 
shrinks 
to zero as the system size 
increases. The return to stability occurs through reentrant instabilities, 
completely analogous to the scenario for the DNLS 
approximation mentioned in Sec.\ \ref{CSW}. The shrinking of the stability 
regime with system size is illustrated in Fig.\ \ref{fig_stab}.

\subsection{The Morse potential}
\label{sec_Morse}

For the soft Morse potential, the nonlinear SWs with wave vector $Q$ have 
frequencies $\omega_b<\omega_l(Q)$, and the continuation from the 
anticontinuous to the linear limit must be performed in two steps as 
illustrated in Fig.\ \ref{fig_planeCw}: first by increasing $C_K$ for fixed 
$\omega_b<1$, and then by increasing $\omega_b$ for fixed $C_K$. Even though 
the on-site potential (\ref{morse}) now is asymmetric, we can use the 
symmetry properties 
of the SW coding sequences discussed in Sec.\ \ref{swams} to restrict the 
space of solutions for the numerical Newton scheme. To find the 
time-reversible SWs of type $H$  with 
coding sequences antisymmetric around a lattice site we impose the conditions 
$u_n(T_b/2)=u_{-n}(0)$, $\dot{u}_n(T_b/2)=\dot{u}_n(0)=0$ in the Newton 
algorithm, while for SWs antisymmetric around a bond center the first 
condition is replaced by $u_n(T_b/2)=u_{-n+1}(0)$. For type $E$ SWs, the 
symmetry condition $u_n(t)=u_{-n+1}(t)$ can always be employed. 

In the case of a soft potential, the $p$th harmonic of a SW with wave vector 
$Q$ enters the linear phonon band if (c.f. (\ref{3omega}))
\begin{equation}
\frac{1}{p}<\omega_b<\min\left\{\sqrt{1+4C_K\sin^2{\frac{Q}{2}}}, 
 \frac{1}{p}\sqrt{1+4C_K}\right\} .
\label{pomegasoft}
\end{equation}
Thus, in contrast to the hard case, we expect to find higher order resonances 
for all $Q$ if, for fixed $C_K$, the frequency $\omega_b$ is small enough, 
\ie, if the SW is sufficiently nonlinear. In particular, when the conditions 
(\ref{Qlimit}) are fulfilled, the resonances occur immediately from the linear 
limit. As the potential is non-symmetric already the second-harmonic 
resonances yield generally bifurcations of the SWs, and thus we find 
qualitatively different behaviour for small-amplitude SWs depending on 
whether the condition (\ref{Qlimit}) with $p=2$,
\begin{equation}
Q < \frac{\pi}{3} \quad \mbox{and} \quad
C_K > \frac{3}{4\left(1-4\sin^2{\frac{Q}{2}}\right)} ,
\label{Qlimit2}
\end{equation}
is fulfilled or not. In order to avoid complications resulting from 
higher-order resonances ($p>2$), which occur only for highly nonlinear SWs, 
we will mainly concentrate the discussion here to the regime $\omega_b>1/2$ 
(more details concerning the behaviour for smaller $\omega_b$ are given in 
\cite{AnnaPhD}). 
\footnote{In fact, the ansatz (\ref{code})-(\ref{hull}) 
for the SW 
coding sequences is valid only for $\omega_b>1/2$; a smooth continuation of 
SWs generated from (\ref{code})-(\ref{hull}) for $\omega_b>1/2$ towards 
smaller  $\omega_b$ shows \cite{AnnaPhD} that they appear with a different 
anticontinuous 
coding sequence when  $\omega_b<1/2$, with the $0$'s in the original code 
replaced by $\pm2$'s. Solutions generated from the anticontinuous 
limit at $\omega_b<1/2$ using (\ref{code})-(\ref{hull}) are lost through  
bifurcations and not continuable to the linear limit \cite{AnnaPhD}.}

\begin{figure}
\centerline{\includegraphics[height=11cm,angle=270]{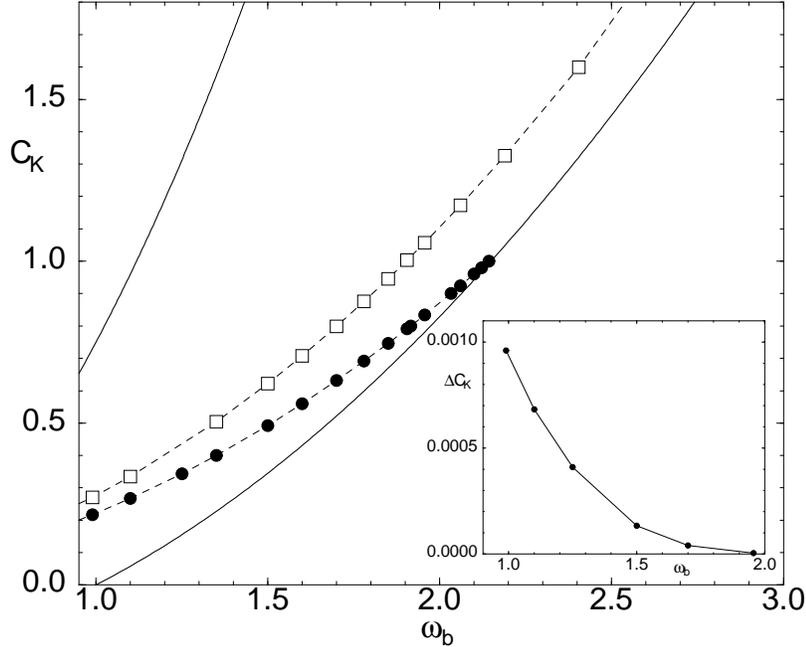}}
\caption{Phase diagram for SWs with $Q=4\pi/5$ in the Morse KG 
model. The lower solid line is the linear dispersion curve (\ref{lin_disp}), 
the filled circles indicate the line (which in fact are two closely 
separated lines, see text) of inversion of stability between the 
SWs of types $E$ and $H$ (characterized by eigenvalue collisions of 
${\bf T_0}$ at +1), and the open squares indicate the line of amplitude 
divergence. The upper solid line is the right boundary of the second-harmonic 
band. Inset shows the difference between the values of the coupling where 
the stability changes for the $H$ and $E$ waves for fixed frequency.}
\label{fig_Morsebif}
\end{figure}

\begin{figure}
\centerline{\includegraphics[height=7cm,angle=270]{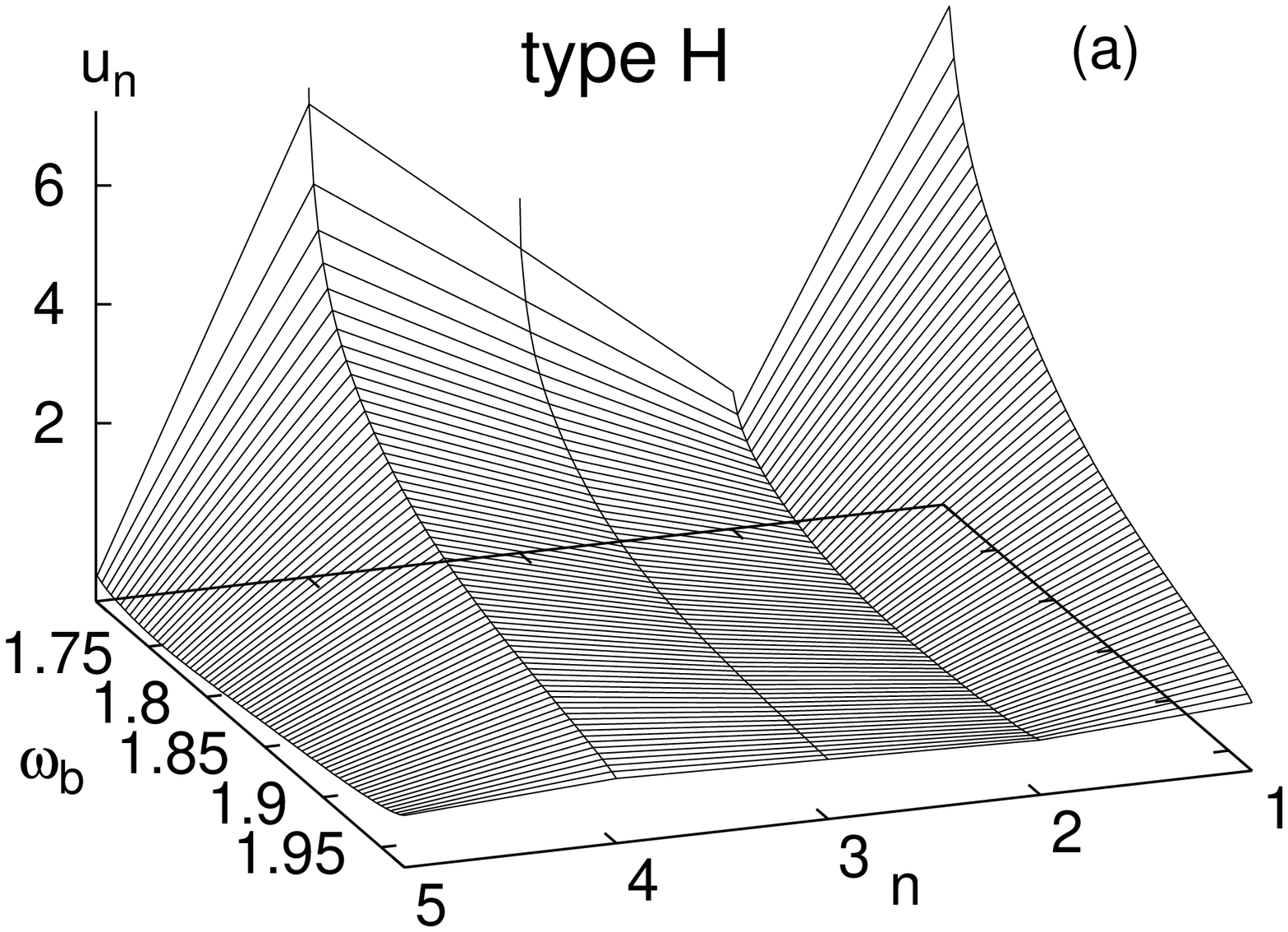}
\includegraphics [height=7cm,angle=270] {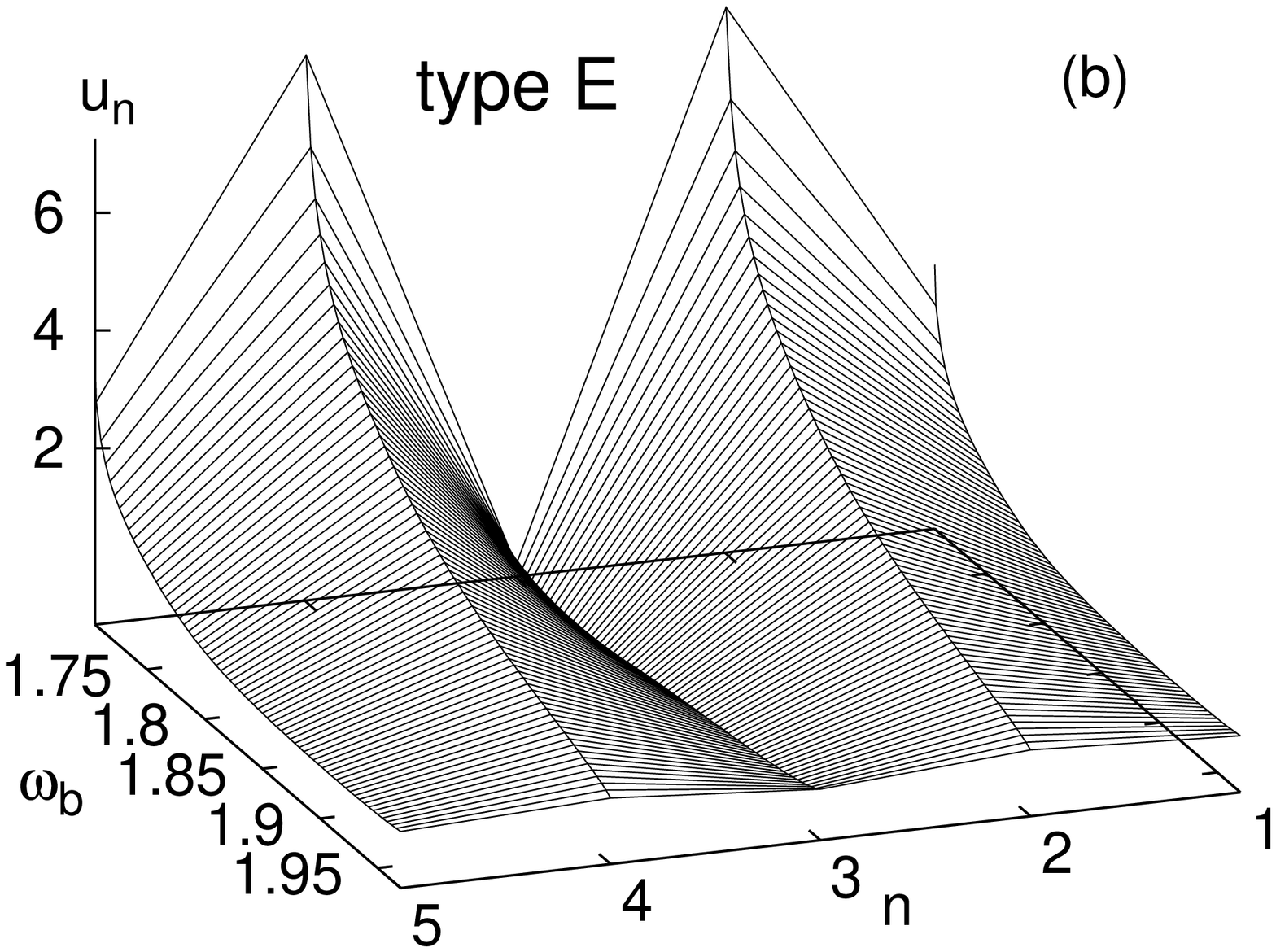}}
\centerline
{\includegraphics[height=7cm,angle=270]{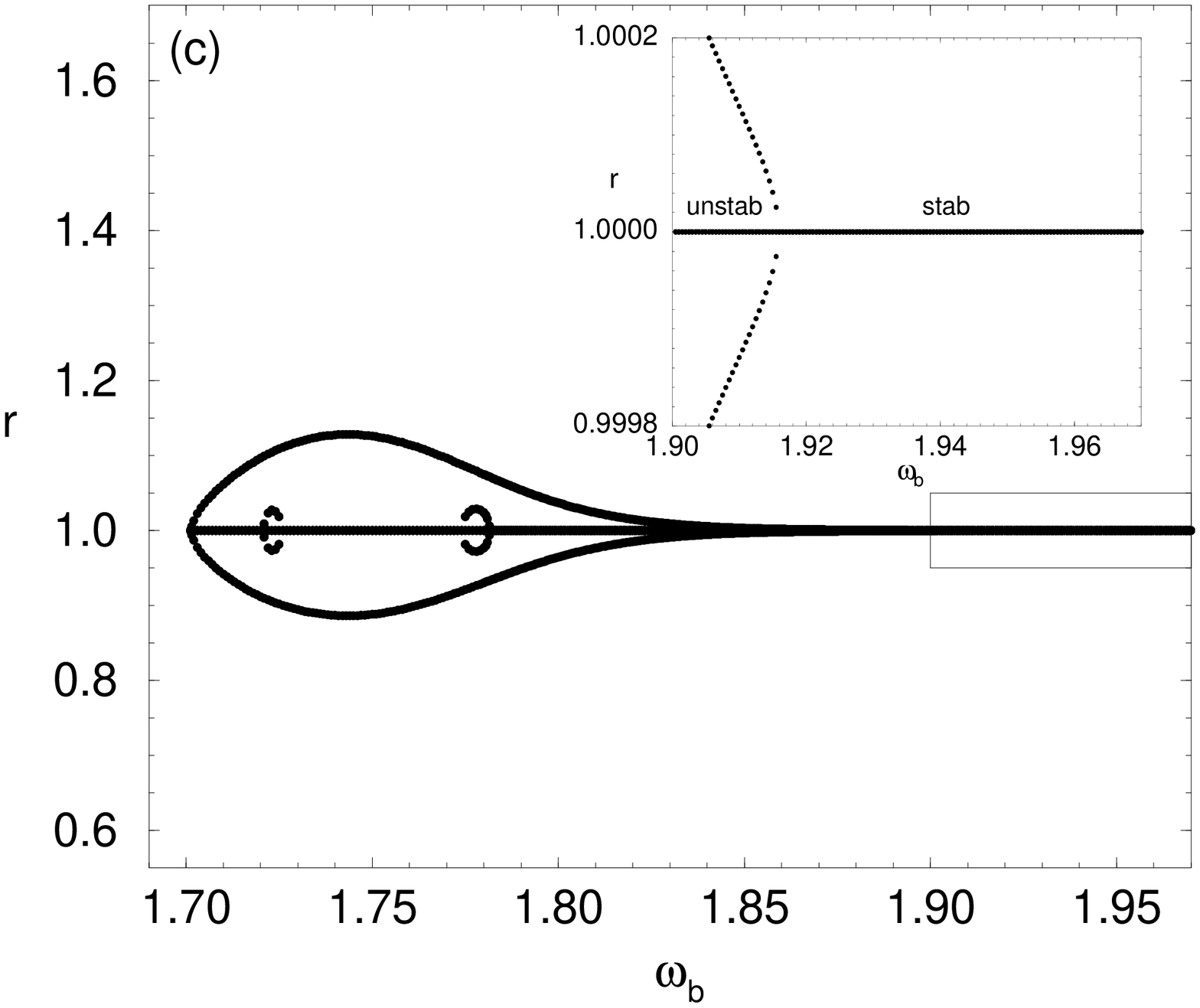}
\includegraphics[height=7cm,angle=270]{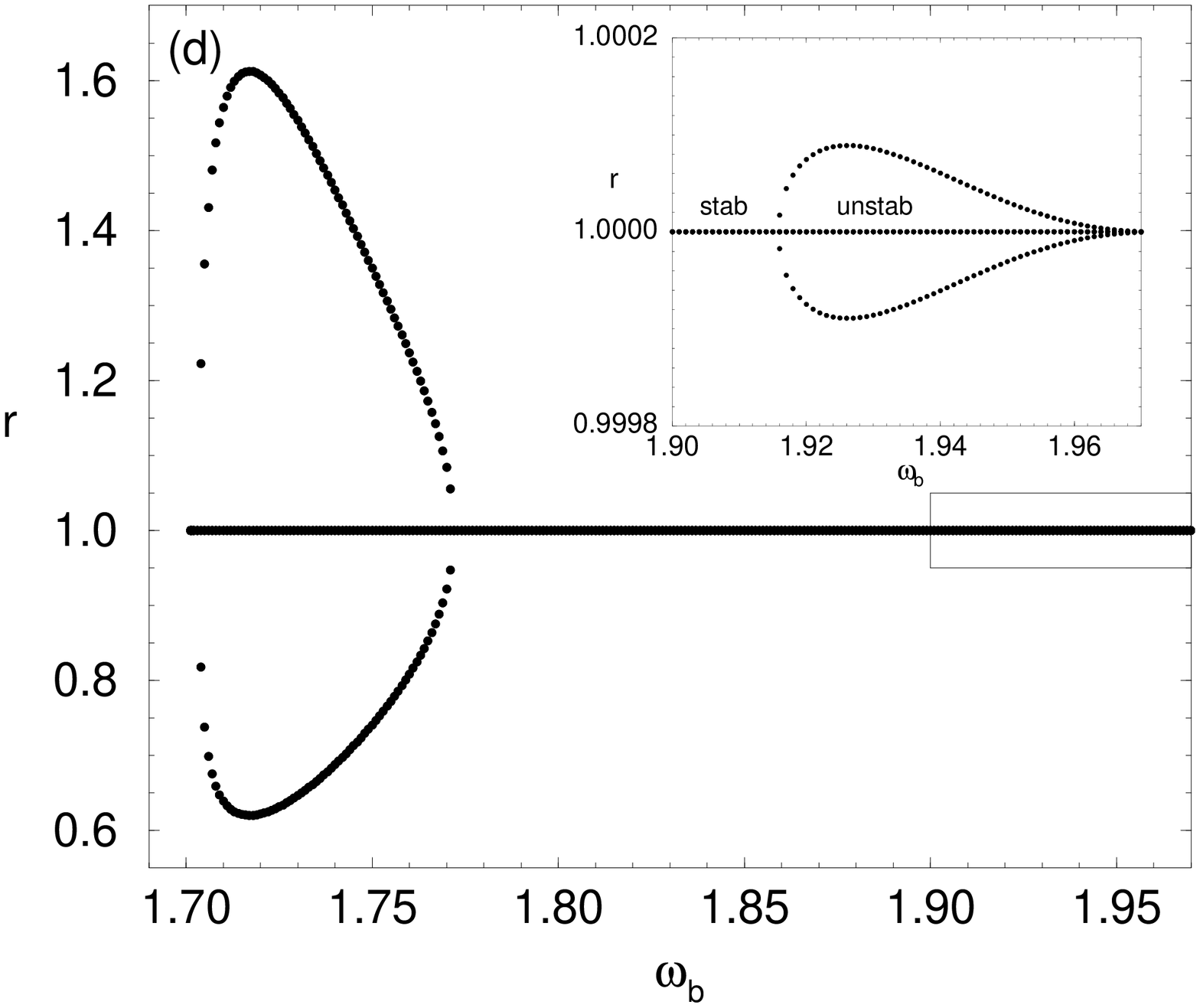}}
\caption{Continuation from the linear limit at fixed coupling $C_K=0.8$ for 
decreasing frequency for (a),(c) the type $H$ and (b),(d) the type $E$ SW with 
$Q=4\pi/5$ in the Morse KG model (corresponding to a horizontal line in the 
phase diagram in 
Fig.\ \ref{fig_Morsebif}). The wave amplitude diverges in both cases at 
$\omega_b\approx 1.701$. (c) and (d) show the magnitude of the 
eigenvalues $r\e^{i\theta}$ of ${\bf T_0}$ corresponding to one unit cell 
($N=5$) of the 
$H$ and $E$ waves, respectively. Insets are magnifications of the region 
between the linear limit and the point of inversion of stability, which occurs 
at $\omega_b\approx 1.9155$. The other instabilities are oscillatory 
instabilities resulting from Krein collisions.}
\label{fig_Morse2}
\end{figure}

The typical behaviour when continuing SWs with $Q>\pi/3$ is summarized in 
Figs.\ \ref{fig_Morsebif}-\ref{fig_Morse2}, for a rather simple case with $Q=4\pi/5$. 
Let us first concentrate on the part closest to 
the linear limit. Then, we find that increasing $C_K$ for a fixed $\omega_b$ 
close to 1 
(which means that the wave amplitude also increases), the type $H$ SWs will 
suffer additional real instabilities, not present neither in the DNLS 
approximation nor in the quartic model. These instabilities occur  
through collisions at $+1$ at some critical value of the coupling. Very 
close to this value, also the type $E$ SW with the same $Q$ and $\omega_b$ 
changes its stability properties, in that the real eigenvalues which are 
normally present collide at $+1$ and go out on the unit circle, leaving only 
Krein instabilities and second-harmonic collisions at -1 as possible remaining 
instability mechanisms. This '{\em inversion of stability}' is analogous to 
the scenario occurring between the single-site (code ...0001000...) and 
two-site (code ...0011000...) breathers in the Morse and other soft asymmetric 
potentials \cite{MAF98,ThierryPhD}, and is generally associated with a 
highly increased mobility \cite{ThierryPhD,CAT96,AC98}. In the case shown in 
Figs.\ \ref{fig_Morsebif}-\ref{fig_Morse2}, the change of stability for the 
$E$-wave with a 
given 
frequency occurs for slightly smaller values of $C_K$ than that of the 
corresponding $H$-wave (see inset in Fig.\ \ref{fig_Morsebif}), so that in the 
absence of other instability 
mechanisms (which is the case for small systems close to the linear limit, 
c.f. Fig.\ \ref{fig_Morse2}), 
there is a small regime in the $(\omega_b,C_K)$-plane of simultaneous 
stability for both waves\footnote{By contrast, for the single breather there 
is a small regime of simultaneous instability \cite{ThierryPhD}.}. When 
increasing 
$\omega_b$ and $C_K$ the regime of simultaneous stability decreases, and 
moreover at a certain point 
($\omega_b\approx 2.1$ and $C_K\approx 1.0$ for the case in 
Fig.\ \ref{fig_Morsebif}) the two lines defining the inversion of stability 
converge and intersect the linear dispersion curve (\ref{lin_disp}). 
Thus, for values of $C_K$ above this 
point the stability scenario close to the linear limit is {\em opposite} to 
what 
we have found previously in the DNLS-approximation: the $H$-wave becomes 
immediately unstable through real eigenvalues, while the $E$-wave remains 
stable in an interval close to the linear limit for finite-size systems. 

When further increasing the coupling, we find typically that the region of 
existence for the SWs is limited by a curve in the $(\omega_b,C_K)$-plane 
where 
the wave amplitude diverges as illustrated in Fig.\ \ref{fig_Morse2} (a), (b) 
(as the Morse potential (\ref{morse}) becomes flat 
for large $u$). However, for smaller $\omega_b$ the scenario becomes more 
complicated with additional bifurcations, and we will not attempt to give an 
exhaustive analysis of this 
regime here (see \cite{AnnaPhD} for more details). Let us just mention that 
for the case illustrated in Fig.\ \ref{fig_Morsebif}, when 
$\omega_b\lesssim 0.72$ ($\omega_b\lesssim 1.1$) for the $H$-wave ($E$-wave), 
the continuation from 
the anticontinuous limit at 
fixed $\omega_b$ is interrupted\footnote{However, it can be restored 
\cite{AnnaPhD} by 
following a curvilinear, 'Z-shaped' path similarly as described 
for the quartic model in footnote \ref{note13}. This 
behaviour is likely to be related to a strong effective long-range 
interaction in the extended DNLS equation (\ref{eqp=1}); 
cf.\ \cite{Mingaleev}.} by a 
bifurcation before reaching the point of 
amplitude divergence. Let us also mention that for SWs with more complicated 
unit cells than that of Figs.\ \ref{fig_Morsebif}-\ref{fig_Morse2} (which is 
somewhat special 
since the coding sequence for one unit cell of the $H$-wave [1-101-1] has 
only one 0 per unit cell, while that of the $E$-wave [-11-11-1] has no 0 and 
only one pair of consecutive -1), there 
are sometimes more than one 'stability-inverting' bifurcation, and the phase 
diagram of 'stable' (neglecting other instability mechanisms) and unstable 
regimes in the  $(\omega_b,C_K)$-plane may contain more than two parts 
\cite{AnnaPhD} (as 
was also found for the single breather \cite{ThierryPhD}). 

\begin{figure}
\centerline{\includegraphics[width=7cm,angle=270]{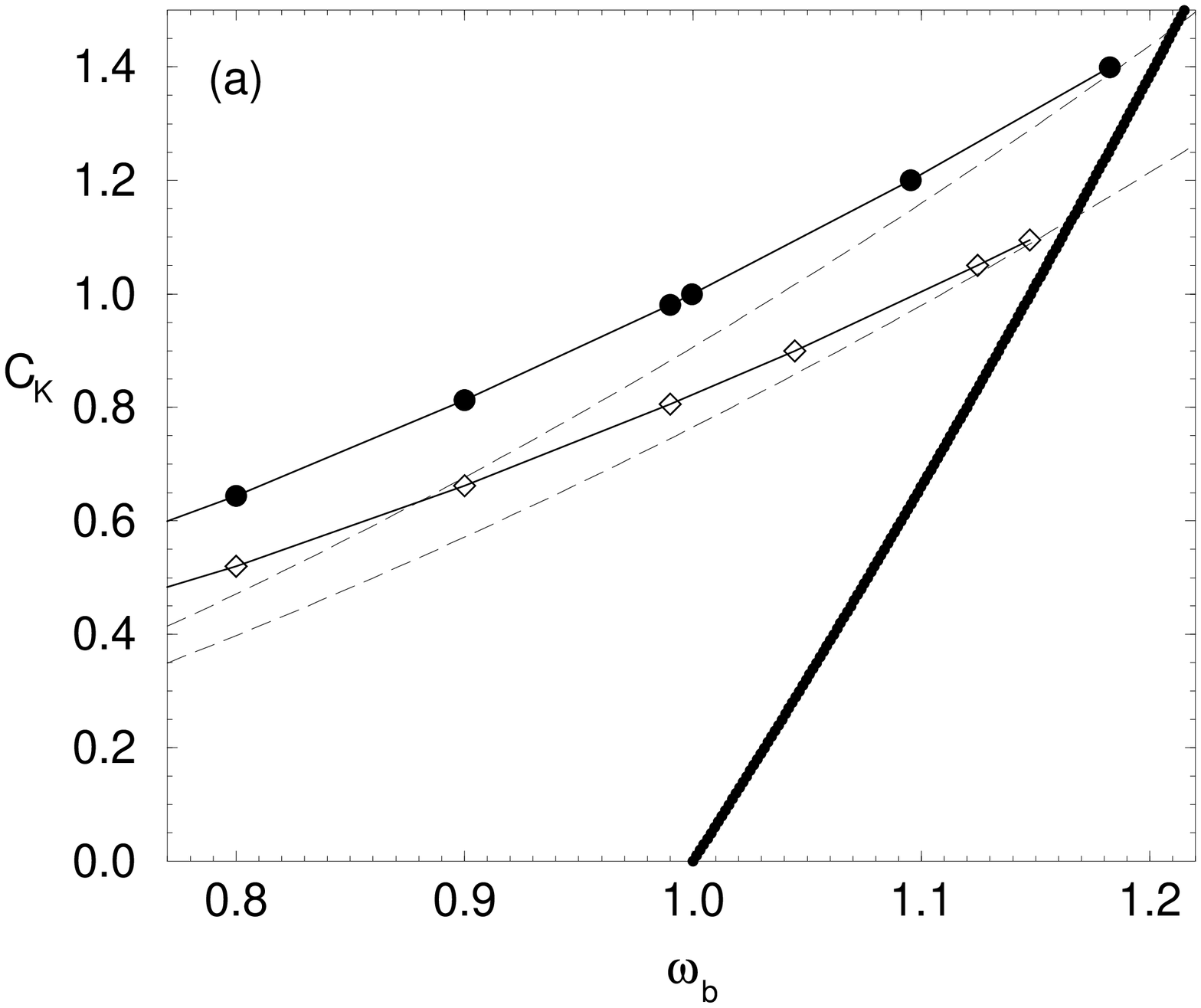}
\includegraphics [width=7cm,angle=270] {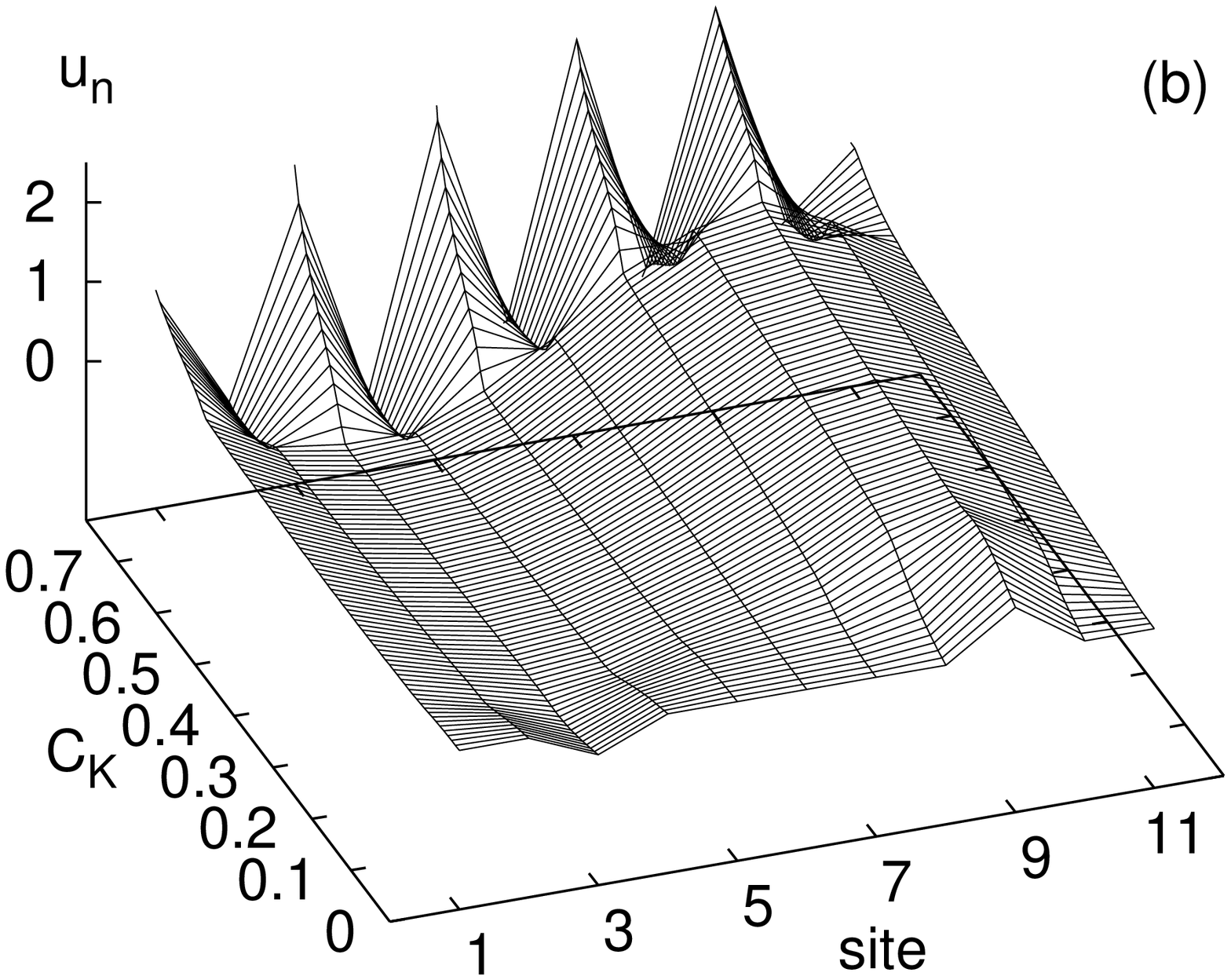}}
\centerline
{\includegraphics[width=5.5cm,angle=270]{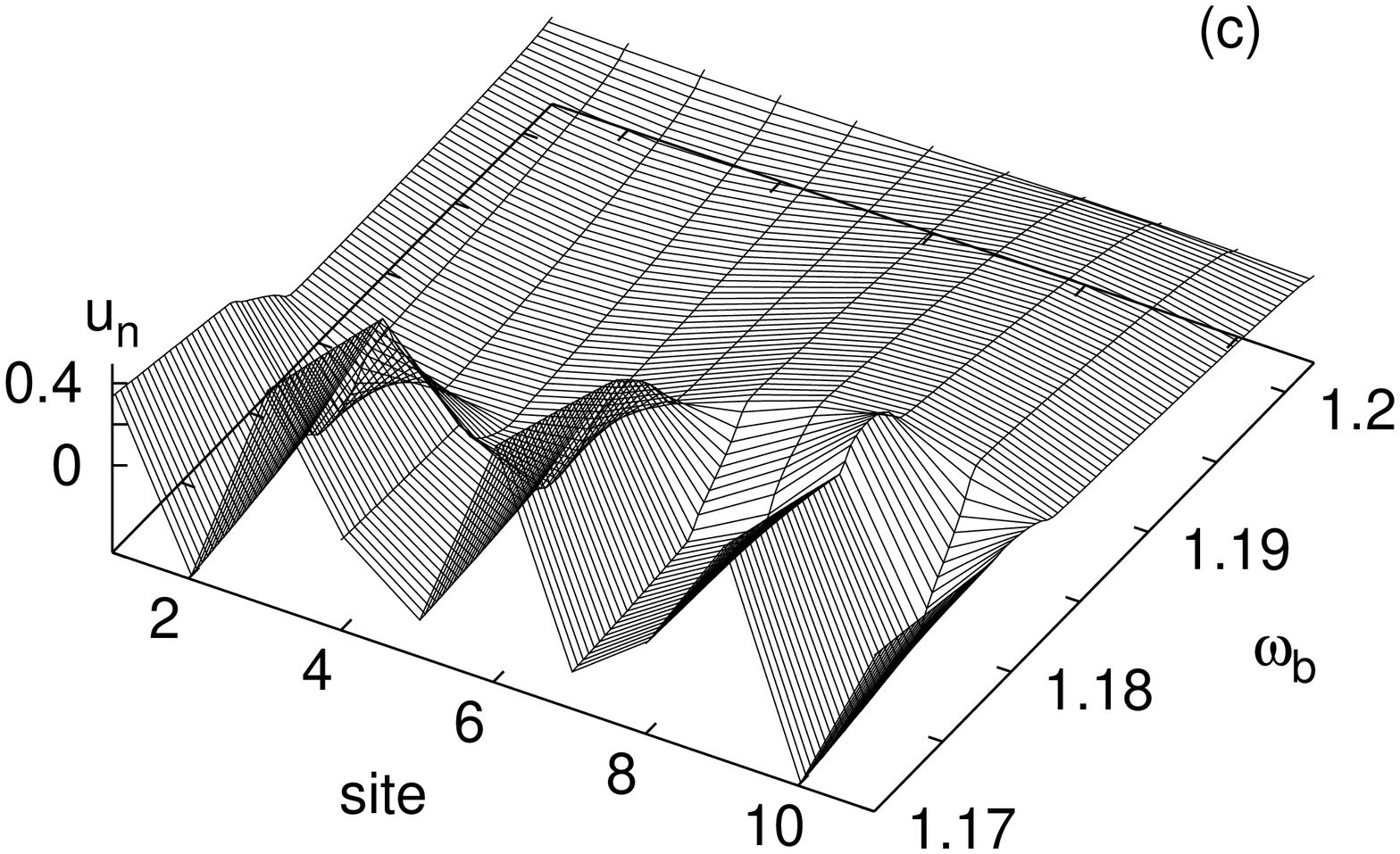}
\includegraphics[width=5.5cm,angle=270]{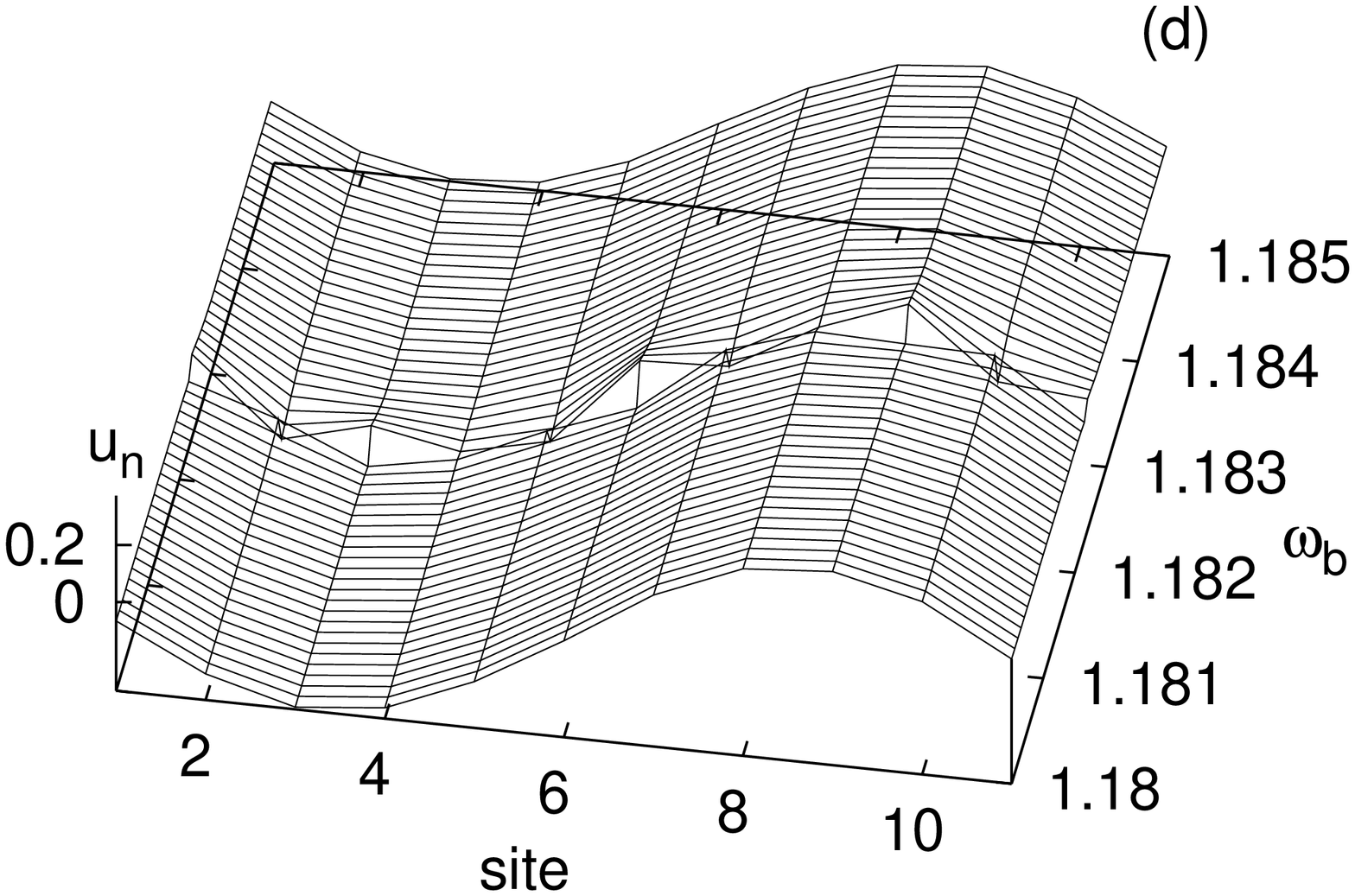}}
\caption{Second-harmonic resonances for a type $H$ SW with $Q=2\pi/11$ in the 
Morse KG model. (a) shows the bifurcation lines for the first two resonances; 
dashed lines are the linear second-harmonic dispersion curves for 
$Q=10\pi/11$ (lower) and $Q=8\pi/11$ (upper); fat solid line is the 
fundamental linear dispersion curve for $Q=2\pi/11$. 
(b) illustrates the smooth 
continuation versus coupling from the anticontinuous limit at fixed 
frequency $\omega_b=0.9$ 
through the first resonance (open diamonds in (a)) into a type $E$ SW with 
frequency $2\omega_b$ and
wave vector $Q^\prime=10 \pi/11$. (c) shows smooth continuation versus 
frequency from the linear limit at constant coupling $C_K=1.4$ (\ie, between 
the first and second resonances) through the second resonance (filled circles 
in (a)) into a type $E$ 
SW with frequency $2\omega_b$ and wave vector $Q''=8 \pi/11$. (d) shows 
intentionally careless continuation which jumps over the resonance in (c); if 
a careful continuation versus larger $\omega_b$ is attempted, it stops at 
$\omega_b\approx 1.18235$ (filled circle in (a)) 
similarly as in Fig.\ \ref{fig_3res} (d) and (e). 
}\label{fig_Morse3}
\end{figure}

Let us now turn to the case with $Q<\pi/3$, in which case we find 
second-harmonic resonances also in the linear regime for large enough 
coupling. The scenario in the small-amplitude regime is then similar to the 
scenario for the third-harmonic resonances in the quartic model illustrated in 
Fig.\ \ref{fig_3res}. An example for a type $H$ SW with  $Q=2\pi/11$ is 
shown in 
Fig.\ \ref{fig_Morse3}. Continuing from the anticontinuous limit for 
increasing coupling then yields (Fig.\ \ref{fig_Morse3} (b)), via an 
eigenvalue collision at +1 for a 
value of $C_K$ 
on the line of open diamonds in (a), a transition to an intermediate state,  
which at a second bifurcation point becomes a SW with frequency $2\omega_b$ 
and wave vector $Q^\prime$, where $Q^\prime$ is the wave vector closest to 
$\pi$ giving a SW whose symmetry is compatible with that of the original wave 
(here, the new wave is the type $E$ SW with 
$Q^\prime=10\pi/11$)\footnote{Similarly to the quartic model 
(footnote \ref{note12}), a necessary 
resonance condition is $Q^\prime=mQ$ mod $2\pi$, but with no restriction on 
the integer $m$.}. 
This new SW can 
then be continued for larger coupling  until its amplitude 
diverges at some finite value of $C_K$ (similarly as in 
Fig.\ \ref{fig_Morse2}). 

The same scenario is observed by continuation from 
the linear limit for fixed coupling towards lower frequency, as long as  $C_K$ 
is below the value where the first resonance curve (which in the linear limit 
coincides with the linear second-harmonic dispersion curve for $Q=Q^\prime$, 
see Fig.\ \ref{fig_Morse3} (a)) 
intersects the linear dispersion curve. For $C_K$ larger than this value the 
SWs are also continuable from the linear limit towards lower frequency, but 
attempts to continue these waves to the anticontinuous limit by 
subsequently decreasing $C_K$ fail since, similarly as for the quartic 
model (Fig.\ \ref{fig_3res} (d), (e)), the continuation is interrupted when  
reaching the first resonance curve. Performing instead the continuation for 
fixed coupling, the scenario is completely analogous to that of smaller $C_K$: 
as illustrated in Fig.\ \ref{fig_Morse3} (c) for $C_K=1.4$, the wave 
bifurcates at a  
point on the line of filled circles in (a) ($\omega_b \approx 1.1816$) into an 
intermediate state, which in a second bifurcation 
(at $\omega_b\approx 1.1789$) turns into the second-harmonic type $E$ SW with 
wave vector $Q''= 8\pi/11$, which then can be continued until its amplitude 
becomes infinite. Similarly, if the continuation from the 
linear 
limit is performed for $C_K$ above the intersection point with this second 
resonance curve, the continuation towards smaller coupling will be 
interrupted by the second 
resonance curve, while the continuation towards larger coupling or 
smaller frequency meets a third resonance curve where the wave changes 
smoothly into 
a SW with wave vector $Q'''=6\pi/11$, etc. However, as illustrated by 
Fig.\ \ref{fig_Morse3} (d), these resonances may be jumped and an apparently 
smooth continuation of the SW with the original wave vector $Q$ observed if 
the numerical continuation is performed taking large steps. 

Finally, we remark that also in the case of second-harmonic resonances close 
to the linear limit, the inversion of stability between the $E$ and 
$H$ waves can be observed. For example, in the case considered in 
Fig.\ \ref{fig_Morse3} (c), the first eigenvalue collision at +1 in the 
continuation from the linear limit occurs at $\omega_b\approx 1.1822$ 
(\ie, very close to the collision corresponding to the second-harmonic 
resonance), and corresponds to inversion of stability. Let us also remark 
that, as mentioned above, the bifurcation scenario for smaller $\omega_b$ 
becomes more complicated, and for example the monotonous continuation for 
fixed frequency versus coupling from the 
anticontinuous limit of the SW in Fig.\ \ref{fig_Morse3} is interrupted 
before reaching the first second-harmonic resonance for 
$\omega_b\lesssim 0.7$. 

\section{Dynamics} \label{sec_dyn}

Let us now discuss the dynamics resulting from the oscillatory instabilities 
of the SWs over different time ranges. We will here emphasize generic rather 
than model-dependent properties, and therefore we consider KG systems in 
the regime of small amplitude and small coupling, where the DNLS approximation 
is expected to be good and the model-dependent phenomena discussed in the 
previous section (\ie,  
higher-harmonic resonances and inversion of stability) do not play any role. 
As was illustrated already by Fig.\ 4 
in 
\cite{MJKA00}, and is further exemplified by the figures in this Section, 
the dynamics in the regime of oscillatory instability can then typically be 
divided into roughly three regimes. 
Initially, the small perturbation grows exponentially leading to 
oscillatory dynamics with small but increasing amplitude. As the 
oscillation amplitude reaches some threshold, an intermediate regime appears 
which typically is characterized by an inhomogeneous translational motion, 
where the wave remains locally coherent but with its different parts moving 
with respect to each other. Gradually, the wave loses coherence and 
a final, apparently chaotic state appears, the statistical properties of which 
depending  crucially on the SW wave vector as discussed below.

\begin{figure}
\centerline{\includegraphics[width=11cm,angle=0]{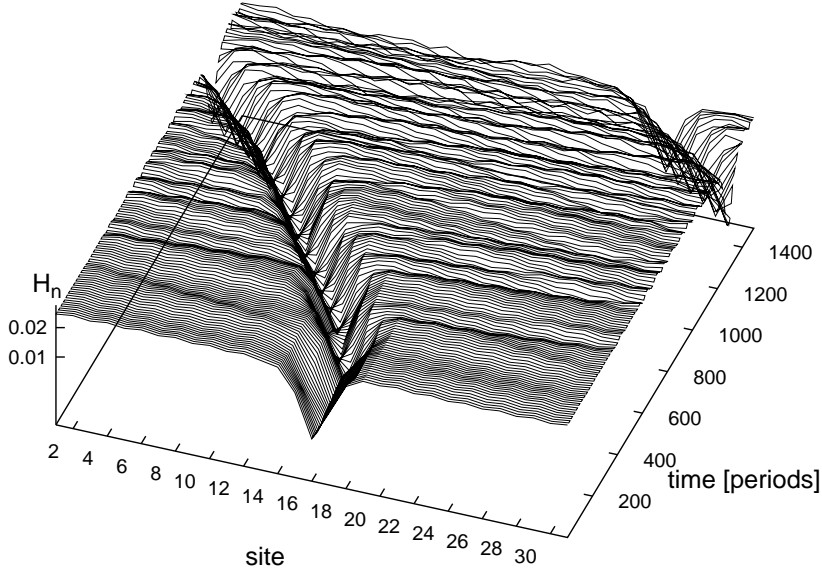}}
\caption {Spontaneous motion (resulting only from numerical truncation 
errors) of a single discommensuration in a Morse potential KG chain 
(\ref{DKG}) with $C_K=0.03$. The initial frequency is $\omega_{b}=1.036$.}
\label{fig_KGdisc}
\end{figure}

\begin{figure}[htbp]
\centerline{\includegraphics[width=5.5cm,angle=270]{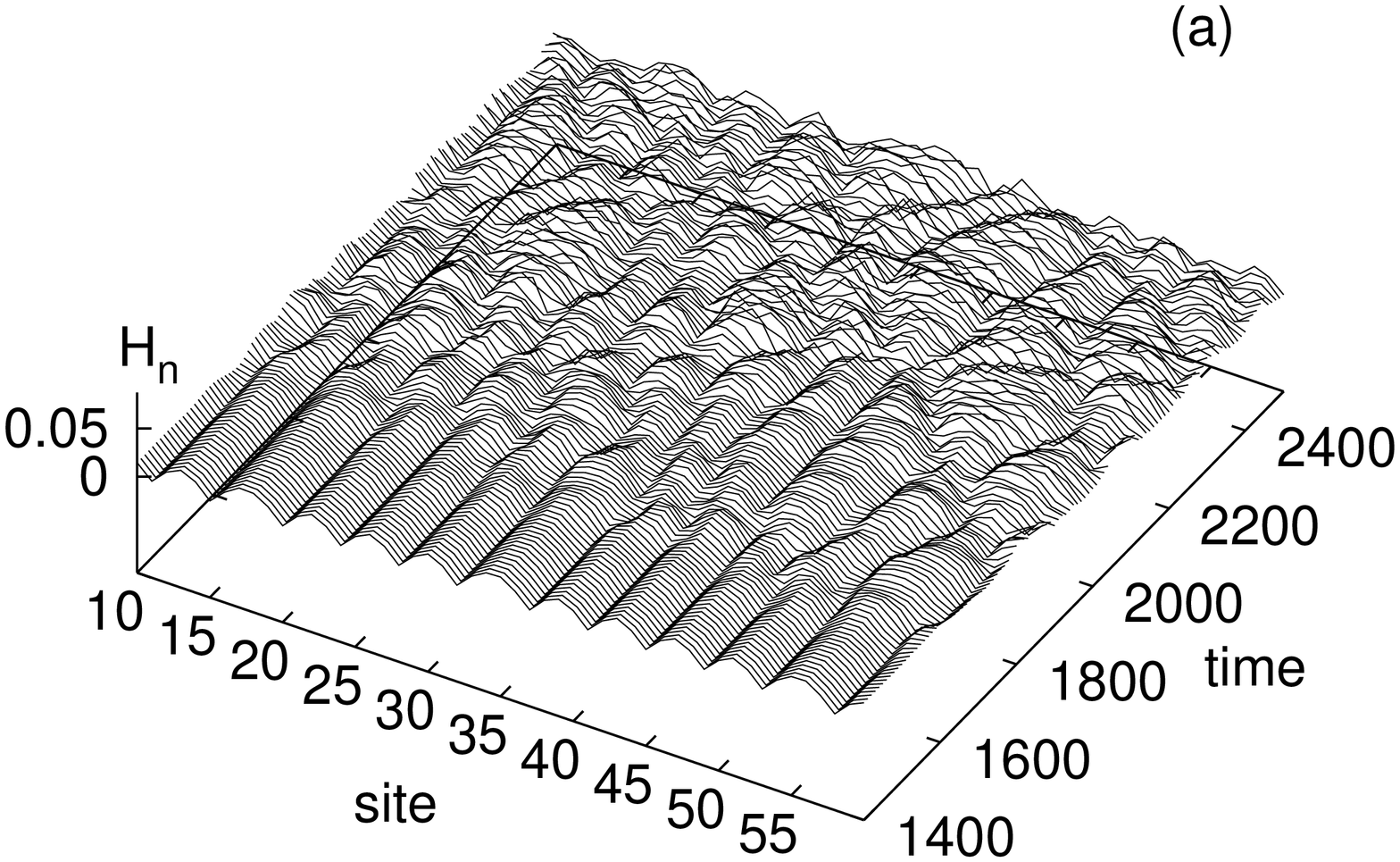}
\includegraphics[width=5.5cm,angle=270]{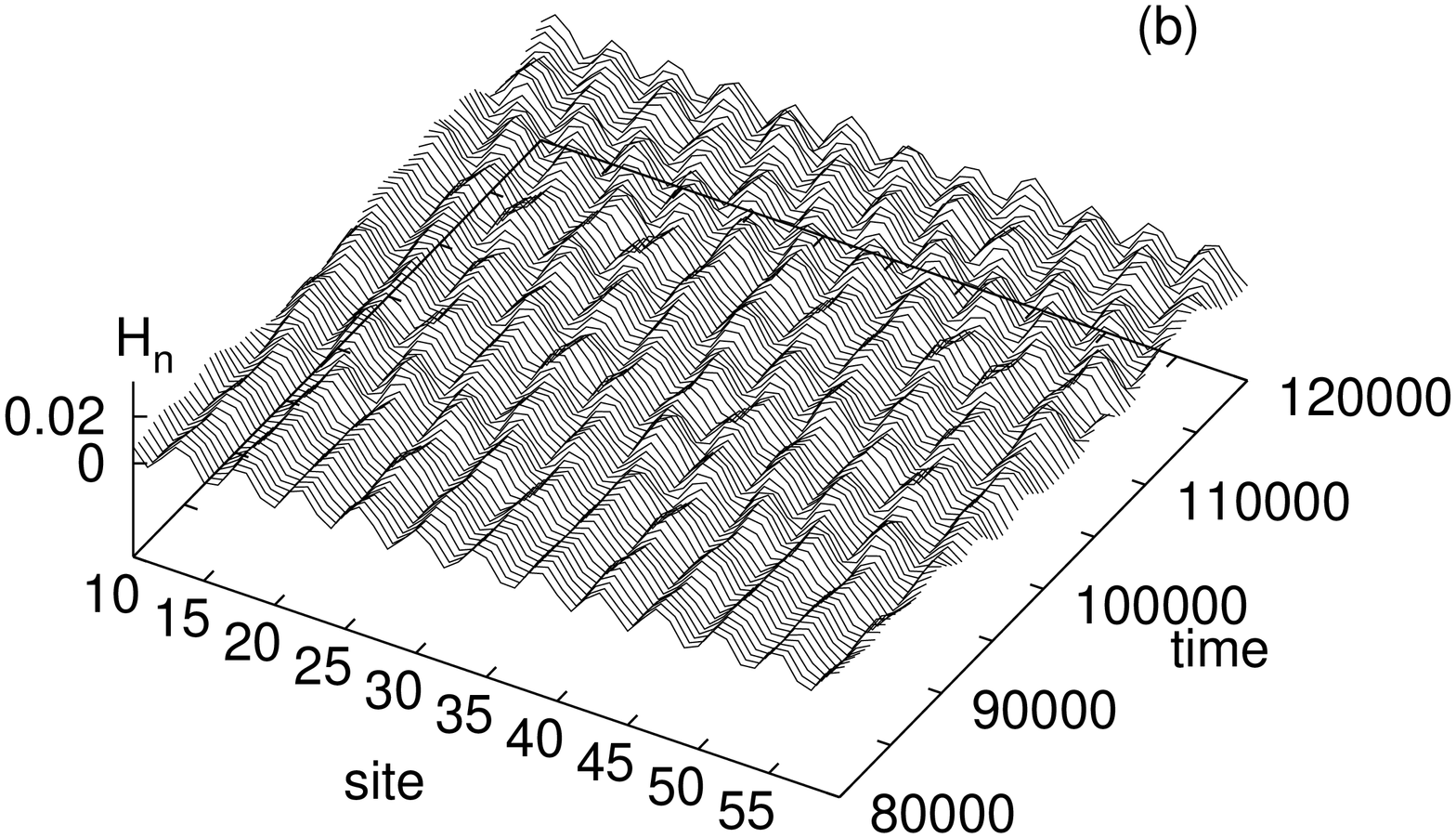}}
\centerline{\includegraphics[height=8.5cm,angle=270]{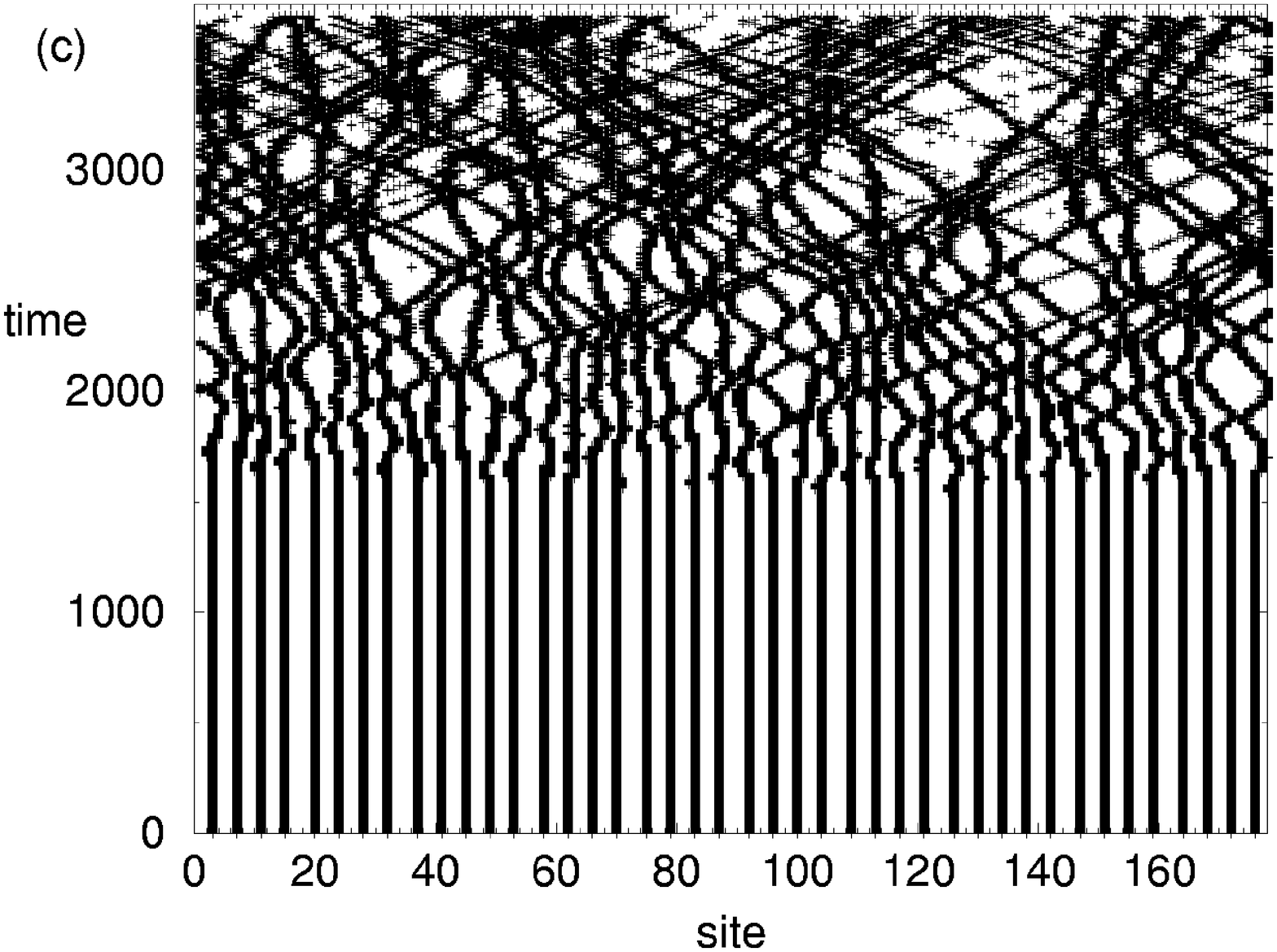}
\includegraphics[height=8.5cm,angle=270]{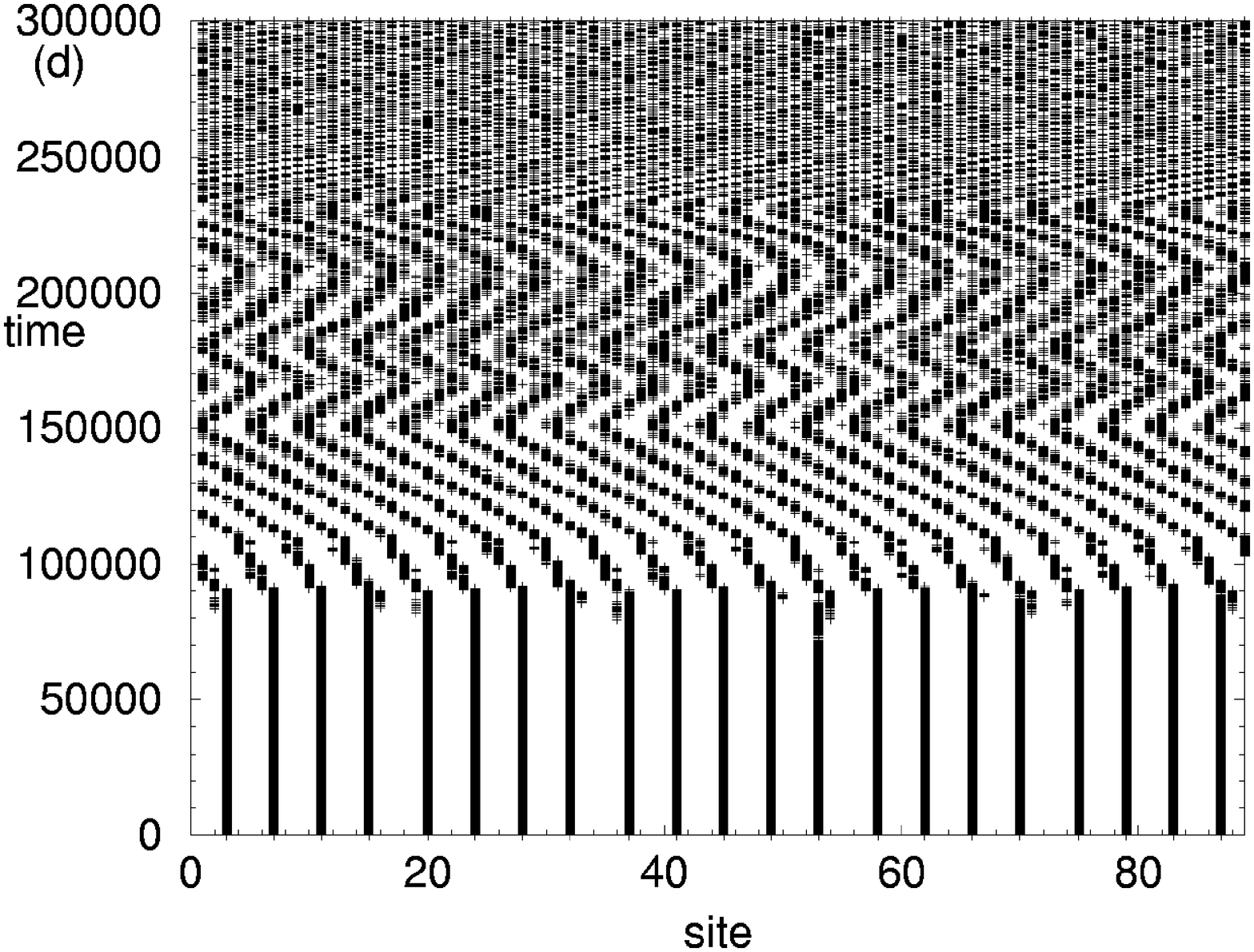}}
\caption{Time evolution over short and intermediate time scales of (parts of) 
slightly perturbed SWs with 
$Q/2\pi=34/89\approx\sigma_G$ in a Morse 
potential KG chain (\ref{DKG}) with (a),(c) $C_K=0.03$ and $\omega_{b}=1.034$ 
(nonanalytic case, type $H$), and (b),(d) $C_K=0.05$ and $\omega_{b}=1.075$ 
(analytic case). (c) and (d) illustrate the time evolution and interaction 
of 'discommensurations', defined as sites where the energy is (c) $H_n<0.01$ 
respectively (d) $H_n<0.00232$ (corresponding to sites with code 0 for the 
initial SWs). Note the very slow evolution of the 
instability in the analytic case, which can be considered as 'quasistable' as 
discussed in Sec.\ \ref{ISW}: the strongest instability appears only to order
$p=4$ according to Eq.\ (\ref{interEeq0}).
 }
\label{fig_discint}
\end{figure}

To understand the appearance of the intermediate regime of translational 
motion, it is useful to first consider the case $Q\rightarrow \pi$ with one 
single 
discommensuration in the $Q=\pi$ phonon (for a soft potential). As was found 
in \cite{JK99,KKC94}} 
for the 
DNLS model, the oscillatory instability leads in this case to the motion 
of the discommensuration (for the DNLS model interpreted as a discrete 'grey' 
soliton) with a rather well-defined velocity for long times, although it 
continuously emits radiation and decays. The same type of behaviour occurs 
also in KG models for small $C_K$ as illustrated in Fig. \ref{fig_KGdisc},
\footnote{In fact, it is 
possible \cite{unpub,AnnaPhD}
to calculate numerically exact solutions corresponding to moving 
discommensurations with given velocities, although they typically acquire 
an oscillatory extended tail and could therefore be described as 
'dark nanopterons' (for the DNLS model, moving dark solitary waves close to 
the continuum limit have been  numerically calculated in \cite{Fed91}).} 
and is basically a 
consequence of the fact that the unstable eigenvector is spatially 
antisymmetric with respect to the background $Q=\pi$ wave around the 
discommensuration site. Thus, if we regard a SW 
with $Q$ close to $\pi$ as an array of interacting discommensurations 
(where each discommensuration corresponds to a site with code 0), the 
intermediate regime of mainly translational motion can be anticipated. 
As 
illustrated in Fig.\ \ref{fig_discint}, this description is at least 
qualitatively 
relevant also for analytic waves (where generally discommensurations can no 
more be regarded as localized objects as their range of interaction diverges 
\cite{AGARQ}) with wave vectors not 
necessarily close to $\pi$ (so that the discommensuration sites may be 
close). Initially, the collective movement of the discommensurations is 
well-ordered and they remain clearly separated from each other, but as time 
evolves their motion becomes more and more erratic and they start to collide 
and possibly merge with each other. Thus, the SW eventually loses its 
coherence. 

\begin{figure}[htbp]
\centerline{\includegraphics[width=5.5cm,angle=270]{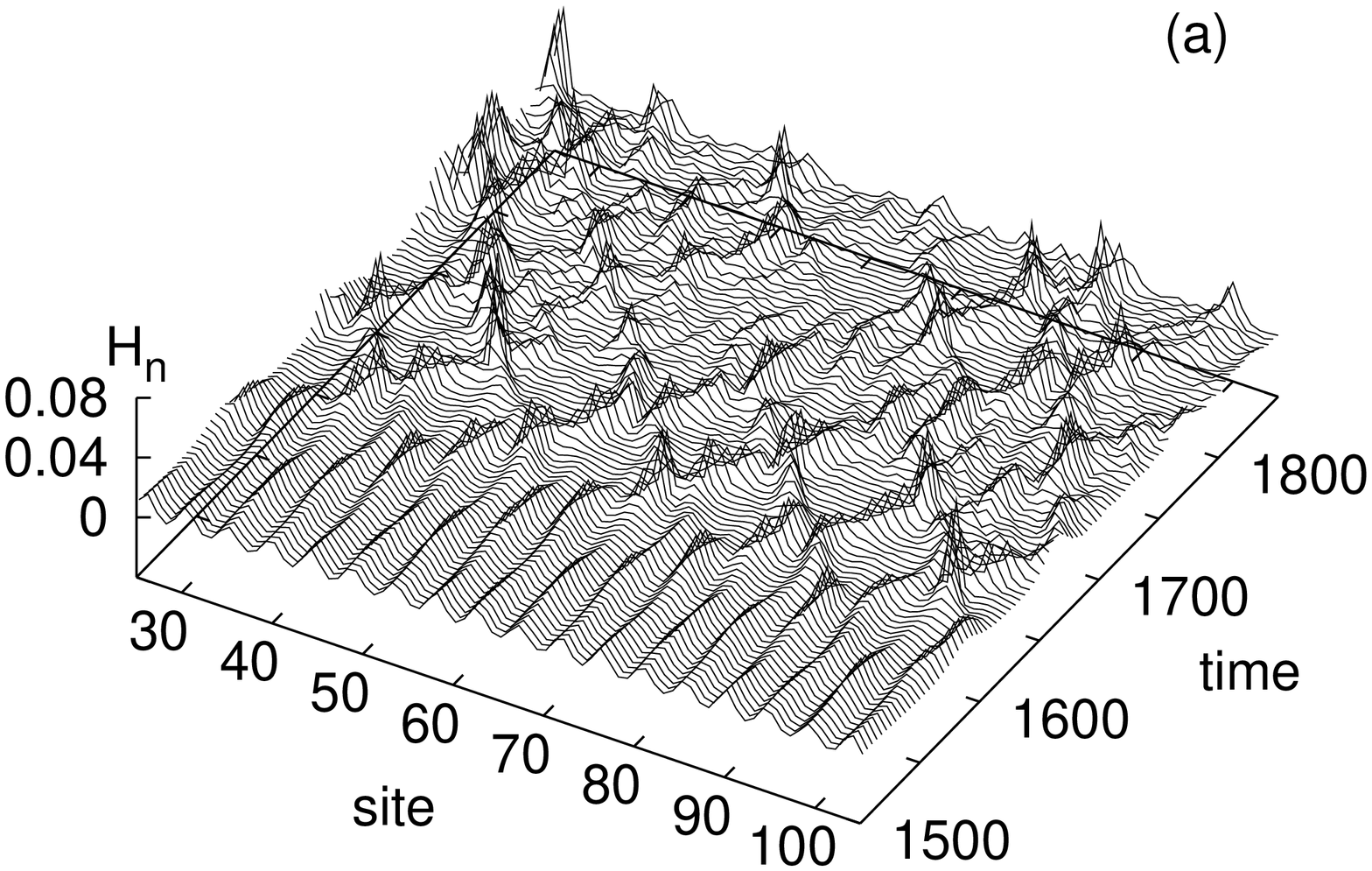}
\includegraphics[width=5.5cm,angle=270]{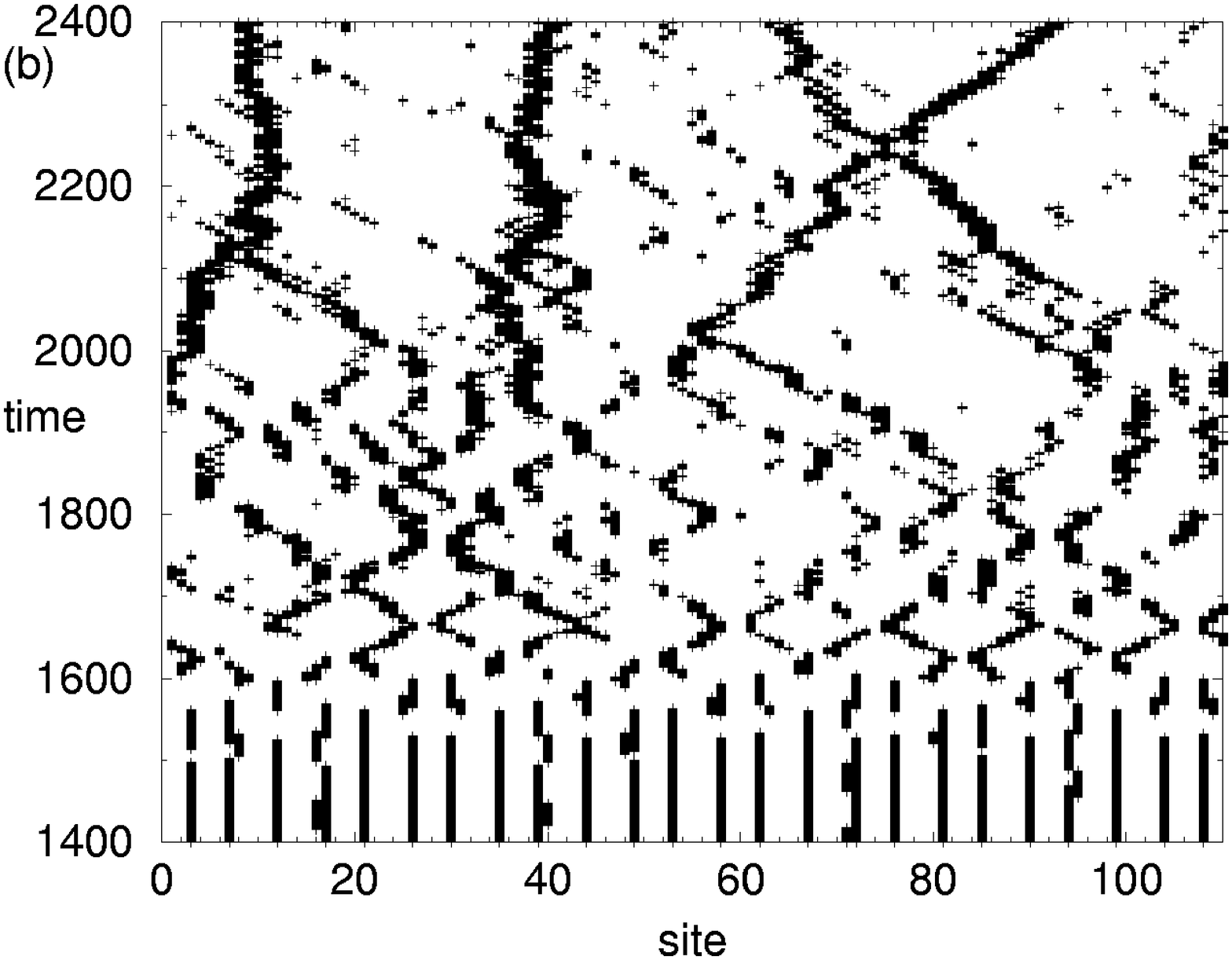}}
\caption{Time evolution over short and intermediate time scales of (parts of) 
a slightly perturbed SW with 
$Q=12\pi/55$ ($<\pi/2$) and $\omega_b=1.0015$ (corresponding to an analytic 
wave in the incommensurate limit) in a Morse 
potential KG chain (\ref{DKG}) with $C_K=0.05$. (b) illustrates the time 
evolution and interaction of high-energy sites ('breathers'), 
defined as sites where the energy is $H_n>0.012$ 
(corresponding to sites with codes $\pm1$ for the 
initial SW). 
 }
\label{fig_breatherint}
\end{figure}

When (for a soft potential) $Q<\pi/2$, the coding sequences contain 
consecutive zeros, and therefore the discommensurations in the initial SWs 
are not isolated. Thus, for small $Q$ it is more natural to regard the SWs 
as composed of individual breathers, where each breather 
corresponds to a site with code $\pm 1$, and the distance between the 
breathers increases as $Q$ decreases. Then, as illustrated in 
Fig.\ \ref{fig_breatherint}, the intermediate regime can be interpreted as 
consisting of moving breathers which collide inelastically with each other. 

\begin{figure}
\includegraphics[height=13cm,angle=270]{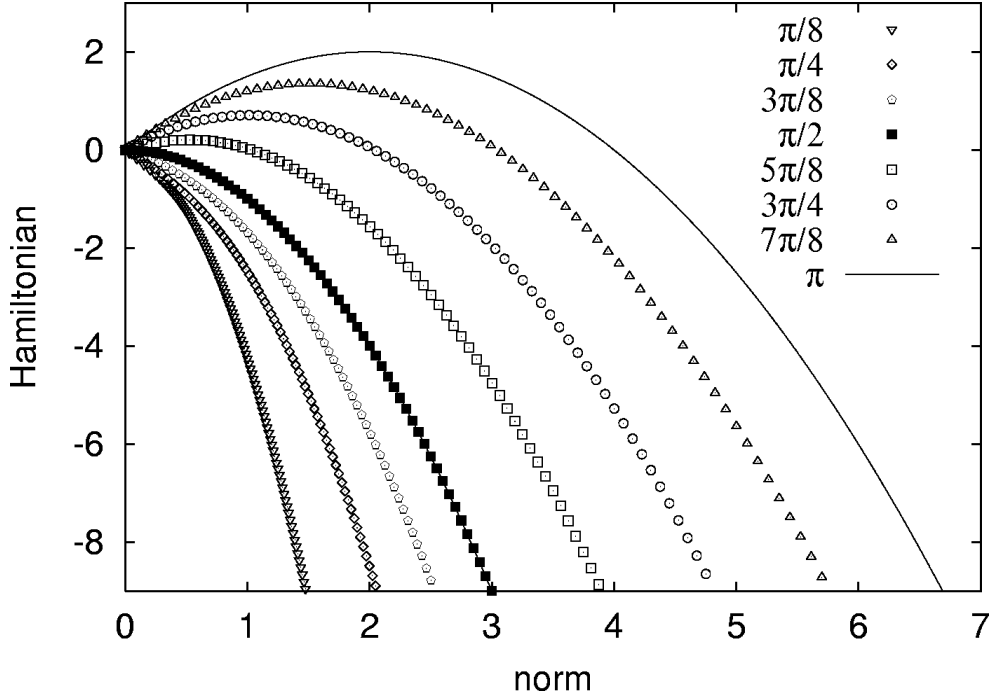}
\caption{Average Hamiltonian $\mathcal{H}^\prime/N$ versus average norm 
$\mathcal{N}/N$ for type $H$ DNLS SWs ($\sigma=-1, C=1$) having wave vectors 
(increasing from left to right) 
$Q=\pi/8$, $\pi/4$, $3\pi/8$, $\pi/2$, $5\pi/8$, $3\pi/4$, $7\pi/8$, and 
$\pi$. The points 
corresponding to $Q=\pi/2$ (filled squares) coincide exactly with the 
phase transition line (\ref{phase}).}
\label{fig_phase}
\end{figure}

Let us now discuss the nature of the asymptotic long-time states for 
different $Q$. For the DNLS equation (\ref{DNLS}), which is non-generic 
among nonlinear lattice equations in the sense 
that it has in addition to the Hamiltonian 
\begin{equation}
\label{DNLSham}
{\mathcal H} \left( \left\{\psi_n\right\}, 
\left\{-\ii \psi_n^\ast \right\}\right) = 
\sum_{n=1}^N \left ( -\delta|\psi_n|^2+ \frac{\sigma}{2}|\psi_n|^4
+C |\psi_{n+1}- \psi_n |^2  \right ) 
\end{equation}
also a second conserved quantity, the {\em excitation norm}
\begin{equation}
\label{norm}
{\mathcal N } = \sum_{n=1}^N |\psi_n|^2, 
\end{equation}
it has recently been shown 
\cite{RCKG00} that the nature of the asymptotic dynamics for typical 
initial conditions depends critically 
on the values of these two conserved quantities. Redefining the 
Hamiltonian (\ref{DNLSham}) as 
${\mathcal H}^\prime={\mathcal H}+(\delta-2C){\mathcal N }$ (equivalent to 
applying the gauge transformation 
$\psi_n\rightarrow \psi_n \e^{-\ii(\delta-2C)t}$ to the conjugated variables 
$\left\{\psi_n\right\}, \left\{-\ii \psi_n^\ast \right\}$), a phase transition 
was found to occur along the line
\begin{equation}
\label{phase}
\frac{{\mathcal H}^\prime}{N } = 
\sigma \left(\frac{\mathcal N }{N}\right)^2
\end{equation}
in the thermodynamic limit $N\rightarrow \infty$. On one side of this line
(below for $\sigma > 0$ and above for $\sigma < 0$), the system thermalizes 
in a standard Gibbsian sense with well-defined temperature and chemical 
potential, and the equilibrium probability distribution function (PDF) for 
$|\psi_n|^2$ decays rapidly for large $|\psi_n|^2$ so that 
large-amplitude excitations are highly improbable. However, on the other side 
of the line (\ref{phase}) the system exhibits a negative-temperature type 
behaviour where the stationary PDF for $|\psi_n|^2$ shows a positive 
curvature, corresponding to the creation of persistent localized 
high-amplitude standing breathers. The phase transition line 
corresponds to the limit of infinite temperature in the Gibbsian regime, and 
was obtained by assuming independent thermalization at each site, \ie, by 
neglecting the 
coupling term in the Hamiltonian. Now, we note that for a type $H$ SW with 
$Q=\pi/2$ the coupling term is exactly zero, since according to the code 
sequence (\ref{hull}) such a wave must have the structure 
$\psi_{2m+1}=(-1)^m \sqrt{\frac{\delta}{\sigma}}$, $ \psi_{2m}=0$ for all 
$\delta$. Thus, {\em the $Q=\pi/2$ SW lies exactly on the phase transition 
line} (\ref{phase}). Moreover, as illustrated in Fig.\ \ref{fig_phase}, this 
implies that type $H$ SWs with a given $Q$ always belong to the same 
well-defined phase regardless on the value of $\delta^\prime$, and that 
{\em SWs with $Q<\pi/2$ and those with $Q>\pi/2$ always belong to different 
phases}.\footnote{Note that this is not in general true for propagating waves, which 
for a given 
wave vector $Q$ can cross the transition line (\ref{phase}) as the amplitude 
is varied \cite{RCKG00}.} By regarding a SW with wave vector 
$Q=\pi r^\prime/s^\prime$ as a multibreather with $r^\prime$ 
breathers (or, equivalently, $s^\prime-r^\prime$ discommensurations) in 
each (half-)period of $s^\prime$ sites, we obtain close to the anticontinuous 
limit (\ie, asymptotically for large $\mathcal{N}/N$):\footnote{This 
expression is exact for $Q=\pi/2$ and $Q=\pi$.}
\begin{equation}
\frac{\mathcal{H}^\prime}{N}\approx\sigma\frac{s^\prime}{2r^\prime}
\left(\frac{\mathcal N }{N}\right)^2 + 2C \frac{2r^\prime-s^\prime}{r^\prime}
\Theta(2r^\prime-s^\prime)\left(\frac{\mathcal N }{N}\right) ,  
\label{NH}
\end{equation}
where $\Theta(x)$ is the usual step function. 

\begin{figure}
\centerline{\includegraphics[height=8.5cm,angle=270]{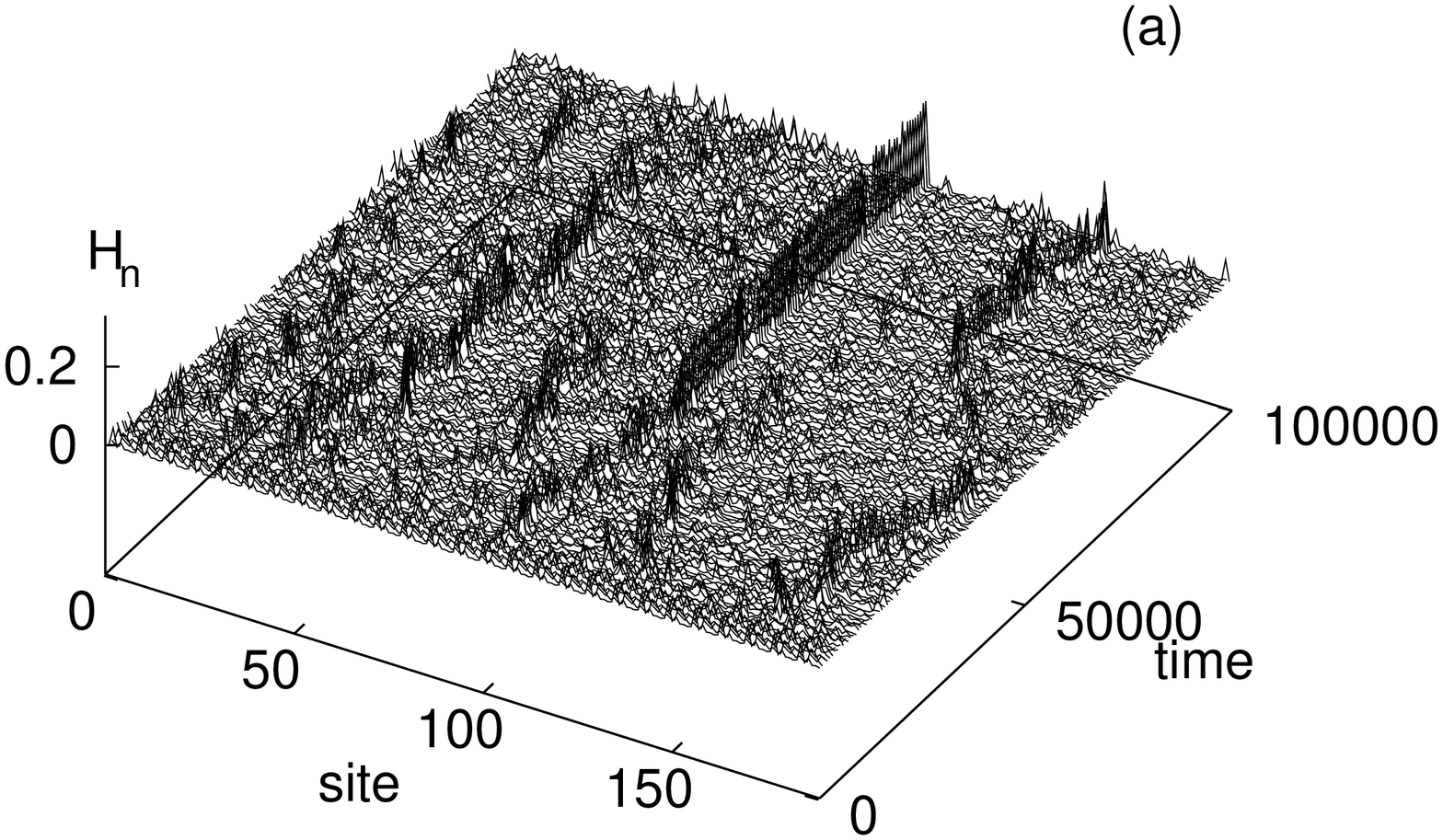}
\includegraphics[height=8.5cm,angle=270]{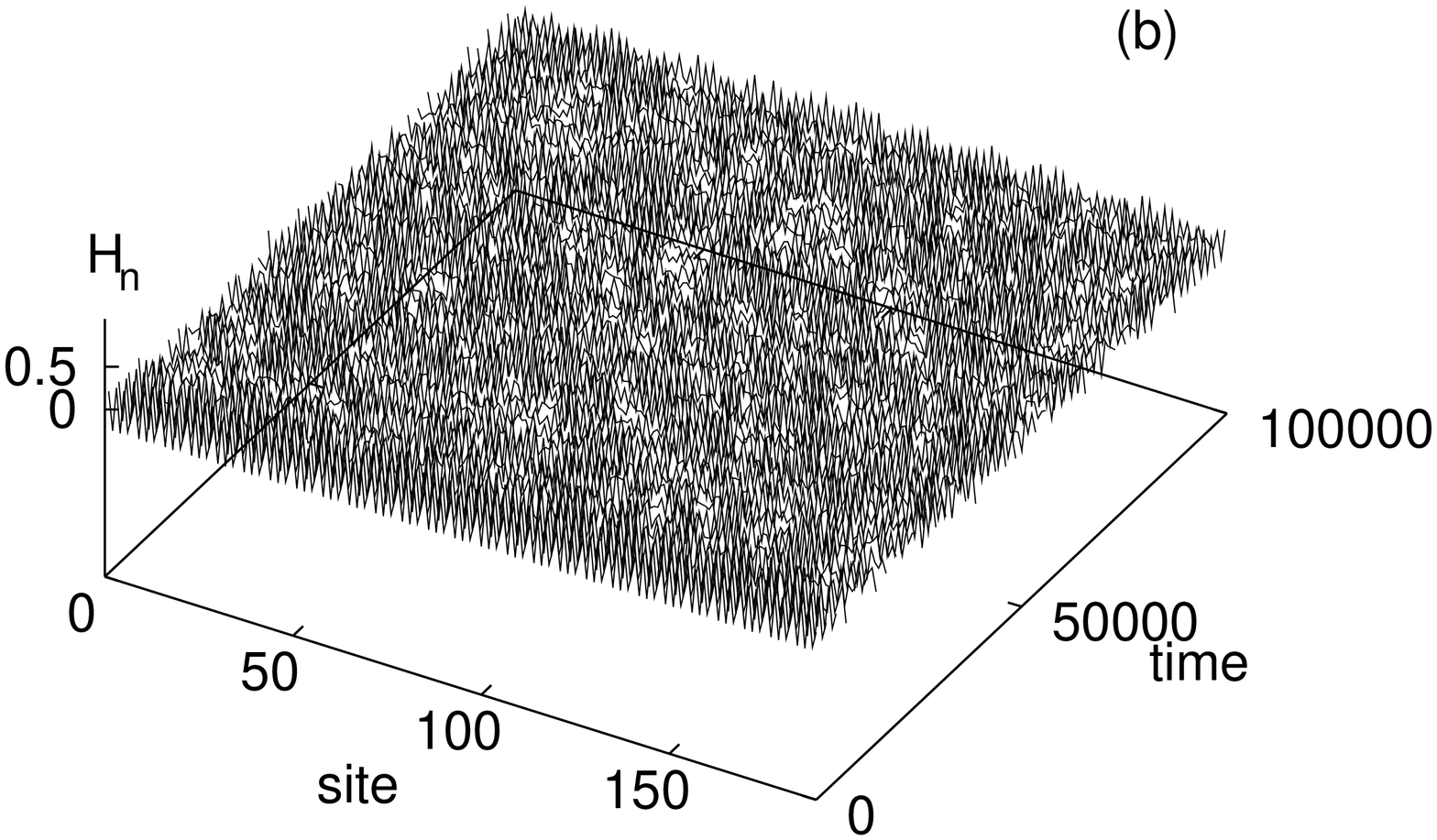}}
\caption{Long-time dynamics of slightly perturbed unstable type $H$ SWs in a 
Morse KG lattice with $C_K=0.035$ and $N=189$. (a) shows breather formation 
for $Q=2\pi/7 < \pi/2$ and $\omega_b=0.995$, while (b) shows thermalization 
for $Q=6\pi/7 > \pi/2$ and $\omega_b=1.035$. }
\label{fig_formation}
\end{figure}

Thus, within the DNLS approximation and assuming a soft 
potential ($\sigma=-1$), we should expect for large 
systems to always observe an equipartitioned 
asymptotic state for all initially slightly perturbed unstable type $H$ SWs 
with 
$Q>\pi/2$, while persistent breathers should be created from SWs with 
$Q<\pi/2$ (with the opposite conclusion for hard potentials). We confirmed 
this by numerical simulations, although for $Q$ close to but smaller than 
$\pi/2$ the system 
size $N$ needs to be rather large to allow breather formation. 
Moreover, we confirmed numerically that also for the 
original KG lattices there appears to be a phase transition at $Q=\pi/2$ 
between a thermalized phase and a 'breather phase' 
(Fig.\ \ref{fig_formation}).  
However, it should be stressed that the results derived in \cite{RCKG00} 
depend crucially on the existence of a second conserved quantity, and 
it is therefore questionable whether such a transition strictly speaking 
occurs also in other models, or if the 'breather phase' observed here for 
KG models only describes the dynamics over a long but finite time range, 
while 
asymptotically thermalization will occur also there. In any case, 
over the time scales that we have been able to follow numerically we have 
seen no signs of breather decay, and thus it is clear that the phase 
transition obtained in the DNLS approximation is of large relevance also for 
the long-time dynamics in the KG models. 

\section{Conclusions}

In conclusion, we have shown that for a general class of anharmonic 
lattices, different types of SWs with spatial period commensurate or 
incommensurate with 
the lattice exist,  but are generically unstable through oscillatory 
instabilities for small amplitudes. Thus, in discrete systems, the 
nonlinearity-induced coupling between counter-propagating waves 
leads to wave  break-down for arbitrarily weak anharmonicity {\em even if 
the individual propagating waves are modulationally stable}. The physical 
mechanisms of the oscillatory instabilities can be related to a two-component 
structure of the SWs, which however is of different nature close to the 
anticontinuous and to the linear limits, respectively. In the first case, 
the natural interpretation is in terms of the multibreather description 
(\ref{code})-(\ref{hull}) of 
the SW, with 
(quasi-)periodically repeated sites of large (code $\pm1$) and small 
(code $0$) amplitude, which defines a two-sublattice structure. 
Then, for small inter-site coupling the eigenmodes of the linearized 
equations describing the 
small-amplitude oscillations around the exact SW are mainly localized on one 
of these two sublattices, and the instabilities appear when, as the coupling 
is increased, eigenfrequencies belonging to the different sublattices start 
to coincide. On the other hand, close to the linear limit, the oscillatory 
instabilities arise from overlap between positive- and negative-frequency 
plane-wave solutions to the linearized equations (\ref{eigeneqns}) in the 
DNLS-approximation, which are coupled by the SW amplitude $\psi_n^2$. 
\footnote{The existence of a regime of two-channel linear phonon scattering 
in the DNLS equation was also discussed in Ref.\ \cite{Kim}; however, in 
contrast to what was claimed in these papers this regime does not exist for 
a single localized breather (the overlap described by Eq.\ (\ref{linspect}) 
vanishes as $Q\rightarrow 0$).}

It is instructive to point out some aspects where the stability analysis of 
SWs considerably differs from that of propagating waves. Firstly, for 
propagating waves there is no two-component structure at the anticontinuous 
limit, as their amplitude is homogeneous with all sites having codes 
$|\tilde{\sigma}_n|=1$ \cite{CA97}. Secondly, propagating waves are 
non-time-reversible with a complex amplitude $\psi_n=\e^{\ii Q n}$ in the 
DNLS approximation, which modifies the eigenvalue problems 
(\ref{eigeneqns})-(\ref{eigen_prob}) \cite{Eilbeck}. Thirdly, the spatial 
Fourier expansion (\ref{psi_esp}) of a propagating wave in the DNLS 
approximation trivially contains only one term ($p=1$), so that 
instabilities resulting from higher-order resonances (which are the only 
instabilities appearing for analytic incommensurate DNLS SWs close to the 
linear limit) are excluded. 

We should also stress that in general, there is no direct relation between 
the frequency of the SW and the frequencies of the oscillatory instabilities, 
as the instabilities are related to resonances between two different classes 
of solutions to the linearized equations (with opposite Krein signature), and 
not to the SW frequency itself. Also, as the instabilities for the 
infinite chain result from overlap between continuous bands and not isolated 
eigenvalues, the wavelengths of the unstable modes generally belong to a 
broad spectrum. However, for commensurate waves close to the linear limit, 
we proved in the Appendix that the first-order (strongest) instabilities 
correspond to frequencies and wavelengths close to, but different from, 
those of the SW itself ($\omega\approx 0, |q| \approx Q$, 
cf.\ Eq.\ (\ref{coninst})). Moreover, we showed by perturbation theory that 
the strength of the oscillatory instability (\ie, its growth rate) close to 
the linear 
limit is proportional to the corresponding spatial Fourier component $|f_p|$ 
of its squared DNLS amplitude $\psi_n^2$, and thus initially increases with 
the amplitude. The instability persists until, in the highly nonlinear 
regime, the frequencies of the two subclasses of linearized modes become out 
of resonance (through an inverse Krein bifurcation), so that close to the 
uncoupled limit the instability is suppressed (cf. the similar scenario for 
the discrete dark soliton corresponding to $Q\rightarrow\pi$ in \cite{JK99}). 

Obviously, this paper has been mainly concerned with fundamental properties 
of the SWs and their instabilities in generic but simple models, and more 
work needs to be done in order to elucidate their role in real physical 
systems, \eg, whether they have some influence on thermal transport 
coefficients. However, one result obtained here (Sec.\ \ref{sec_dyn}) 
that should be emphasized in 
this context is, that the partition of the phase space for the asymptotic 
dynamics of the DNLS model 
(due to the norm conservation) in two 
regimes found in Ref.\ \cite{RCKG00}, 
corresponding respectively to a Gibbsian thermal equilibrium and 
formation of persisting breathers, is given a clear physical interpretation 
in terms of the SWs: the transition line (\ref{phase}) 
is defined by a (type $H$) SW with wave vector $Q=\pi/2$, and SWs with 
given $Q$ always belong to the same regime regardless of their amplitude. 
More work should be done in order to elucidate the relevance of this 
transition for the KG system, \eg, to see if the partition of the phase space 
is fully broken in the absence of norm conservation, and to determine typical 
lifetimes of 'special' configurations like breathers and SWs. In this 
context, we also want to draw the attention to the very recent Ref.\ 
\cite{Lichtenberg}, where the processes of breather formation, coalescence 
and evolution to energy equipartition  were analyzed for the FPU chain. 
There, it was found that exciting initially a single linear mode, the initial 
state first relaxes to an anharmonic spatially periodic solution, which 
we interpret as the FPU counterpart of our SWs. Later, this solution breaks 
up and the periodicity is lost, which we believe is related to instabilities 
similar to those obtained in our work. Thus, the SWs
 and their instabilities 
apparently provide an important link in the understanding of the pathway 
to energy localization and equipartition in anharmonic lattices.

Concerning possible experimental observation, the general 
nature of our results suggests that these instabilities should be observable 
in macroscopic as well as microscopic contexts. In particular, as was 
recently experimentally verified \cite{waveguides}, nonlinear 
optical waveguide arrays provide a direct application of the DNLS equation, 
and thus these systems are also good candidates for detection  of SW 
instabilities. 

\begin{ack}
G. Iooss and K. Kirchg\"assner are acknowledged for sending us their 
preprints prior to publication. We also thank A.J.\ Lichtenberg for 
discussions, and the two referees for useful remarks. GK thanks CNRS and LLB 
for their hospitality. 
MJ acknowledges a Marie Curie Research Training Grant from the European 
Community. 
\end{ack}

\appendix

\section{Linear stability analysis of small-amplitude standing waves 
close to double resonances} 

\subsection{Points of simultaneous resonances}

Consider the ellipses (\ref{ellips}) describing, for each $p$, 
the points in the $(\omega,E$)-plane where $p$th order resonances of the 
second kind (\ref{degen2}) occur. On each such ellipse, there are special 
points where also the first resonance condition (\ref{eq:qres}) is fulfilled, 
so that both the above resonances exist simultaneously.
These points are obtained for special values of $q$ in the parametrized
representation  of the ellipse $(\omega_{p}(q), E_{p}(q))$
defined by Eqs.\ (\ref{omegcond}) and (\ref{Econd})
with parameter $q=p^{\prime}Q$  ($p^{\prime}$ integer). 
Then, a gap with a width of the order of
$|f_{p^{\prime}}|$ opens in the ellipse, which makes this perturbed ellipse 
discontinuous. Thus, even if the ellipse (\ref{ellips}), describing the center 
of the $p$th order gap openings of the second kind, intersects the real 
axis $E=0$ so that an instability should be expected (\ie, if  condition 
(\ref{interEeq0}) is fulfilled), this disconnected curve might not intersect 
the real axis if $E=0$ is close to a point of simultaneous resonance, 
so that there is not necessarily an instability.
However, since $f_{p}$ decreases exponentially
as a function of $p$, only the gaps generated for $|p^{\prime}| \leq |p|$ 
makes a significant discontinuity in the ellipse of $p$th order gaps. Thus, as 
the number of significant discontinuities in each ellipse is finite, 
we expect that only in exceptional cases a predicted gap opening around $E=0$ 
might be prohibited, and consequently the condition (\ref{interEeq0}) for 
$p$th order instability should be generic. 

The values of $Q$ for which simultaneous resonances of orders $p^\prime$ 
and $p$ occur 
on the axis $E=0$ are obtained by vanishing 
Eq.\ (\ref{Econd}):
\begin{equation}
	2C- \delta -2f_{0}-C \cos{(p^{\prime}+2p) Q}-
	C \cos{ p^{\prime} Q} =0 . 
	\label{eq:dblres}
\end{equation}
In the zero-amplitude limit $f_0=0$, $\delta=\delta_0(Q)$ 
(Eq.\ (\ref{ellipt})), this condition reduces to
\begin{equation}
	\cos{Q} = \cos{pQ}
\cos{(p+p^{\prime})Q} ,  
	\label{eq:dblres2}
\end{equation}
which is only fulfilled for a discrete set of values of ${Q}$
when $|p|>1$. 
However, for the ellipse (\ref{ellipeq}) with $p=1$, the condition 
(\ref{eq:dblres2}) 
is fulfilled for all $Q$ when $p^{\prime}=-1$. Thus, in the zero-amplitude 
limit there is always a point of double first order resonances at $\omega=0$ 
where the corresponding ellipse is tangent to the axis $E=0$. Consequently, we 
should treat this case separately as done below. 

We will here discuss both commensurate and incommensurate SWs. 
In the commensurate case with $Q=\pi r^\prime/s^\prime$, 
$\{\psi_n^2\}$ is  an $s^\prime$-periodic sequence, 
and $2pQ$ modulo $2\pi$ takes only $s^\prime$ different values. 
Then, the Fourier expansion (\ref{psi_esp}) of  $\psi_{n}^{2}$ will 
be a finite sum, where  $|p|$ runs only from $0$ to $s^{\prime}/2$
when $s^\prime$ is even, and to $(s^\prime-1)/2$ when $s^\prime$ is odd. 
Consequently, there is only a finite  number of ellipses (\ref{ellips}) 
in the commensurate case. Furthermore, it follows from Eq.\ (\ref{hull}) and 
the model symmetries that the sequence $\{\psi_n^2\}$ for a periodic cycle 
will be symmetric either around a lattice site, in which case we choose 
$\psi_n^2=\psi_{-n}^2$, or around a bond center, so that we choose 
$\psi_n^2=\psi_{-n+1}^2$.\footnote{Both symmetries may also exist 
simultaneously, c.f. the discussion in Sec.\ \ref{swams}.} 
In the first case, we have the phase $\phi=0$ 
in Eq.\ (\ref{psi_esp}), so that $f_{p}=f_{-p}$ are real. In the second case, 
$\phi=-Q/2$ in (\ref{psi_esp}) so that $f_{p}=f_{-p}^\ast$ in general would be 
complex, but by redefining the Fourier transforms (\ref{uq}) and 
(\ref{psi_esp}) by changing $n\rightarrow n-\frac{1}{2}$ we again have 
$\phi=0$ and $f_{p}=f_{-p}$ real. 

\subsection{General properties of the band spectrum at $\omega=0$}
\label{subsec2}

Let us consider the eigenequation (\ref{eigeneqharm2}) for the special 
parameter value 
$\omega=0$.\footnote{For convenience, in 
the rest of this Appendix we assume $C=1$ and $\sigma=-1$ without loss of 
generality.} 
Then, it splits
into two independent eigenequations which can be solved
either with $W_{n} \equiv 0$ or $U_{n}\equiv 0$:
\begin{eqnarray}
{\mathcal{L}}_1 U_n = 
(2-\delta -3 \psi_{n}^{2}) U_{n}-U_{n+1}-U_{n-1} &=& E U_{n}
\label{eigeneqharm3} \\
{\mathcal{L}}_0 W_n =  (2-\delta - \psi_{n}^{2}) W_{n}-W_{n+1}-W_{n-1} &=& 
E W_{n} . 
\label{eigeneqharm4}
\end{eqnarray}
Eq.\ (\ref{eigeneqharm3}) can be related to the nonlinear symplectic  map 
(\ref{Map}), for which the linearized equation is
\begin{equation}
{\mathcal{L}}_1 \epsilon_n = (2- \delta - 3 \psi_{n}^{2}) \epsilon_{n} -
\epsilon_{n+1} - \epsilon_{n-1}=0
\label{linmap}
\end{equation}
where $\epsilon_{n}$ is real and time constant.
As well as for Eq.\ (\ref{DNLSHill}), we can associate with
this linear equation 
a real eigenvalue problem defined by 
${\mathcal{L}}_1 \epsilon_n= E \epsilon_{n}$, 
which is nothing but Eq.\ (\ref{eigeneqharm3}).
It was shown very generally in \cite{AMB92} that a necessary and
sufficient condition for a trajectory $\psi_{n}$ of a symplectic map
to be uniformly hyperbolic is that
$E=0$ does not belong to the spectrum of the operator defined
by the second variation of the action, which in our case is ${\mathcal{L}}_1$. 

When $\psi_{n}$ is space periodic, and thus
represented by a periodic cycle in the map defined by Eq.\ (\ref{Map}), 
the spectrum of Eq.\ (\ref{eigeneqharm3})
consists of bands separated by gaps. Then, if (and only if) the cycle is  
hyperbolic $E=0$ belongs to a gap, if it is 
parabolic $0$ is at the edge of a gap, while for
an elliptic cycle $0$ belongs to the inner of a band. 
When $\{\psi_{n}\}$ is represented by a cantorus in
the map (\ref{Map}),
the corresponding trajectory is uniformly hyperbolic
and then $0$ does not belong to the spectrum of Eq.\ (\ref{eigeneqharm3}).
On the other hand, when it is represented by a KAM torus $0$ belongs to the 
spectrum, 
and moreover there is an explicit eigenvector for this eigenvalue.
In that case, $\psi_{n}$  is
a quasiperiodic function of $n$ with an  analytic hull function $\chi_S(x)$ 
defined by Eq.\ (\ref{chi_S}), 
and  Eq.\ (\ref{eigeneqharm3}) exhibits the solution $U_{n}=
\partial \chi_S(Qn +\phi)/\partial \phi$ (the sliding
mode) for the eigenvalue $E=0$.

The second eigenequation (\ref{eigeneqharm4}) has always the explicit solution
$W_{n}=\psi_{n}$ (the phase mode) for the eigenvalue $E=0$. 
When $\psi_{n}^2$ is space periodic with period $s^\prime$, the 
eigensolutions  of Eq.\ (\ref{eigeneqharm4}) are Bloch states with a
wave vector $q_{s^\prime}$ defined modulo $2 \pi/s^\prime$, 
and the corresponding eigenvalues form bands
separated by gaps. Since the solution  $W_{n}=\psi_{n}$ has the
wave vector $q_{s^\prime}=0$ or $\pi$ modulo $2 \pi/s^\prime$, its 
corresponding eigenvalue 
$E=0$ is at the edge of a gap of the spectrum of Eq. (\ref{eigeneqharm4}). 
When $\{\psi_{n}\}$ is space quasiperiodic and
represented by a KAM torus in the map (\ref{Map}), the
spectrum of Eq.\ (\ref{eigeneqharm4}) is a  Cantor set
with infinitely many gaps. Again, $E=0$ is at an edge of
a gap of this Cantor spectrum.

\subsection{First-order perturbative approach for $\omega$ and $E$ close to 
zero}

We now take advantage of these informations for studying the spectrum
of Eq.\ (\ref{eigeneqharm2}) close to the point of double first-order 
resonances $(\omega,E) \approx (0,0)$  in the limit when the SW 
amplitude $\psi_{n}$ is small and thus close to a cosine function.
Then, at the lowest order $p=1$, it is sufficient to assume that 
$f_{0} \neq 0$,
$f_{1}=f_{-1} \neq 0$  and $f_{p}=0$ for $|p| \geq 2$. In analogy with 
Eq.\ (\ref{epsilon}) we set 
\begin{equation}
f_{0}=2f_{1}+\alpha ,
\label{alpha}
\end{equation}
where $|\alpha| \ll f_{0}$ but in general not strictly zero. 
Then, for describing consistently at leading order the band structure in
the neighbourhood of the origin in the  $(\omega,E)$-plane, we should
essentially consider the reduction of Eq.\ (\ref{eigeneqfour}) to 
the subspace involving the four almost degenerate
eigenvectors which are the two components with wave vector $q$ and
the two components with wave vector $q + 2Q$ with 
$q \approx - Q$.
Then, close to the origin $\omega=0,E=0$, the bands
of eigenvalues $E_\lambda(q; \omega)$ are well described by
diagonalization of the $4 \times 4$-matrix

\begin{equation}
\mathbf{M}(q;\omega) = \left(
  \begin{array}{cccc}
a(q)-f_{0} & -\omega & - 3 f_{1} & 0  \\
-\omega & a(q)+f_{0} & 0 & -f_{1}  \\
- 3 f_{1} & 0  &a(q+2Q)-f_{0} &  -\omega  \\
0 & -f_{1} &  -\omega & a(q+2Q)+f_{0}
\end{array} \right)
\label{matrix4-4}
\end{equation}
where
\begin{equation}
	a(q)=2-\delta-2f_{0}-2 \cos{q}
	\label{eq:aq} .
\end{equation}

\subsubsection{Spectrum at $\omega=0$}

When $\omega=0$, we should recover from Eq.\ (\ref{matrix4-4}) the 
properties of the spectrum of Eq.\ (\ref{eigeneqharm3}) 
associated with the tangent map (\ref{linmap}), as well as the 
phase mode of Eq.\ (\ref{eigeneqharm4}). Then, the $4 \times 4$ matrix
(\ref{matrix4-4}) splits into two diagonal blocks of
$2\times 2$ matrices, 
\begin{equation}
\mathbf{L}_{S}(q) = \left(
  \begin{array}{cc}
a(q)-f_{0} &  - 3 f_{1}   \\
- 3 f_{1} &a(q+2Q)-f_{0}
\end{array} \right)
\label{matrix2-2a}
\end{equation}
which yields the spectrum of (\ref{eigeneqharm3}) calculated in
perturbation for $q$ close to $-Q$ as
\begin{eqnarray}
	E_{S\pm}(q)= 2-\delta-3f_{0}- \cos{q}- \cos{(q+2Q)}
\nonumber\\
\pm \sqrt{(\cos{q}- \cos{(q+2Q)})^{2}+9f_{1}^{2}} , 
\label{eq:spect1}
\end{eqnarray}
and
\begin{equation}
\mathbf{L}_{P}(q) = \left( \begin{array}{cc}
a(q)+f_{0} &  - f_{1}   \\
-  f_{1} &a(q+2Q)+f_{0} \end{array} \right)
\label{matrix2-2b}
\end{equation}
which yields the spectrum of (\ref{eigeneqharm4}) as
\begin{eqnarray}
	E_{P\pm}(q)= 2-\delta-f_{0}- \cos{q}- \cos{(q+2Q)}
\nonumber\\
\pm \sqrt{(\cos{q}- \cos{(q+2Q)})^{2}+f_{1}^{2}} . 
\label{eq:spect2}
\end{eqnarray}

The phase mode at $E=0$  corresponds to the eigenvector $(1,1)$
of the matrix (\ref{matrix2-2b}) at $q=-Q$,
which implies
\begin{equation}
	E_{P-}(-Q) = 2-\delta - f_{0}-f_{1}- 2 \cos{Q}=0 ,  
	\label{eq:phsm}
\end{equation}
or 
\begin{equation}
\delta= \delta_0(Q) - f_{0}-f_{1}. 
	\label{deltashift}
\end{equation}
Then, we have 
\begin{equation}
E_{P+}(-Q) = 2 f_{1} >0, 
	\label{eq:phsm2}
\end{equation}
\begin{equation}
E_{S+}(-Q) =  - 2 (f_{0}-2f_{1}) \equiv - 2 \alpha, 
	\label{eq:sldm}
\end{equation}
and 
\begin{equation}
E_{S-}(-Q) = -2(f_{0}+f_{1}) < 0. 
	\label{eq:sldm2}
\end{equation}

For an analytic incommensurate SW,  the sliding mode at $E=0$ 
corresponds to the eigenvector $(1,-1)$ of the matrix (\ref{matrix2-2a}) 
at $q=-Q$. Then, vanishing Eq.\ (\ref{eq:sldm}) 
yields $\alpha=0$ so that $f_{0}=2 f_{1}$, 
and Eq.\ (\ref{eq:phsm}) then yields $2 \cos{Q} = 2-\delta - 3f_{0}/2$.

Otherwise, as explained in Sec.\ \ref{subsec2}, for commensurate SWs 
represented by hyperbolic periodic cycles $E=0$ does not belong to the 
spectrum of the matrix (\ref{matrix2-2a}) for any  $q$ close to $-Q$, 
while for elliptic cycles $E=0$ is in the spectrum of (\ref{matrix2-2a}) 
for some values of $q$ close to (but away from) $-Q$ (as $E=0$ is not at a 
band edge). In both cases, Eq.\ (\ref{eq:sldm}) yields that $\alpha\neq 0$.
 Defining $\xi$ as $q=\xi-Q$, 
$\xi$ small and real, it follows from Eq.\ (\ref{eq:spect1}) that $E=0$ is in 
the spectrum of (\ref{matrix2-2a}) if and only if the equation
\begin{eqnarray}
E_{S+}(\xi-Q)E_{S-}(\xi-Q)=4 \cos^{2} \xi - 4 (2-\delta-3f_{0}) \cos Q 
\cos \xi \nonumber \\
+(2-\delta-3f_{0})^{2}-9f_{1}^{2}+4 \cos^{2}Q -4 =0
\label{eq:xi}
\end{eqnarray}
has a solution for $\xi$ real. 
With  $\cos Q$ taken from Eq.\ (\ref{eq:phsm}),
this equation becomes using (\ref{alpha})
\begin{eqnarray}
(1-\cos \xi)\left( 2+	2 \cos  \xi  -  (2-\delta-3f_{0})
(2-\delta - 3f_{0}/2+\alpha/2)\right) \nonumber \\
= \alpha(3f_{0}-\alpha) .
\label{eq:xi2}
\end{eqnarray}
At leading order for $\xi$ small and  $f_{0}$ small, 
Eq.\ (\ref{eq:xi2}) yields
\begin{equation}
		 \frac{1}{2}\xi^{2}(4-(2-\delta)^{2}) =
	\alpha (3f_{0} - \alpha) . 
	\label{eq:xilo}
\end{equation}
Since  $0< \delta <4$ for all $Q\neq 0$ if $f_0$ is small enough, there is 
a real solution for small $\xi$ only when $\alpha \geq 0$. 
\footnote{Eq.\ (\ref{eq:xi2}) may also have solutions for $\xi \approx 2Q$ 
($q \approx Q$), but these should be discarded since we have assumed 
$q\approx-Q$ in the perturbation calculation leading to 
Eq.\ (\ref{eq:spect1}).}
Consequently, as a perturbative result valid at small amplitude,
$E=0$ belongs to the spectrum of (\ref{matrix2-2a}) for some real $|\xi|>0$ 
when $\alpha > 0 $,  and does not when $\alpha < 0$.
Thus, the commensurate SW is represented by an elliptic
periodic cycle when $f_{0} > 2 f_{1}$ and a hyperbolic periodic cycle when
$f_{0} < 2 f_{1}$.

\subsubsection{Band intersection with $E=0$}

The determinant of the matrix (\ref{matrix4-4}) is
\begin{eqnarray}
D(q;\omega)=\omega^{4} &-&\omega^{2}
\left(a^{2}(q)+a^{2}(q+2 Q)-2f_{0}^{2}
+6 f_{1}^{2}\right) \nonumber \\
&+&\left((a(q)-f_{0})(a(q+2
Q)-f_{0})-9f_{1}^{2}\right) \times \nonumber \\
&& \qquad \left((a(q)+f_{0})(a(q+2
Q)+f_{0})-f_{1}^{2}\right) . 
\label{eq:detmat4-4}
\end{eqnarray}
Vanishing $D(q;\omega)$ yields the intersections with the axis $E=0$ 
of the perturbed bands at wave vector $q$ close to $-Q$. Thus, stability 
requires the  zeros $\omega_s(q)$ of 
$D(q;\omega)$ to be real for all wave vectors $q$ close to $-Q$. At 
$q=-Q$ Eq.\ (\ref{eq:phsm}) yields $a(-Q)=a(Q)=-(f_{0}-f_{1})$, so that 
$D(-Q;\omega)$ has a degenerate zero at $\omega=0$, while 
the second pair of zeros is given by
\begin{equation}
\omega_s^{2}(-Q)= -4f_{1}(f_{0}-2f_{1})=-4f_1\alpha .
\end{equation}
Thus, they are real only when $\alpha<0$, so that for small-amplitude SWs 
represented by elliptic periodic cycles there is always a non-oscillatory 
instability (with purely imaginary $\omega_s$) at wave vector $q=-Q$.

When $q=-Q+\xi$ with $\xi$ small, Eq.\ (\ref{eq:aq}) becomes using again 
(\ref{eq:phsm}):
\begin{equation}
a(\mp Q+\xi) \approx a(Q)\mp 2 \xi \sin Q
	=-(f_{1}+\alpha)\mp 2 \xi \sin Q , 
	\label{eq:1}
\end{equation}
which yields for the determinant (\ref{eq:detmat4-4}) when also $f_1$ is 
small and $\alpha\ll f_1$:
\begin{equation}
D(-Q+\xi;\omega) \approx \omega^4 - 4(2\xi^{2}\sin^{2}Q-f_{1}\alpha) \omega^2
- 48 \xi^{2} \alpha f_{1}\sin^{2} Q .
\label{eq:sumomeg2}
\end{equation}
Thus, at order $\xi^{2}$, Eq.\ (\ref{eq:detmat4-4}) has no real
zeros when 
\begin{equation}
f_1 \alpha \left(f_{1} \alpha
  +  8 \xi^{2}  \sin^{2}Q \right) <0 .
	\label{eq:coninst}
\end{equation}
For commensurate SWs represented by hyperbolic periodic cycles,
$\alpha<0$. Then this condition becomes
\begin{equation}
\xi^{2} > \frac{f_{1} }{8 \sin^{2}Q} |\alpha|, 
	\label{coninst}
\end{equation}
which proves that at small amplitude, the commensurate SWs represented by
hyperbolic periodic cycles  are unstable by oscillatory modulational
instabilities with wave vectors $q$ close to but not equal to $\pm Q$.

Note that this proof does not hold for analytic incommensurate SWs because 
then $\alpha=0$ at order $f_{0}$. In that case, the first-order instabilities 
found in Sec.\ \ref{ISW} for $0<Q<\pi/2$ may occur outside 
the immediate neighborhood of $\omega=0$ where the condition 
$q\approx-Q$ assumed here is not valid. 

%%%%%%%%%%%%%%%%%%%%%%%%%%%%%%%%%%%%%%%%%%%%%%%%%%%%%%%%%%%%%%

\end{document}